\documentclass[12pt,preprint]{aastex}

\shorttitle{Silicate Evolution in Brown Dwarf Disks}
\shortauthors{Riaz, B.}

\begin{document}

\title{Silicate Evolution in Brown Dwarf Disks}

\author{Riaz, B.\altaffilmark{1} }
\altaffiltext{1}{Instituto de Astrof'sica de Canarias, E38205 La Laguna, Tenerife, Spain}

\email{basmah@iac.es}

\begin{abstract}

We present a compositional analysis of the 10 $\micron$ silicate spectra for brown dwarf disks in the Taurus and Upper Scorpius (UppSco) star-forming regions, using archival {\it Spitzer}/IRS observations. A variety in the silicate features is observed, ranging from a narrow profile with a peak at 9.8 $\micron$, to nearly flat, low-contrast features. For most objects, we find nearly equal fractions for the large-grain and crystalline mass fractions, indicating both processes to be active in these disks. The median crystalline mass fraction for the Taurus brown dwarfs is found to be 20\%, a factor of $\sim$2 higher than the median reported for the higher mass stars in Taurus. The large-grain mass fractions are found to increase with an increasing strength in the X-ray emission, while the opposite trend is observed for the crystalline mass fractions. A small 5\% of the Taurus brown dwarfs are still found to be dominated by pristine ISM-like dust, with an amorphous sub-micron grain mass fraction of $\sim$87\%. For 15\% of the objects, we find a negligible large-grain mass fraction, but a $>$60\% small amorphous silicate fraction. These may be the cases where substantial grain growth and dust sedimentation has occurred in the disks, resulting in a high fraction of amorphous sub-micron grains in the disk surface. Among the UppSco brown dwarfs, only usd161939 has a S/N high enough to properly model its silicate spectrum. We find a 74\% small amorphous grain and a $\sim$26\% crystalline mass fraction for this object.  

\end{abstract}

\keywords{stars: abundances --- circumstellar matter --- stars: low-mass, brown dwarfs}    

\section{Introduction}

Several surveys in recent years in the near- and mid-infrared have confirmed the presence of circum(sub)stellar disks around brown dwarfs (e.g., Muench et al. 2001; Liu et al. 2003; Luhman et al. 2005), and indicate a range in disk properties, similar to T Tauri disks. The disk fractions for stellar and sub-stellar members in young star-forming regions are found to be comparable (e.g., Liu et al. 2003; Luhman et al. 2005), and may even exceed the fractions found for higher mass T Tauri stars (e.g., Scholz et al. 2007; Riaz \& Gizis 2008). A key question in the study of disk evolution is the extent of dust processing that has occurred in the disk, as indicated by processes such as grain growth, crystallization and dust settling. In this regard, a detailed compositional analysis of the 10 $\micron$ silicate emission feature has been immensely valuable in gaining insight into such processes, that may eventually lead to planet formation. The most dominant dust species in the circumstellar material are oxygen-rich silicates, that are either olivines (Mg$_{2x}$Fe$_{2(1-x)}$SiO$_{4}$), ranging from fayalite ($x = 0$) to forsterite ($x = 1$), or pyroxenes (Mg$_{x}$Fe$_{(1-x)}$SiO$_{3}$), ranging from ferrosillite ($x = 0$) to enstatite ($x = 1$) (e.g., van Boekel et al. 2005; Kessler-Silacci et al. 2005). Silicate dust in the interstellar medium (ISM) is dominated by amorphous olivine, the 10 $\micron$ spectrum of which shows a narrow peak at 9.8 $\micron$ (e.g., Bouwman et al. 2001). In more evolved systems such as the comet Hale-Bopp or the debris disk around $\beta$ Pictoris, the silicate spectrum is dominated by crystalline enstatite and forsterite (e.g., Crovisier et al. 1997; Bouwman et al. 2001; Okamoto et al. 2004) that show prominent emissivity peaks near 9.3 and 11.3 $\micron$, respectively. An evolutionary sequence from ISM-like dust to cometary dust thus involves a composition change from amorphous to crystalline silicates, indicated by a shift in the peak wavelength of the 10$\micron$ spectrum from $\sim$9.8 to 11.3$\micron$ (e.g., Bouwman et al. 2001; Kessler-Silacci et al. 2005). This is also accompanied by a broadening and flattening of the observed spectrum, as smaller grains coagulate into larger particles and settle to the optically thick disk midplane. Detailed studies on the extent of grain processing in Herbig Ae/Be systems have indicated a lack of ISM-like dust, and a correlation between the grain growth and crystallization processes (e.g., Bouwman et al. 2001; van Boekel et al. 2005). Bouwman et al. (2001) also report on an increase in the abundance of silica (SiO$_{2}$) with increasing crystallinity, another important species formed when amorphous silicates are thermally annealed to form forsterite (e.g., Fabian et al. 2000). Among lower mass ($M_{*}\la1M_{\sun}$) T Tauri systems, a trend of decreasing small and crystalline grain abundances is observed with age, and both the large-grain and crystalline mass fractions are found to increase with the degree of dust settling in the disk (e.g., Sargent et al. 2009; Watson et al. 2009; Bouwman et al. 2008). Honda et al. (2006) report crystallization of 5\%-20\% amorphous silicate grains in T Tauri systems, regardless of the stellar age. Dust compositional analyses of sub-stellar objects indicate a wide range in the large-grain and crystalline mass fractions, with a degree of crystallinity as high as that observed among Herbig Ae/Be stars (e.g., Apai et al. 2005). A more flattened disk structure is observed for the disks around cool stars, suggesting a higher degree of dust processing in these disks compared to higher mass stars (e.g., Apai et al. 2005; Kessler-Silacci et al. 2006; Pascucci et al. 2009).

We present {\it Spitzer} Infrared Spectrograph (IRS) archival observations of brown dwarfs in the Taurus-Auriga  ($\sim$1-2 Myr; Kenyon \& Hartmann 1995) and the Upper Scorpius (UppSco) ($\sim$5 Myr, Preibisch \& Zinnecker 1999) star-forming regions. Combining our work with the results from Apai et al. (2005) in the Chamaeleon I (Cha I) star-forming region, we compare the mass fractions for crystalline and large silicates among brown dwarfs at different ages, as well as with higher mass stars, and discuss any possible correlations observed.

\section{Targets}

Our criteria for selecting targets in Taurus was to search for spectroscopically confirmed members with spectral types (SpT) later than M6, making these more likely to be substellar objects (e.g., Luhman 2004), and for which excess emission has been detected in the {\it Spitzer}/IRAC bands, thus confirming the presence of disks around them. We therefore mainly searched for objects discussed in the works by Luhman et al. (2006) and Guieu et al. (2007). We were able to obtain publicly released {\it Spitzer}/IRS observations for 20 objects. Five of these do not show any emission in the 10$\micron$ silicate feature and have flat spectra, i.e. the emission levels are consistent with the underlying continuum. The 15 objects with emission features are listed in Table 1. Fig.~\ref{cont-subt} shows the normalized continuum-subtracted spectra in units of [($F_{\nu} - F_{c})/F_{c}$] for these objects, along with the 5 flat spectra shown in the bottom panel. The method for normalizing the silicate feature to the continuum level is discussed in $\S\ref{shape}$. In UppSco, Scholz et al. (2007) have reported the presence of a solid-state 10 $\micron$ emission feature in only 3 out of the 13 brown dwarfs with confirmed disk excess. These are also listed in Table 1, with the continuum-subtracted spectra shown in Fig.~\ref{cont-subt}. The {\it Spitzer} PIDs for these observations are 30540, 2, 248, and 20435. We note that four of these objects have SpT of M5.5 (with an uncertainty of $\pm$0.25), but since these have been discussed earlier in the literature (e.g., Luhman et al. 2006; Luhman 2006; Scholz et al. 2007) as brown dwarf members, we have included them in our study.

The IRS spectra presented here were obtained in the low spectral resolution ($\lambda$/$\delta\lambda$ $\sim$ 90) Short-Low [SL] module that covers 5.2-14.5 $\micron$ in 2 orders. The observations were obtained at two nod positions along each slit. The Basic Calibrated Data (BCD) were produced by the S15 pipeline at the Spitzer Science Center (SSC). The sky background was removed from each spectrum by subtracting observations taken in the same module, but with the target in the other order. The spectra were extracted and calibrated using the Spitzer IRS Custom Extraction (SPICE) software provided by the SSC. The extraction was performed using a variable-width column extraction that scales with the width of the wavelength-dependent point spread function. The spectra at the two nod positions were then averaged to obtain a single spectrum. The signal-to-noise (S/N) ratio for the spectra varies strongly with the wavelength, and the noise levels are found to be higher at the red end. For the spectra presented here, the S/N varies between $\sim$8 and $>$50; the more noisy spectra such as 2M04554801 and usco112 have S/N values that are quite low ($\la$10), while it is between $\sim$30 and 40 for MHO 5 and 2M04141188.

\section{Method}

Five dust species were considered in order to obtain a compositional fit to the observed spectra: amorphous olivine and pyroxene, crystalline enstatite, crystalline forsterite, and silica. The characteristics of these are outlined in Table 2, and the spectral profiles plotted in Fig.~\ref{species}. The bulk densities listed in Table 2 have been obtained from Bouwman et al. (2001). The amorphous grains show little dependence on shape (e.g., Bouwman et al. 2001), and thus can be assumed to be homogeneous and spherical so that the standard Mie theory can be applied to determine their spectroscopic properties. The properties of crystalline silicates however are sensitive to the grain shape which makes it important to consider shapes other than homogeneous spheres. Min et al. (2003) have studied the absorption and scattering properties of particles that are inhomogeneous in structure and composition, based on the assumption that the characteristics of such irregularly shaped particles can be simulated by the average properties for a distribution of shapes, such as ellipsoids, spheroids or hollow spheres. We considered both the distribution of hollow spheres (DHS) and the continuous distribution of ellipsoids (CDE) from Min et al. (2003), and found a better match both in terms of the location and width of the spectral profiles using the CDE routine. Therefore the absorption efficiencies of crystalline silicates (enstatite and forsterite) and silica have been calculated using ellipsoidal grains. Since the CDE method has been used, the grains are treated in the ``Rayleigh limit'', i.e. the grain sizes are assumed to be much smaller than the wavelength of radiation. Our model is therefore biased towards sub-micron (0.1 $\micron$) sized crystalline grains, and the effects of grain growth could not be studied for these species. For the case of amorphous silicates, two grain sizes of 0.1 and 2.0 $\micron$ have been used. As discussed in Bouwman et al. (2001), the optical properties of dust grains with a range in sizes between 0.01 and 5 $\micron$ can be characterized by two typical grain sizes: 0.1 $\micron$ for ``small'' grains with sizes $<$ 1 $\micron$, and 2.0 $\micron$ for ``large'' grains with sizes $>$1 $\micron$. 

To construct a model spectrum, we have followed the method outlined in Honda et al. (2003), wherein a power-law source function is assumed and the model flux $F_{\lambda}$ is the sum of the emission from a featureless power-law continuum and the emission from the different dust species. The model spectrum can be written as 

\begin{eqnarray*}
\lambda F_{\lambda} = a_{0} \left(\frac{\lambda}{9.8 \micron}\right)^{n} +~ (a_{1} Q_{0.1ao} ~+~ a_{2} Q_{2.0ao} ~ +~a_{3} Q_{0.1ap} ~+~ a_{4} Q_{2.0ap} \\
~+~ a_{5} Q_{ens} ~+~ a_{6} Q_{fors} ~+~ a_{7} Q_{sil})~ \left(\frac{\lambda}{9.8 \micron}\right)^{m},
\end{eqnarray*}

\noindent where $Q_{0.1ao}$ and $Q_{2.0ao}$ are the absorption efficiencies for 0.1 and 2.0 $\micron$ amorphous olivine, $Q_{0.1ap}$ and $Q_{2.0ap}$ are the absorption efficiencies for 0.1 and 2.0 $\micron$ amorphous pyroxene, and  $Q_{ens}$, $Q_{fors}$, and $Q_{sil}$ are the absorption efficiencies for enstatite, forsterite and silica, respectively. There are 10 free parameters; eight multiplicative factors $a_{0}$-$a_{7}$ in $10^{-14}$ W $m^{-2}$, and two spectral indices of the source function, {\it m} and {\it n}. To fit the spectrum, we used the $\chi^{2}$-minimization method outlined in van Boekel et al. (2005). The best-fit values thus obtained for the multiplicative factors are listed in Table 1, along with the derived mass fractions. Also listed is the reduced-$\chi^{2}$ value ($\chi^{2}$/dof; dof=651) of the fit. The uncertainties for the fit parameters were obtained by considering the errors on the observed fluxes, as well as the S/N ratio. We used the Monte Carlo method for error estimation as outlined in van Boekel et al. (2005), and discussed further in Juhasz et al. (2009). In this method, random Gaussian noise is added to the spectrum, with 1000 synthetic spectra generated and fitted at each noise level. We considered four different noise levels with a $F_{\nu}^{error}/F_{\nu}^{observed}$ of 0.2, 0.1, 0.01 and 0.001, corresponding to a S/N ratio of 5, 10, 100 and 1000, respectively. The mean of the resulting distribution of all fit parameters then corresponds to the best-fit value, and the errors are derived from the standard deviation.

In order to validate our method, we modeled some known low-mass young objects and compared our results with those published in the literature. Fig.~\ref{validate} shows the observed and modeled spectra for two such objects at SpT$\sim$M6, V410 Anon 13 in Taurus and SSTc2d J161159.9-382337 (SST-Lup3-1) in the Lupus III dark cloud. V410 Anon 13 shows a high abundancy of crystalline silicates but a negligible large grain mass fraction (Sargent et al. 2009). We find a small amorphous, large amorphous and crystalline silicate mass fraction of $40.2\%^{+6}_{-4}$, $0.002\%\pm0.1\%$, and $44.3\%^{+6}_{-5}$ for this object, which is consistent within the uncertainties with the fractions of 46.3\%$\pm$17.2\%, 10.4\%$\pm$20.1\%, and 37.2\%$\pm$11.3\% found by Sargent et al. (2009). Sargent et al. (2006) had earlier reported a 0\% large-grain mass fraction for V410 Anon 13. For SST-Lup3-1, Mer\'{i}n et al. (2007) have reported a $\sim$33\% crystalline mass fraction, and an equal mix of small and large grains from the hot component ($T \sim$ 300 K; high-continuum case) of their compositional fit. We find a small and large amorphous grain mass fraction of $40.3\%^{+3}_{-4}$ and $37.4\%^{+4}_{-4}$ for this object, with a crystalline (sub-micron) mass fraction of $28.3\%^{+4}_{-3}$, similar to the one reported by Mer\'{i}n et al. (2007). Our derived crystalline mass fractions for some of the higher mass T Tauri stars from Sargent et al. (2009) work, as well as the Cha I brown dwarfs from Apai et al. (2005), were also found to be consistent within the uncertainties (better than 8\%) with the fractions reported by these authors. We note that since our model does not include opacities of large crystalline grains, it underestimates the contribution to the total crystalline mass fraction due to large-sized grains. This makes a direct comparison of our 28.3\% crystalline mass fraction for SST-Lup3-1 with the 33\% reported by Mer\'{i}n et al. difficult, since they find the crystalline silicates to be a mixture of small and large grains (in the hot component). We refer to the discussion in Sargent et al. (2009) on the opacities of large crystalline grains. These authors have presented a compositional analysis of T Tauri stars in Taurus, and have also considered only sub-micron crystalline silicates in their model. As explained by these authors, large grains should be considered as heterogenous aggregates, made up of sub-micron sized amorphous silicates, forsterite or enstatite, with the highest abundance of amorphous grains. The opacities of the forsterite components of such large heterogenous grains have been shown by Min et al. (2008) to resemble the opacity of small homogenous forsterite grains, whereas the opacity of the amorphous component of the aggregate is found to resemble that of large amorphous homogenous grains, with a size comparable to that of the heterogenous aggregate. Therefore modeling the crystalline component of a large heterogenous aggregate with small crystalline grains should not result in largely erroneous mass fractions for the crystalline silicates. This, as Sargent et al. (2009) indicate, is consistent with the findings of Bouwman et al. (2008), who have considered three different grain sizes of 0.1, 1.5 and 6.0 $\micron$ in their models, but find the typical sizes from the model fits to be sub-micron for the crystalline silicates, and 6 $\micron$ for the amorphous grains.

\section{Results}
\label{results}

Fig.~\ref{taurus} shows the models fits obtained for the Taurus and UppSco objects, along with the contribution from the individual components. Among the Taurus brown dwarfs, 2M04141760 shows a narrower feature compared to other objects, with a peak in emission near 9.8 $\micron$ (Fig.~\ref{cont-subt}). It exhibits the least signs of dust processing, with a small amorphous silicate mass fraction of $\sim$87$\pm$10\% and large grain and crystalline mass fractions of $\la$10\%. This object is similar to the case of Cha H$\alpha$ 1 that shows a smooth narrow feature with a peak at 9.8 $\micron$ (Apai et al. 2005). CFHT-BD-Tau 8, V410 X-ray 6, CFHT-BD-Tau 6, 2M04400067 and 2M04230607 also show peak positions near 9.8 or 9.3 $\micron$, and have small amorphous silicate mass fractions of $\ga$50\%. V410 X-ray 6 also has the lowest crystalline mass fraction (0.002\%) in the sample. The object 2M04141188  shows a clear peak in emission near 11.3 $\micron$ (Fig.~\ref{cont-subt}) and a broad, flat-peaked feature; we find a $\sim$20\% crystalline mass fraction for this object and roughly equal fractions between 30 and 40\% for the small and large amorphous grains. MHO 5, 2M04414825 and CFHT-BD-Tau 9 also exhibit broad, flat features. MHO 5 and 2M04414825 show the strongest signs of grain growth in their disks, with large grain mass fractions of 64\% and 73\%, respectively, while a higher degree of crystallinity (32\% by mass) must be responsible for the flattening observed in CFHT-BD-Tau 9. High crystalline mass fractions between 40\% and 50\% are found for 2M04290068 and CFHT-BD-Tau 20. The case for  CFHT-BD-Tau 20 however is dubious given its noisy spectrum. It shows the weakest feature among the Taurus members, and has a negligible large grain mass fraction. Among the UppSco objects, only usd161939 has a S/N high enough to obtain good estimates on the mass fractions. Modeling of its silicate spectrum indicates that the disk may still be dominated by small amorphous grains (74\% by mass). A high crystalline mass fraction is also found for usco112 ($\sim$82\%), but similar to CFHT-BD-Tau 20, the spectrum is too weak and noisy which makes the derived mass fractions highly uncertain. 

\subsection{Additional Model Fits}

For the three UppSco objects and the Taurus brown dwarf CFHT-BD-Tau 20, we constructed additional model spectra considering only amorphous silicates (both small- and large-sized) in the model, and excluding all crystalline grains (Fig.~\ref{nolarge}a). This case `(a)' therefore highlights the importance of crystalline silicates in these disks, particularly in the case of usco112 and CFHT-BD-Tau 20 that are found to have high crystalline mass fractions ($\ga$50\%), despite the absence of any clear crystalline features in their observed spectra. In addition, these objects show a negligible amount of large grains in their disks (mass fraction $<$0.1\%). We therefore constructed another set `(b)' of model spectra considering only large-sized amorphous grains (Fig.~\ref{nolarge}a; third panel). This set (b) was also constructed for the objects CFHT-BD-Tau 8, 2M04230607 and 2M04400067 that have large grain mass fractions between 0.002\% and 0.003\% (Fig.~\ref{nolarge}b). We have listed for comparison in Table 1 the reduced-$\chi^{2}$ values for these additional model-fits. The first panels in these figures show the best-fits obtained (based on the lowest reduced-$\chi^{2}$ value) using all five species. The fits for cases (a) and (b) for usco112 result in an increase in the reduced-$\chi^{2}$ value of only 0.03 and 0.08, respectively, indicating that all three could fit the observed spectrum well. The crystalline fraction for this object therefore could be as low as 0\%, and the large grain fraction as high as 100\%. For the case of usd160958, the reduced-$\chi^{2}$ value is the same for the best-fit and the case (a) fit, indicating crystalline silicates to be insignificant in this disk. The fit for case (b) though has a larger reduced-$\chi^{2}$ value of 1.7, and shows that some fraction of small grains is required to fit the observed spectrum. For the case of usd161939, the difference in the reduced-$\chi^{2}$ values for the best-fit and case (a) fit is again very small (0.04), though the case (b) model is not a good fit. Inspecting the fits indicates that some fraction of crystalline silicates should be included to better fit the spectrum near 11.3 $\micron$. The small amorphous fraction for this object could therefore lie between $\sim$74 and 100\%, and the crystalline fraction between 0 and 26\% (Table 1). CFHT-BD-Tau 20 has the flattest feature among Taurus objects. Fits for the cases (a) and (b) result in an increase in the reduced-$\chi^{2}$ value of 0.09 and 0.23, respectively. Inspecting the fits closely again indicates that some crystalline fraction is required to better fit the observed spectrum near 11.3 $\micron$, though the feature is so flat that large grains alone could fit the spectrum. In the case of CFHT-BD-Tau 8, 2M04230607 and 2M04400067 (Fig.~\ref{nolarge}b), the case (b) fits result in an excess emission around $\sim$12 $\micron$, and are also bad fits near $\sim$9 $\micron$ for 2M04230607 and 2M04400067. The differences in the reduced-$\chi^{2}$ values are between $\sim$0.3 and 0.7. A model with large grains only therefore is inadequate to fit these spectra, and the negligible large grain mass fractions as deduced from the best-fits must then be good estimates for these objects.

\subsection{Shape vs. Strength of the Silicate Features: Dependence on Grain Growth, Crystallization and Dust Settling}
\label{shape}

In the discussion that follows, we have excluded the objects usco112, usd160958 and CFHT-BD-Tau 20 due to the large uncertainties in their model-fits and the derived mass fractions. Figure~\ref{plots1}a compares the `shape' and `strength' in the 10 $\micron$ silicate feature for Taurus and UppSco brown dwarfs with sun-like T Tauri stars in Taurus (Pascucci et al. 2009) and Herbig Ae/Be stars (van Boekel et al. 2005). The feature strength can be estimated from the peak normalized flux, $F_{peak}$, above the continuum, while the shape can be estimated by the parameter $F_{11.3}/F_{9.8}$, the ratio of the normalized fluxes at 11.3 and 9.8 $\micron$ (e.g., Bouwman et al. 2001). In order to normalize the silicate feature to the continuum, we first approximated the underlying continuum by connecting the two end points of the spectrum, i.e. the observed fluxes at 8 and 13 $\micron$ (averaging over 0.2 $\micron$ at the end points; Kessler-Silacci et al. 2005). The normalized silicate spectrum is then defined in units of [($F_{\nu}-F_{c}$)/$F_{c}$], where $F_{\nu}$ and $F_{c}$ are the observed and the continuum normalized fluxes, respectively. The continuum-subtracted normalized spectra are shown in Fig.~\ref{cont-subt}, and the values for $F_{peak}$ and  $F_{11.3}/F_{9.8}$ plotted in Fig.~\ref{plots1}a have been calculated from these normalized spectra. Typical uncertainties in $F_{peak}$ are $\pm$0.1 for the brown dwarfs, $\pm$0.3 for the T Tauri stars, and $\pm$0.6 for the Herbig Ae/Be stars, while the uncertainties in the $F_{11.3}/F_{9.8}$ values are $\pm$0.1 for the brown dwarfs and T Tauri stars, and $\pm$0.15 for the Herbig Ae/Be stars. van Boekel et al. (2003) first noted a correlation between the shape and strength of the emission features for a sample of Herbig Ae/Be stars. The correlation was such that spectra with a larger value for the 11.3 to 9.8 flux ratio were found to have a lower peak-over-continuum flux. This was explained by the extent of dust processing in the disk; spectra that are dominated by more processed dust due to grain growth and/or crystallization processes were found to be flatter with a peak wavelength close to 11.3 $\micron$, while spectra dominated by unprocessed ISM-like dust showed more peaked, narrower profiles with a peak wavelength closer to 9.8 $\micron$. Similar correlation have been observed among T Tauri stars (e.g., Pryzgodda et al. 2003; Kessler-Silacci et al. 2005; Pascucci et al. 2009) and sub-stellar objects (Apai et al. 2005; Pascucci et al. 2009). In a strength vs. shape diagram then, the narrow, peaked profiles would occupy the space near the bottom right corner, while with increasing dust processing the feature would move towards the upper left region (e.g., Kessler-Silacci et al. 2006). Apai et al. (2005) further noted a `reversal' in this trend for Cha I brown dwarfs, such that for highly processed dust the feature strength may continue to increase with increasing crystallinity. 

Fig.~\ref{plots1}a shows the Taurus and UppSco brown dwarfs clustered around $F_{peak} \sim$ 1.3 and $F_{11.3}/F_{9.8} \sim$ 1.0, with the features being weaker in these objects compared to most other T Tauri and Herbig Ae/Be stars. The brown dwarf 2M04141760 lies to the right of all other objects in the sample, as expected given the stronger, more peaked feature observed for this object (Fig.~\ref{cont-subt}). We do not observe any of the extreme cases as in Cha I with strong features for highly processed grains. The presence of weaker features in our objects is consistent with the findings from previous works (e.g., Apai et al. 2005; Kessler-Silacci et al. 2006; Sargent et al. 2009; Pascucci et al. 2009): disks around cool stars exhibit flatter silicate emission features, indicating a higher degree of dust processing compared to the disks around higher mass stars. Kessler-Silacci et al. (2007) have explained this by the differences in the location of the silicate emission zone for stars of different luminosities. The 10 $\micron$ feature probes smaller radii ($R_{10} <$ 0.001-0.1 AU) in disks around brown dwarfs than in disks around T Tauri stars ($R_{10} >$ 0.1-3 AU). In a recent compositional analysis of T Tauri disks in Taurus, Sargent et al. (2009) have reported higher abundances of large grains in the ``inner'' disk regions probing radii of $\sim$0.6 AU, than in the ``outer'' regions that probe larger disk radii around 10 AU, suggesting that the grain growth process occurs more rapidly at smaller radii. For the crystalline silicates, the abundances were found to be similar in both regions. Thus due to the smaller radii probed by the 10 $\micron$ feature around brown dwarfs, stronger signatures of dust processing would be observed, that may or may not represent the properties of the bulk silicate dust in the disk (Kessler-Silacci et al. 2007).   

In Figs.~\ref{plots1} b-e, we investigate whether the observed flattening and the shift in the peak position of the silicate features is due to an increase in the grain sizes or a higher degree of crystallinity in the disk. No clear rise in the grain growth or the crystallinity levels is observed towards decreasing $F_{peak}$ values (Figs.~\ref{plots1} b-c), and the fractions are largely dispersed for $F_{peak}$ between $\sim$1.5 and 1.3. Thus neither processes show any correlation with the strength in the silicate feature. One exception is 2M04141760 that shows the narrowest silicate profile, consistent with a small $\sim$8\% and 5\% large and crystalline mass fraction, respectively. In Figs.~\ref{plots1} d-e, both the large grain and crystalline mass fractions are found to increase as the 11.3 $\micron$ flux increases relative to the 9.8 $\micron$ flux. Thus the feature shape seems to be correlated with both processes. This is contrary to the results seen among higher mass objects in Taurus, as Watson et al. (2009) report both the shape and strength to be strongly linked to the crystallinity in these disks, but no strong correlation is found for the large grain abundances. Watson et al. have discussed these spectral indices not to be good tracers of the large grain mass fractions, since the shape-strength correlation is mainly due to the trend of sedimentation and crystallinity, and is thus a good tracer of grain growth only in objects with small crystalline mass fractions. The correlations observed in Figs.~\ref{plots1} d-e nevertheless suggest both the grain growth and crystallization processes to be active in these brown dwarf disks, as both seem to affect the shape of the observed features.

Fig.~\ref{plots2}a shows the extent of flaring in the brown dwarf disks in our sample. Here, we plot the flux densities at 8 and 13 $\micron$, normalized to the flux at 8 $\micron$. These wavelengths mark the end points of the 10 $\micron$ silicate emission feature, and have been used to determine the continuum in the feature. As discussed in Apai et al. (2005), the continuum of the silicate spectrum is defined by the disk geometry, and therefore the normalized flux density at 13 $\micron$ will be higher for a highly flared disk, and will decrease as the disk geometry gets flatter. Disks dominated by pristine ISM-like dust show flared structures, while flatter structures are observed for the ones that have gone through substantial grain processing. The flattening is mainly caused by gravitational settling of dust grains from the optically thin upper atmosphere to the optically thick disk midplane. The processing of dust into crystalline silicates as well as growth to larger sizes both effect the vertical structure of the disk. Sargent et al. (2009) and Watson et al. (2009) have reported an increase in the mass fractions for both crystalline and large silicates as the degree to which the disks are sedimented increases. Among the Taurus brown dwarfs, 2M04141760 shows the most flared geometry, consistent with an amorphous sub-micron grain mass fraction of $\sim$87\%. The extent of flaring in this disk is similar to Cha H$\alpha$ 1 that is also dominated by unprocessed dust grains ($<$10\% crystalline mass fraction; Apai et al. 2005). Most other Taurus brown dwarfs, along with the UppSco object usd161939, show intermediate flaring, with the flattest structures observed for MHO 5 and GM Tau. Comparing Figs.~\ref{plots2}a and b, MHO 5 flat geometry is consistent with a high large-grain mass fraction of $\sim$64\%. However, 2M04414825 that shows the strongest signs of grain growth in the sample has a more flared disk than MHO 5 or GM Tau. Similar is the case for 2M04290068 that has the highest crystalline mass fraction in the sample but shows intermediate flaring. The brown dwarf GM Tau has the flattest structure but shows nearly equal fractions for large and crystalline silicates. No strong correlation is therefore observed for these fractions with the extent of dust settling in the disk. 

The sedimentation process however is effected by the level of turbulence in the disk. In disks with higher accretion rates, turbulent mixing may occur alongside grain growth and thus prolong the timescales over which large grains settle to the optically thick disk midplane (e.g., Dullemond \& Dominik 2004). As discussed in Dullemond \& Dominik (2008), with decreasing level of turbulence, the level of sedimentation would increase, resulting in an under-abundance of large grains in the upper disk layers. In Fig.~\ref{plots2}c, we observe the opposite for the case of MHO 5 and 2M04414825 that have weaker accretion rates than the rest but show high large-grain mass fractions of $\sim$60-70\%. There are however two more objects in Fig.~\ref{plots2}c that show small fractions of $\sim$20\% at similar accretion rates. The more actively accreting disk 2M04141760 shows a flared geometry, but has a small 8\% large-grain mass fraction. In Fig.~\ref{plots2}d, no clear correlation is observed between crystallization and the disk mass accretion rate, as similar levels of crystallinity are found for $log\dot{M}$ between $\sim$-9 and -11  $M_{\sun}~  yr^{-1}$. Among higher mass T Tauri stars in Taurus, while Sargent et al. (2009) have noted higher abundances for the crystalline and large silicates in the more settled disks, Pascucci et al. (2009) have found similar large-grain fractions at all accretion rates, thus suggesting that turbulence in the disk may have little or no effect on the dust processing mechanisms. A larger number of data points would be valuable to investigate any dependencies on the mass accretion rate. 

Returning to Fig.~\ref{plots2}b, most objects show roughly equal fractions for the large and crystalline grains, again indicating both processes to be important in these disks. However, no clear correlation is observed between the mass fractions, suggesting that the two processes may proceed independently from each other at different rates (e.g., Bouwman et al. 2001; Sargent et al. 2006). In the case of the three Taurus brown dwarfs, CFHT-BD-Tau 8, 2M04230607 and 2M04400067, that have negligible large grain mass fractions, substantial grain growth may have already occurred in these disks, with the larger grains settled towards the disk midplane leaving only the small crystalline dust in the optically thin regions (e.g., van Boekel et al. 2005). Increasing dust sedimentation would thus result in an increase in the fraction of small amorphous grains in the disk surface layers (e.g., Honda et al. 2006; Dullemond \& Dominik 2008). This is consistent with the $>$60\% small amorphous grain fractions found for these objects. These disks may thus be more evolved systems than the rest in the sample. The three brown dwarfs however show intermediate flaring, with the normalized 13 $\micron$ flux densities between 1.03 and 1.18, and thus do not have particularly flat disk structures which would be expected if significant dust settling had occurred in the disks. 
 
 \subsection{Dependence on X-ray Luminosity and Disk Masses}
 
Figures~\ref{plots2}e-f compare the X-ray luminosity with the large-grain and crystalline mass fractions. The X-ray luminosities in the 0.3-10 keV range have been obtained from G\"{u}del et al. (2007). These observations were part of the XMM-Newton Extended Survey of the Taurus Molecular Cloud (XEST) project. Five of these objects have a `variability flag' of 0, i.e. they exhibit only low-level variability with no flares visible in their X-ray spectra. For the brown dwarf V410 X-ray 6, two measurements are plotted. The measurement from XEST exposure number 23, made on March 11, 2001, shows low-level variability (a variability flag of 0), and is plotted as a red open diamond. The measurement from XEST exposure number 24, made on March 12, 2001, has a flag of 2, i.e. clear flaring is observed in the X-ray spectrum (G\"{u}del et al. 2007). The X-ray luminosity is higher by a factor of $\sim$2 for this second measurement. The full range in the X-ray emission for these brown dwarfs are marked by black open diamonds in Figs.~\ref{plots2}e-f. For two other objects, CFHT-CD-Tau 8 and 2M04141188, the 95\% upper limits are plotted. 

Both the large-grain and crystalline mass fractions show some dependence on the strength in the X-ray emission. Fig.~\ref{plots2}e shows an increase in the large-grain mass fractions with increasing X-ray luminosities, while higher crystallinity levels are observed among the weaker X-ray sources (Fig.~\ref{plots2}f). Several works have concentrated on the effects of X-rays as a source of ionisation in protosellar disks (e.g., Glassgold et al. 1997; Ilgner \& Nelson 2006). In order to explain the angular momentum transport mechanism that drives accretion in protostellar disks, Glassgold et al. found that X-ray ionization produces an accreting surface layer that is coupled to the magnetic field, and lies over a deeper, quiescent layer. Ilgner \& Nelson (2006) have considered X-ray irradiation from the central star to be the main source of ionisation, and have studied the vertical diffusion of chemical species, that mimics the effects of turbulent mixing in the disk. These authors have shown that X-ray ionisation induces turbulence that results in a decrease or a complete removal of the ``dead'' zones in the disk. Such zones are defined to be the regions that are too neutral and decoupled from the disk magnetic field for MHD turbulence to be maintained, as opposed to the ``active'' zones that are sufficiently ionised for the gas to be well-coupled to the magnetic field, and thus able to maintain turbulence. The main criteria for turbulent mixing to be effective is a significant presence of heavy metals (magnesium) in the disk surface layers, since the recombination time between metal ions and electrons is longer than the turbulent diffusion timescale. Increasing X-ray irradiation should thus result in more active turbulent mixing in the disk. This would decrease the level of sedimentation and a higher fraction of large-grains would be detected in the upper disk layers, as probed by the silicate feature (e.g., Dullemond \& Dominik 2008). For the case of crystalline silicates, stronger X-ray emission seems to remove crystaline silicates from the disk surface layers. V410 X-ray 6, for which even the quiescent emission is stronger than the rest in Fig.~\ref{plots2}f, shows a nearly 0\% crystalline mass fraction. However, X-ray flaring may be responsible for varying its crystalline structure, as noted recently for the young solar-type star EX Lupi (\'{A}brah\'{a}m et al. 2009). A similar anti-correlation between X-ray emission and crystalline fractions has been reported by Glauser et al. (2008) among higher mass T Tauri stars in Taurus. Amorphization by ion irradiation may be a possible explanation. Such processes have been discussed to explain the absence of crystalline silicates in the ISM (e.g., Bringa et al. 2007; J\"{a}ger et al. 2003; Kemper et al. 2004). Some laboratory experiments have shown low-energy (keV) ions to be efficient in amorphizing silicate dust (e.g., Demyk et al. 2001; J\"{a}ger et al. 2003). Though the amophization timescales estimated by some of these studies are quite long. In the ISM, Kemper et al. (2004) suggest that the amorphization process occurs on a timescale much shorter than the grain destruction timescale, and it would take $\sim$9 Myr to achieve a crystallinity level of 0.4\%. Bringa et al. (2007) estimate a $\sim$70 Myr timescale for amorphization of ISM silicates by heavy-ion cosmic rays at GeV energies. On the other hand, Glauser et al. (2008) suggest that coronal X-ray emission may correlate with other high energetic radiation in effectively destroying the crystalline silicates. This may reduce the timescale over which such processes take place. 

For three objects in our sample, 2M04141188, CFHT-BD-Tau 6 and 2M04414825, we were able to find disk mass measurements of 0.49$M_{JUP}$, 0.85$M_{JUP}$ and 0.79$M_{JUP}$, respectively, from Scholz et al. (2006). The disk masses were derived from the millimeter fluxes for these objects. The estimate for 2M04141188 is a 2-$\sigma$ upper limit. Honda et al. (2006) have noted a weak trend (5\% significance level) of high crystallinity in objects that are weakly accreting and have less massive disks. In Fig.~\ref{plots2}d, CFHT-BD-Tau 6 is a weak accretor but has a more massive disk, and similar crystalline levels, as the more actively accreting 2M04141188. We also do not find a higher abundance of large-grains in the more massive disks; the large-grain fractions for CFHT-CD-Tau 6 and 2M04414825 are 23\% and 73\%, respectively, at a similar disk mass of $\sim$0.8$M_{JUP}$. Disk mass measurements for a larger sample are required to investigate any such trends.

\subsection{Dependence on Spectral Type}

Figs.~\ref{plots3} a-b show a comparison of the large grain and crystalline mass fractions among the Taurus stellar and sub-stellar objects. Data for T Tauri stars in Taurus has been obtained from Sargent et al. (2009) and are the ``warm'' large and crystalline mass fractions, i.e. the fraction in the inner regions of the disk at radii around 0.6 AU. We note here that Sargent et al. have also considered sub-micron crystalline silicates in their models. Therefore, a comparison with their estimated fractions is appropriate. These authors have reported a weak anti-correlation between the large grain mass fraction and the stellar mass, with higher abundances for lower mass stars. However, inspecting their Fig. 23 shows a large spread in the large-grain mass fractions for stellar masses between $\sim$1 and 0.1 $M_{\sun}$. While we do see a similar spread in the large-grain fractions between 0 and $\sim$80\% at all spectral types, the mean and median fractions for brown dwarfs are found to be 23.4\% and 18.2\%, respectively, and are a factor of $\sim$2 smaller than that found for the earlier type stars (44\% and 45\%, Sargent et al. 2009). The mean and median crystalline mass fractions, on the other hand, are found to be comparable among the two sets, 17\% and 20\% for the brown dwarfs, and 17\% and 11\% for the higher mass objects. The spectra shown in Fig.~\ref{nolarge}a have not been included in calculating the mean and median values. Brown dwarfs are expected to show stronger signs of grain growth, as discussed in many previous works (e.g., Kessler-Silacci et al. 2006; Apai et al. 2005). As noted above in $\S\ref{shape}$, we have three extreme cases in our sample, the objects 2M04230607, 2M04400067 and CFHT-BD-Tau 8, that show less than 0.01\% large-grain mass fractions, and these may be in a more advanced stage of disk evolution than the rest. Excluding these objects results in a mean large-grain mass fraction of $\sim$30\%, which is more comparable to the 44\% for higher mass stars. Both the crystallization and grain growth processes contribute towards flattening of the observed features in brown dwarfs. The flatter features do not necessarily imply a higher degree of grain growth in the disks, and thus higher large-grain abundances compared to higher mass stars. In Fig.~\ref{plots3}b, no decline in the crystallinity levels is observed towards the later types, as was earlier noted by van Boekel et al. (2005) for Herbig Ae/Be stars. The median crystalline mass fractions are actually higher by a factor of $\sim$2 in brown dwarfs compared to earlier type stars. As discussed in Kessler-Silacci et al. (2007), while the silicate emission zone is larger and lies farther away from the central star in higher luminosity objects, the more luminous stars are also more capable of heating amorphous silicates located farther away in the disk. Thus the crystallization process should proceed independent of the stellar luminosity, resulting in similar percentages of crystalline silicates across all spectral types.

\subsection{Age Dependence}

At an age of $\sim$1 Myr, 5 out of the 20 Taurus brown dwarfs studied here show flat silicate spectra, resulting in a 75\% detection rate for emission in the 10 $\micron$ feature. A comparable detection rate of 82\% (14/17) was reported recently by Pascucci et al. (2009) for cool star disks (SpT $\sim$ M5-M9) in the $\sim$2-3 Myr Cha I star-forming region. Among the older UppSco systems ($\sim$5 Myr), Scholz et al. (2007) have reported a $\sim$23\% (3/13) detection rate. As discussed by these authors, brown dwarf disks by this age have undergone significant dust evolution, i.e. substantial grain growth followed by dust settling, resulting in featureless silicate spectra for nearly 80\% of the disks. Objects such as usco112 and some others which have flat, very low contrast 10 $\micron$ features could then be transitional objects. The above comparison suggests a decline in the 10 $\micron$ emission detection rate in brown dwarf disks with age. However, given the uncertainty on the ages, it is difficult to confirm any such age dependence. At an age of $\sim$4 Myr, Sicilia-Aguilar et al. (2007) have found 6/19 low-mass stars (SpT G, K, and M) in the Tr37 cluster with marginal emission in the 10 $\micron$ feature, resulting in a detection rate of $\sim$68\%. At a similar age, the detection rate for UppSco brown dwarfs is lower by a factor of $\sim$3. This may suggest a shorter timescale over which grain growth and dust settling occurs in cool star disks. However, this could again be explained by the smaller disk radii probed by the 10 $\micron$ feature around brown dwarfs, that would show stronger dust processing signatures in these objects compared to higher mass stars (Kessler-Silacci et al. 2007; $\S\ref{shape}$). Flat features do not necessarily imply grain growth to sizes larger than 10 $\micron$ in the disk, followed by sedimentation. Other factors such as strong stellar winds could result in a dearth of optically thin material in the upper disk layers, and thus flat silicate spectra (e.g., Bouwman et al. 2001). 

The mean and median crystalline mass fractions for the Taurus brown dwarfs are found to be 17\% and 20\%, respectively. For the Cha I brown dwarfs, the mean and median are 37\% and 46\% (Apai et al. 2005). This suggests an increase in crystallinity with age among sub-stellar objects. For the UppSco brown dwarf usd161939, the crystalline mass fraction is 26\%, higher than the Taurus objects. The trend observed among brown dwarfs is contrary to the declining trend noted for higher mass T Tauri systems, where the median crystalline mass fraction drops from 11\% in Taurus to just 2\% in the FEPS sample (3-5 Myr; Bouwman et al. 2008). The FEPS sample from Bouwman et al. (2008) work constitutes seven pre-main sequence systems, and these are the only systems among the 328 targets studied under the FEPS {\it Spitzer} Legacy program that show signatures of solid-state dust components in their mid-infrared spectra. Watson et al. (2009) explain processes such as collisional grinding of crystalline grains, or amorphization, that may result in an absence of crystalline silicates at older ages. If so, such processes seem to be less prominent in brown dwarf disks. 

The amount of crystalline silicates in a disk is also found to be dependent on the presence of dense disk material in the inner disk regions (Sargent et al. 2006). As discussed by these authors, if crystalline silicates form due to thermal annealing of amorphous grains in the warm inner disk regions, and then radially transported to the outer disk, then the presence of an inner hole in the disk would result in a reduction in the crystalline silicate production. This may be the case for the brown dwarf V410 X-ray 6 in our sample, that shows a nearly 0\% crystalline mass fraction. The sub-micron amorphous grain fraction for this object however is quite high (85\%), and the large grain fraction is only 14\%, indicating that significant dust processing has not yet occurred in this disk. Among the 65 T Tauri systems in Taurus studied by Sargent et al. (2009) and Watson et al. (2009), 5\% are found to have mass fractions of amorphous sub-micron grains $>$95\%, with no signs of crystalline or large silicates. One out of the 20 Taurus brown dwarfs studied here shows a $\sim$87\%$\pm$10\% sub-micron amorphous silicate fraction, resulting in a similar 5\% fraction of the sample still dominated by ISM-like grains. At a young age of $\sim$1 Myr, we see a variety in the silicate features in Taurus, from a narrow profile with a peak at 9.8 $\micron$, still dominated by unprocessed dust, to a broad, flat feature peaked at 11.3 $\micron$ dominated by both large and crystalline grains, to nearly flat, low-contrast features. None of the objects in our study show {\it both} a negligible ($<$ 1\%) large and crystalline mass fraction, and some form of dust processing has occurred in these disks. For half of the young brown dwarfs studied here, we find crystalline mass fractions $\ga$20\%. Honda et al. (2006) have noted the crystallization of 5\%-20\% amorphous silicates in low-mass pre-main sequence stars at a very early stage of or before the T Tauri phase. Thus not only are such mineral formation processes common among stellar and sub-stellar objects, but the timescale over which these take place are also similar, and seem to occur fairly quickly and at an early evolutionary stage of the system. And if these processes subsequently lead to planet formation, then the building blocks are already present in brown dwarf disks at a young age of $\sim$1 Myr.

\section{Summary}
We have conducted a compositional analysis of the 10 $\micron$ silicate spectra in the Taurus and UppSco brown dwarf disks. A comparison of the crystalline and large grain mass fractions among these two data sets, as well as with brown dwarfs in Cha I and higher mass T Tauri stars in Taurus indicates the following:

\begin{enumerate}

\item In comparison with T Tauri and Herbig Ae/Be stars, we find flatter features for the brown dwarf disks, consistent with previous findings of heightened flattening in lower-mass objects.
\item While the strength in the feature shows no correlation with either the large-grain or crystalline mass fractions, the shape is found to be affected by both processes. 
\item The Taurus brown dwarf 2M04141760 shows the least amount of dust processing in its disk, with a small amorphous grain mass fraction of $\sim$87\%. This is consistent with a flared disk geometry observed for this object.
\item For most objects, we find nearly equal fractions of large-grain and crystalline silicates, indicating both processes to be important in these disks.
\item The large-grain mass fractions are found to increase with an increasing strength in the X-ray emission, while the opposite trend is observed for the crystalline mass fractions. 
\item The median large-grain mass fraction for Taurus brown dwarfs is found to be lower by a factor of $\sim$2 compared to higher mass T Tauri systems in Taurus. The median crystalline mass fraction on the other hand is found to be a factor of $\sim$2 higher. 
\item No strong correlation is observed between the extent of dust settling in the disk, and the large-grain or the crystalline mass fractions. 
\item Among the UppSco brown dwarfs, only usd161939 has a S/N high enough to properly model its silicate spectrum. Our modeling indicates a 74\% small amorphous grain mass fraction, and a $\sim$26\% crystalline mass fraction for this object. 
\item A comparison with the Cha I brown dwarfs suggests an increase in crystallinity with age among sub-stellar objects. The detection rate for emission in the 10 $\micron$ silicate feature shows a decline with age, with only 20\% of the brown dwarfs at $\sim$5 Myr showing any detectable emission in the feature.

\end{enumerate}

\acknowledgements
I wish to thank the anonymous referee for many helpful comments and suggestions, J. Stepan for helpful discussions, and the organizers and speakers at the CONSTELLATION X-ray Astronomy Workshop held at Palermo, in particular Dr. E. Feigelson for his talk on the effects of X-ray emission on circumstellar disks. Support for this work was provided by CONSTELLATION grant \# YA 2007. This work has made use of the SIMBAD database. This work is based in part on observations made with the Spitzer Space Telescope, which is operated by the Jet Propulsion Laboratory, California Institute of Technology under a contract with NASA.

\begin{deluxetable}{cccccccccccccccc}
\tabletypesize{\tiny}
\setlength{\tabcolsep}{0.0001in}
\rotate
\tablecaption{Model Fit Parameters}
\tablewidth{1000pt}
\tablehead{
 \colhead{Name} & \colhead{SpT} &
\colhead{Ref\tablenotemark{a}} & \colhead{$F_{peak}$} & \colhead{$F_{11.3}/F_{9.8}$} & \multicolumn{2}{c}{Olivine\tablenotemark{b}}  & \multicolumn{2}{c}{Pyroxene\tablenotemark{b}}  & \colhead{Enstatite\tablenotemark{b}} & \colhead{Forsterite\tablenotemark{b}} & \colhead{Silica\tablenotemark{b}} & \colhead{$\chi^{2}$/dof\tablenotemark{c}} & \colhead{Amorphous\tablenotemark{d}} & \colhead{Large\tablenotemark{e}} & \colhead{Crystalline\tablenotemark{f}} \\
&&&&&\colhead{0.1$\micron$}&\colhead{2.0$\micron$}&\colhead{0.1$\micron$}&\colhead{2.0$\micron$}&\colhead{0.1$\micron$}&\colhead{0.1$\micron$}&\colhead{0.1$\micron$}&&\colhead{Silicates}&\colhead{Silicates}&\colhead{Silicates} \\
}
\startdata

J04141188+2811535	&	M6.25	&	1	&	1.48	&	1.01	&$	2.5	^{+	1.5	}_{-	1.5	}$ & $	0.13	^{+	0.006	}_{-	0.006	}$ & $	3.1	^{+	0.87	}_{-	1.2	}$ & $	0.3	^{+	0.01	}_{-	0.01	}$ & $	1.66	^{+	0.3	}_{-	0.3	}$ & $	2.7	^{+	0.7	}_{-	0.2	}$ & $	1.8	^{+	1.04	}_{-	1.2	}$ & 	1.69	&$	30.9	^{+	10.1	}_{-	11.2	}$ & $	42.9	^{+	4.8	}_{-	5.1	}$ & $	20.3	^{+	4.3	}_{-	2.9	}$ \\
J04141760+2806096	&	M5.5	&	3	&	1.68	&	0.80	&$	90.2	^{+	9.5	}_{-	9.0	}$ & $	0.3	^{+	0.05	}_{-	0.007	}$ & $	44.6	^{+	3.2	}_{-	4.3	}$ & $	0.4	^{+	0.5	}_{-	0.5	}$ & $	5E-04	^{+	1.07	}_{-	0.002	}$ & $	8.6	^{+	2.25	}_{-	2.3	}$ & $	9E-05	^{+	0.34	}_{-	0.7	}$ & 	1.36	&$	86.7	^{+	10.5	}_{-	10.3	}$ & $	8.3	^{+	7.0	}_{-	6.7	}$ & $	5.0	^{+	1.5	}_{-	1.4	}$ \\
V410 X-ray 6	&	M6	&	2	&	1.36	&	0.90	&$	18.9	^{+	0.9	}_{-	1.4	}$ & $	8E-03	^{+	0.06	}_{-	0.02	}$ & $	2.7	^{+	0.85	}_{-	0.8	}$ & $	0.2	^{+	0.012	}_{-	0.02	}$ & $	5E-04	^{+	0.08	}_{-	0.1	}$ & $	1.7E-04	^{+	0.02	}_{-	0.1	}$ & $	9E-05	^{+	0.004	}_{-	0.005	}$ & 	1.42	&$	85.5	^{+	7.7	}_{-	8.6	}$ & $	14.4	^{+	4.9	}_{-	2.6	}$ & $	0.002	^{+	0.2	}_{-	0.4	}$ \\
J04230607+2801194	&	M6.5	&	2	&	1.31	&	0.91	&$	2E-05	^{+	4E-04	}_{-	0.006	}$ & $	4.8E-06	^{+	0.008	}_{-	0.006	}$ & $	4.8	^{+	0.3	}_{-	0.2	}$ & $	3.5E-06	^{+	0.008	}_{-	0.007	}$ & $	0.53	^{+	0.1	}_{-	0.22	}$ & $	2E-05	^{+	3.8E-04	}_{-	0.06	}$ & $	0.76	^{+	0.05	}_{-	0.1	}$ & 	1.72	&$	84.9	^{+	8.1	}_{-	6.5	}$ & $	0.003	^{+	4.0	}_{-	3.4	}$ & $	7.1	^{+	1.4	}_{-	3.1	}$ \\
 Model (b)\tablenotemark{g}	&		&		&		&		&	0					& $	4.8E-06	^{+	5.2E-07	}_{-	5E-07	}$ & 	0					& $	0.5	^{+	0.04	}_{-	0.04	}$ & 	0					&	0					&	0					&	2.44	&	0					& 	100					&	0					\\
J04242090+2630511	&	M7	&	2	&	1.30	&	0.94	&$	0.15	^{+	0.06	}_{-	0.03	}$ & $	0.02	^{+	0.001	}_{-	0.001	}$ & $	1.3	^{+	0.11	}_{-	0.1	}$ & $	4E-03	^{+	0.02	}_{-	0.02	}$ & $	0.032	^{+	0.02	}_{-	0.05	}$ & $	2E-05	^{+	3E-05	}_{-	0.03	}$ & $	0.32	^{+	0.05	}_{-	0.01	}$ & 	1.92	&$	65.9	^{+	11.7	}_{-	11.2	}$ & $	24.5	^{+	15.2	}_{-	15.6	}$ & $	1.1	^{+	0.6	}_{-	2.1	}$ \\
CFHT-BD-Tau 9	&	M6.25	&	3	&	1.33	&	1.07	&$	2E-05	^{+	0.01	}_{-	0.02	}$ & $	1E-06	^{+	1.5E-03	}_{-	5.4E-04	}$ & $	2.4	^{+	0.14	}_{-	0.2	}$ & $	0.03	^{+	0.02	}_{-	0.02	}$ & $	1.4	^{+	0.14	}_{-	0.18	}$ & $	0.8	^{+	0.07	}_{-	0.08	}$ & $	1.2	^{+	0.12	}_{-	0.1	}$ & 	1.52	&$	44.3	^{+	4.6	}_{-	4.9	}$ & $	10.5	^{+	8.0	}_{-	6.6	}$ & $	32.0	^{+	3.6	}_{-	3.7	}$ \\
J04290068+2755033	&	M8.25	&	3	&	1.40	&	0.96	&$	0.4	^{+	0.008	}_{-	0.01	}$ & $	0.01	^{+	0.005	}_{-	0.003	}$ & $	0.4	^{+	0.01	}_{-	0.004	}$ & $	0.01	^{+	0.003	}_{-	0.004	}$ & $	0.5	^{+	0.05	}_{-	0.04	}$ & $	0.4	^{+	0.06	}_{-	0.05	}$ & $	9E-07	^{+	0.004	}_{-	0.0088	}$ & 	1.66	&$	41.3	^{+	3.1	}_{-	2.6	}$ & $	18.6	^{+	6.6	}_{-	5.6	}$ & $	40.1	^{+	4.6	}_{-	3.7	}$ \\
CFHT-BD-Tau 20	&	M5.5	&	3	&	1.26	&	1.07	&$	2E-04	^{+	2E-04	}_{-	0.3	}$ & $	5E-05	^{+	0.008	}_{-	0.03	}$ & $	5.3	^{+	0.7	}_{-	1.12	}$ & $	4E-05	^{+	0.06	}_{-	0.025	}$ & $	4.3	^{+	1.2	}_{-	1.12	}$ & $	3.5	^{+	1.2	}_{-	0.5	}$ & $	1.8	^{+	0.6	}_{-	0.13	}$ & 	1.43	&$	41.7	^{+	8.6	}_{-	10.7	}$ & $	0.014	^{+	9.9	}_{-	6.2	}$ & $	49.9	^{+	13.6	}_{-	10.0	}$ \\
 Model (a)\tablenotemark{h} 	&		&		&		&		&$	3.5	^{+	1.1	}_{-	1.2	}$ & $	5E-05	^{+	8E-06	}_{-	7.5E-06	}$ & $	10.1	^{+	4.3	}_{-	3.8	}$ & $	4E-05	^{+	6E-06	}_{-	7E-06	}$ & 	0					&	0					&	0					&	1.52	&	100					 & $	0.013	^{+	4.6E-03	}_{-	4.2E-03	}$ & 	0					\\
 Model (b)\tablenotemark{g}	&		&		&		&		&	0					&$	5E-05	^{+	0.03	}_{-	0.06	}$ & 	0					&$	1.5	^{+	0.04	}_{-	0.05	}$ & 	0					&	0					&	0					&	1.66	&	0					&	100					&	0					\\
MHO 5	&	M6.2	&	3	&	1.38	&	1.07	&$	1E-03	^{+	1.8	}_{-	0.02	}$ & $	0.7	^{+	0.2	}_{-	0.12	}$ & $	6.6	^{+	0.6	}_{-	0.86	}$ & $	0.9	^{+	0.15	}_{-	0.13	}$ & $	5.7	^{+	0.3	}_{-	0.9	}$ & $	7.1	^{+	0.66	}_{-	0.8	}$ & $	0.78	^{+	0.2	}_{-	0.09	}$ & 	1.06	&$	13.5	^{+	4.0	}_{-	2.1	}$ & $	63.7	^{+	11.2	}_{-	8.8	}$ & $	21.9	^{+	2.5	}_{-	2.7	}$ \\
GM Tau	&	M6.5	&	1	&	1.49	&	0.98	&$	5.6	^{+	2.0	}_{-	1.2	}$ & $	0.11	^{+	0.05	}_{-	0.03	}$ & $	12.0	^{+	1.03	}_{-	0.8	}$ & $	0.4	^{+	0.02	}_{-	0.003	}$ & $	5E-04	^{+	0.1	}_{-	0.14	}$ & $	5.9	^{+	0.35	}_{-	0.6	}$ & $	9E-05	^{+	0.6	}_{-	0.4	}$ & 	1.35	&$	53.6	^{+	8.0	}_{-	5.2	}$ & $	30.2	^{+	3.9	}_{-	2.5	}$ & $	16.2	^{+	1.6	}_{-	1.8	}$ \\
CFHT-BD-Tau 6	&	M7.25	&	1	&	1.33	&	0.99	&$	3E-05	^{+	0.6	}_{-	0.02	}$ & $	1E-06	^{+	0.008	}_{-	0.008	}$ & $	4.0	^{+	0.4	}_{-	0.66	}$ & $	0.09	^{+	0.015	}_{-	0.011	}$ & $	0.4	^{+	0.12	}_{-	0.05	}$ & $	1.4	^{+	0.03	}_{-	0.14	}$ & $	1.0	^{+	0.4	}_{-	0.14	}$ & 	1.42	&$	50.0	^{+	10.6	}_{-	9.5	}$ & $	22.9	^{+	5.0	}_{-	4.0	}$ & $	19.9	^{+	2.4	}_{-	2.5	}$ \\
J04400067+2358211	&	M6.5	&	2	&	1.29	&	0.91	&$	2.4	^{+	0.3	}_{-	0.2	}$ & $	4.8E-06	^{+	6E-05	}_{-	3E-05	}$ & $	0.9	^{+	0.05	}_{-	0.15	}$ & $	1E-06	^{+	0.007	}_{-	0.007	}$ & $	1.04	^{+	0.12	}_{-	0.12	}$ & $	0.4	^{+	0.07	}_{-	0.04	}$ & $	1.1	^{+	0.1	}_{-	0.1	}$ & 	1.24	&$	64.5	^{+	6.6	}_{-	6.6	}$ & $	0.002	^{+	2.7	}_{-	2.7	}$ & $	23.2	^{+	2.6	}_{-	2.4	}$ \\
 Model (b)\tablenotemark{g}	&		&		&		&		&	0					&$	6.5E-05	^{+	9E-06	}_{-	7.5E-06	}$ & 	0					&$	0.6	^{+	0.04	}_{-	0.033	}$ & 	0					&	0					&	0					&	1.70	&	0					&	100					&	0					\\
J04554801+3028050	&	M5.6	&	1	&	1.27	&	0.89	&$	1E-04	^{+	0.003	}_{-	0.02	}$ & $	0.02	^{+	0.008	}_{-	0.002	}$ & $	1.4	^{+	0.03	}_{-	0.2	}$ & $	4E-03	^{+	0.002	}_{-	0.01	}$ & $	0.8	^{+	0.16	}_{-	0.23	}$ & $	2E-05	^{+	3E-05	}_{-	0.03	}$ & $	0.8	^{+	0.3	}_{-	0.2	}$ & 	1.29	&$	46.8	^{+	4.4	}_{-	8.3	}$ & $	17.9	^{+	5.8	}_{-	6.9	}$ & $	20.0	^{+	4.5	}_{-	6.3	}$ \\
CFHT-BD-Tau 8	&	M6.5	&	1	&	1.37	&	0.89	&$	1.6	^{+	0.1	}_{-	0.2	}$ & $	1E-05	^{+	6E-05	}_{-	8E-05	}$ & $	3.8	^{+	0.3	}_{-	0.1	}$ & $	3.5E-06	^{+	0.02	}_{-	0.03	}$ & $	0.9	^{+	0.22	}_{-	0.5	}$ & $	0.95	^{+	0.25	}_{-	0.23	}$ & $	1.0	^{+	0.12	}_{-	0.15	}$ & 	1.23	&$	72.2	^{+	7.0	}_{-	7.7	}$ & $	0.004	^{+	5.4	}_{-	7.7	}$ & $	20.1	^{+	4.0	}_{-	5.9	}$ \\
 Model (b)\tablenotemark{g}	&		&		&		&		&	0					&$	1E-05	^{+	0.009	}_{-	0.009	}$ & 	0					&$	0.9	^{+	0.088	}_{-	0.086	}$ & 	0					&	0					&	0					&	1.54	&	0					&	100					&	0					\\
J04414825+2534304	&	M7.75	&	1	&	1.36	&	0.98	&$	4.4E-03	^{+	0.1	}_{-	0.2	}$ & $	0.5	^{+	0.006	}_{-	0.003	}$ & $	1.8	^{+	0.08	}_{-	0.1	}$ & $	4E-03	^{+	0.035	}_{-	0.02	}$ & $	0.4	^{+	0.15	}_{-	0.1	}$ & $	1.5	^{+	0.32	}_{-	0.21	}$ & $	0.3	^{+	0.007	}_{-	0.02	}$ & 	1.45	&$	13.3	^{+	1.2	}_{-	1.5	}$ & $	73.1	^{+	6.6	}_{-	3.8	}$ & $	12.3	^{+	2.3	}_{-	1.5	}$ \\
usco112	&	M5.5	&	4	&	1.30	&	0.96	&$	9.5E-02	^{+	0.02	}_{-	0.02	}$ & $	4.8E-06	^{+	0.003	}_{-	0.005	}$ & $	1E-06	^{+	0.014	}_{-	0.02	}$ & $	3.5E-06	^{+	0.001	}_{-	0.003	}$ & $	0.5	^{+	0.1	}_{-	0.09	}$ & $	0.1	^{+	0.022	}_{-	0.01	}$ & $	1E-05	^{+	0.004	}_{-	0.001	}$ & 	1.25	&$	17.8	^{+	5.5	}_{-	7.0	}$ & $	0.03	^{+	13.3	}_{-	20.5	}$ & $	82.1	^{+	22.7	}_{-	24.5	}$ \\
 Model (a)\tablenotemark{h} 	&		&		&		&		&$	0.61	^{+	0.10	}_{-	0.08	}$ & $	1E-06	^{+	0.008	}_{-	9E-03	}$ & $	3E-02	^{+	0.008	}_{-	0.01	}$ & $	1E-06	^{+	0.009	}_{-	0.007	}$ & 	0					&	0					&	0					&	1.28	&	100					 & $	0.006	^{+	37.7	}_{-	35.7	}$ & 	0					\\
 Model (b)\tablenotemark{g}	&		&		&		&		&	0					&$	0.03	^{+	0.02	}_{-	0.01	}$ & 	0					&$	0.04	^{+	0.003	}_{-	0.005	}$ & 	0					&	0					&	0					&	1.33	&	0					&	100					&	0					\\
usd160958	&	M6.5	&	5	&	1.33	&	0.82	&$	0.42	^{+	0.2	}_{-	0.2	}$ & $	4.8E-06	^{+	0.004	}_{-	0.006	}$ & $	1.02	^{+	0.09	}_{-	0.11	}$ & $	3.5E-06	^{+	0.0025	}_{-	0.0035	}$ & $	5E-05	^{+	0	}_{-	0.06	}$ & $	0.2	^{+	0.04	}_{-	0.088	}$ & $	0.3	^{+	0.12	}_{-	0.11	}$ & 	1.38	&$	80.9	^{+	19.4	}_{-	18.9	}$ & $	0.009	^{+	5.5	}_{-	7.5	}$ & $	9.0	^{+	2.3	}_{-	5.3	}$ \\
 Model (a)\tablenotemark{h} 	&		&		&		&		&$	2E-05	^{+	7E-06	}_{-	5.5E-06	}$ & $	3E-06	^{+	7E-06	}_{-	7.5E-06	}$ & $	1.65	^{+	0.4	}_{-	0.3	}$ & $	3E-06	^{+	8E-06	}_{-	9E-06	}$ & 	0					&	0					&	0					&	1.38	&	100					 & $	0.007	^{+	0.013	}_{-	0.01	}$ & 	0					\\
 Model (b)\tablenotemark{g}	&		&		&		&		&	0					&$	4E-06	^{+	2E-05	}_{-	1E-05	}$ & 	0					&$	0.09	^{+	0.009	}_{-	0.01	}$ & 	0					&	0					&	0					&	1.69	&	0					&	100					&	0					\\
usd161939	&	M7	&	5	&	1.56	&	1.00	&$	0.62	^{+	0.1	}_{-	0.4	}$ & $	4.8E-06	^{+	0.01	}_{-	0.008	}$ & $	2.1	^{+	0.4	}_{-	0.4	}$ & $	3.5E-06	^{+	0.0045	}_{-	0.003	}$ & $	0.92	^{+	0.24	}_{-	0.12	}$ & $	0.3	^{+	0.07	}_{-	0.006	}$ & $	1E-05	^{+	0.05	}_{-	0.04	}$ & 	1.64	&$	74.1	^{+	15.6	}_{-	18.2	}$ & $	0.005	^{+	5.3	}_{-	4.5	}$ & $	25.9	^{+	6.3	}_{-	4.7	}$ \\
 Model (a)\tablenotemark{h} 	&		&		&		&		&$	2.04	^{+	1.0	}_{-	0.9	}$ & $	6.5E-06	^{+	7E-06	}_{-	8.5E-06	}$ & $	1.9	^{+	0.2	}_{-	0.4	}$ & $	7.3E-06	^{+	5.8E-06	}_{-	4.8E-06	}$ & 	0					&	0					&	0					&	1.68	&	100					 & $	0.007	^{+	5E-03	}_{-	5.2E-03	}$ & 	0					\\
 Model (b)\tablenotemark{g}	&		&		&		&		&	0					&$	0.3	^{+	0.04	}_{-	0.05	}$ & 	0					&$	0.18	^{+	0.035	}_{-	0.015	}$ & 	0					&	0					&	0					&	2.13	&	0					&	100					&	0					\\

\enddata

\tablenotetext{a}{Reference for spectral type: (1) Luhman 2004, (2) Luhman et al. (2006), (3) Luhman (2006), (4) Ardila et al. (2000), (5) Mart\'{i}n et al. (2004). Uncertainty in spectral type is $\pm$0.25.}
\tablenotetext{b}{Best-fit values for the multiplicative factors in units of $10^{-14}$ W $m^{-2}$.}
\tablenotetext{c}{The reduced $\chi^{2}$-value (dof=651).}
\tablenotetext{d}{Percentage of small amorphous olivine and pyroxene silicates compared to all other silicates.}
\tablenotetext{e}{Percentage of large amorphous olivine and pyroxene silicates compared to all other silicates.}
\tablenotetext{f}{Percentage of sub-micron crystalline silicates (enstatite and forsterite) compared to all other silicates.}
\tablenotetext{g}{Results from remodeling considering large amorphous grains only.}
\tablenotetext{h}{Results from remodeling without any crystalline silicates.}

\end{deluxetable}

\begin{deluxetable}{ccccccc}
\tabletypesize{\scriptsize}
\tablecaption{Characteristics of the grain species used for the fitting procedure.}
\tablewidth{0pt}
\tablehead{
\colhead{Species}  & \colhead{Composition} & \colhead{Size [$\micron$]} & \colhead{Shape} & \colhead{Reference} & \colhead{$\rho_{b}$ [$g ~cm^{-3}$]}
} 
\startdata
Amorphous olivine & MgFeSiO$_{4}$ & 0.1 \& 2.0 & Homogeneous spheres & Dorschner et al. (1995) & 3.71\\
Amorphous pyroxene & Mg$_{0.8}$Fe$_{0.2}$SiO$_{3}$ &  0.1 \& 2.0 & Homogeneous spheres & Dorschner et al. (1995) & 3.71 \\
Crystalline forsterite & Mg$_{1.9}$Fe$_{0.1}$SiO$_{4}$ & 0.1 & Irregular (CDE) & Fabian et al. (2001) & 3.33\\
Crystalline enstatite & MgSiO$_{3}$ & 0.1 & Irregular (CDE) & J\"{a}ger et al. (1998) & 2.80\\
Amorphous silica & SiO$_{2}$ & 0.1 & Irregular (CDE) & Fabian et al. (2000) & 2.21\\
\enddata
\end{deluxetable}

\clearpage

\begin{figure}
 \begin{center}
    \begin{tabular}{ccc}      
      \resizebox{50mm}{!}{\includegraphics[angle=0]{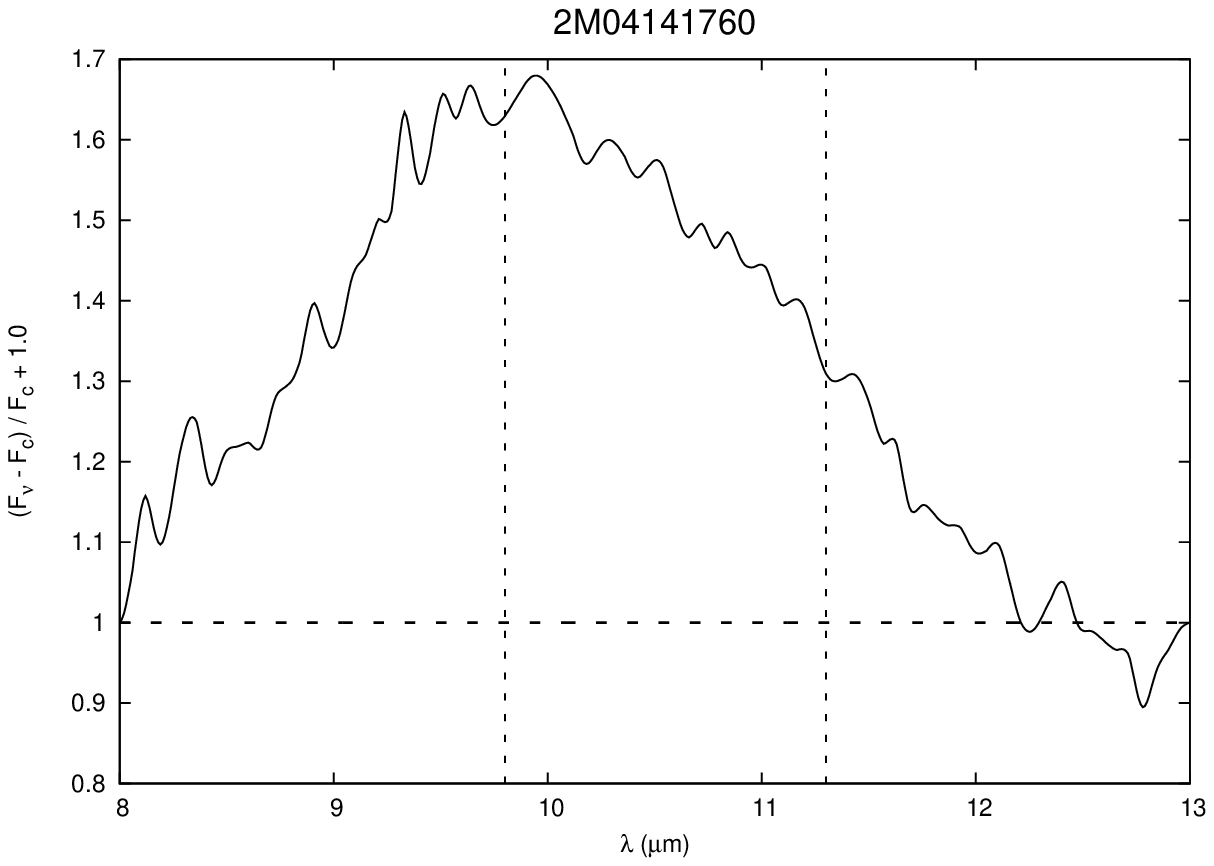}} &    
      \resizebox{50mm}{!}{\includegraphics[angle=0]{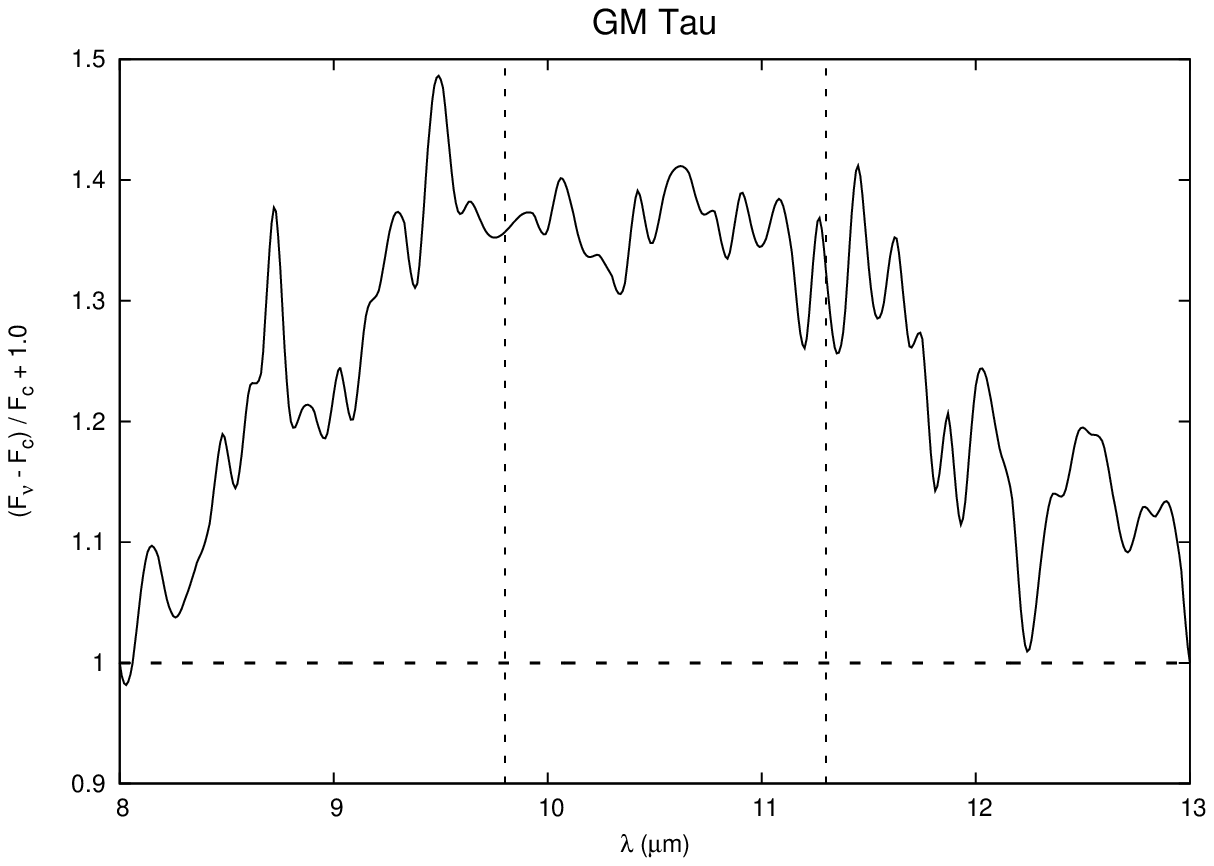}}  &   
      \resizebox{50mm}{!}{\includegraphics[angle=0]{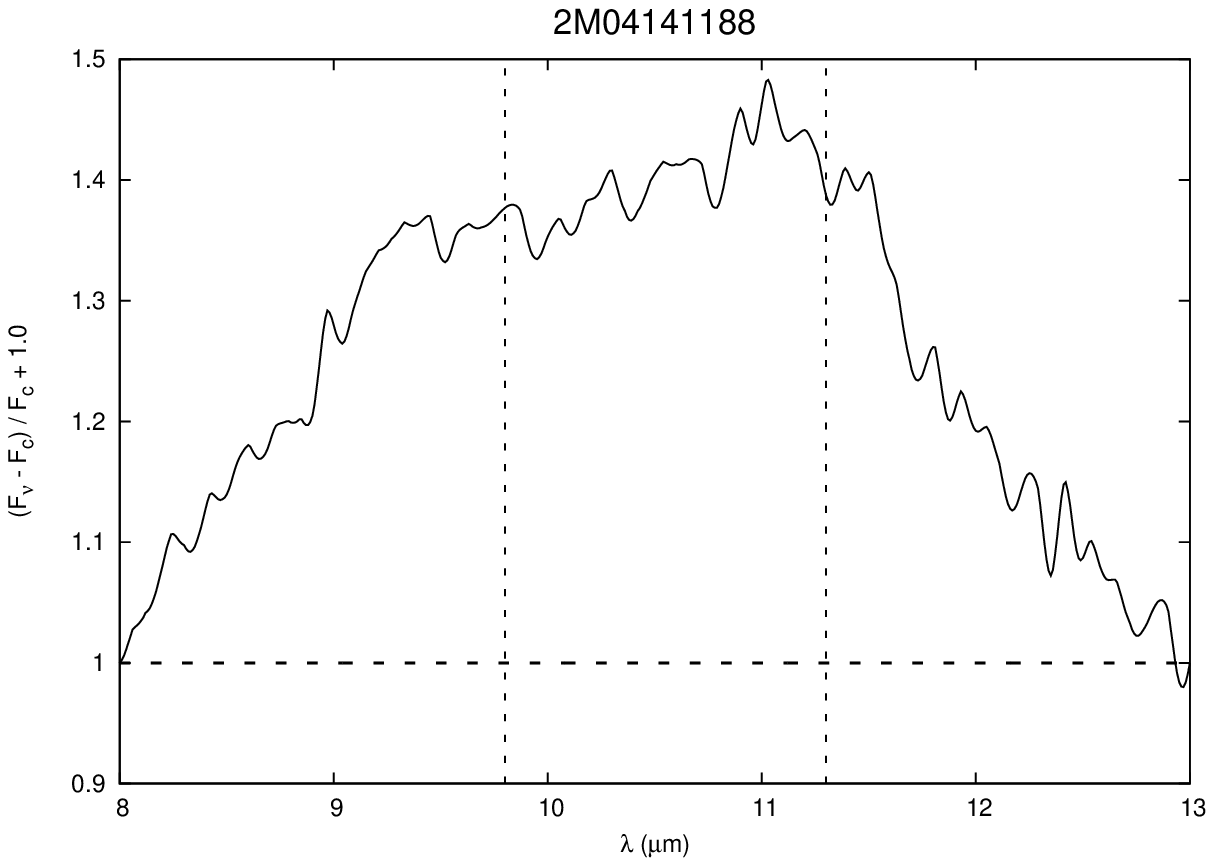}} \\
      \resizebox{50mm}{!}{\includegraphics[angle=0]{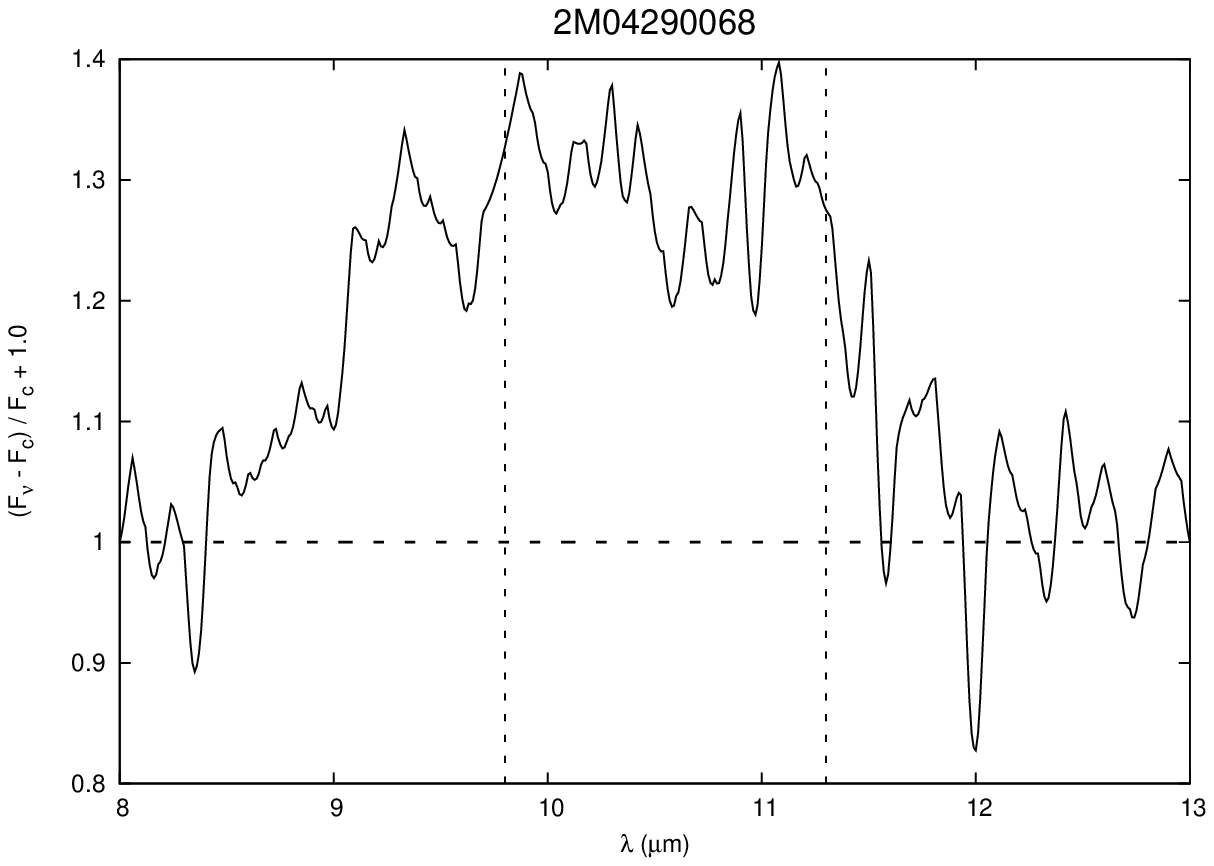}} &
      \resizebox{50mm}{!}{\includegraphics[angle=0]{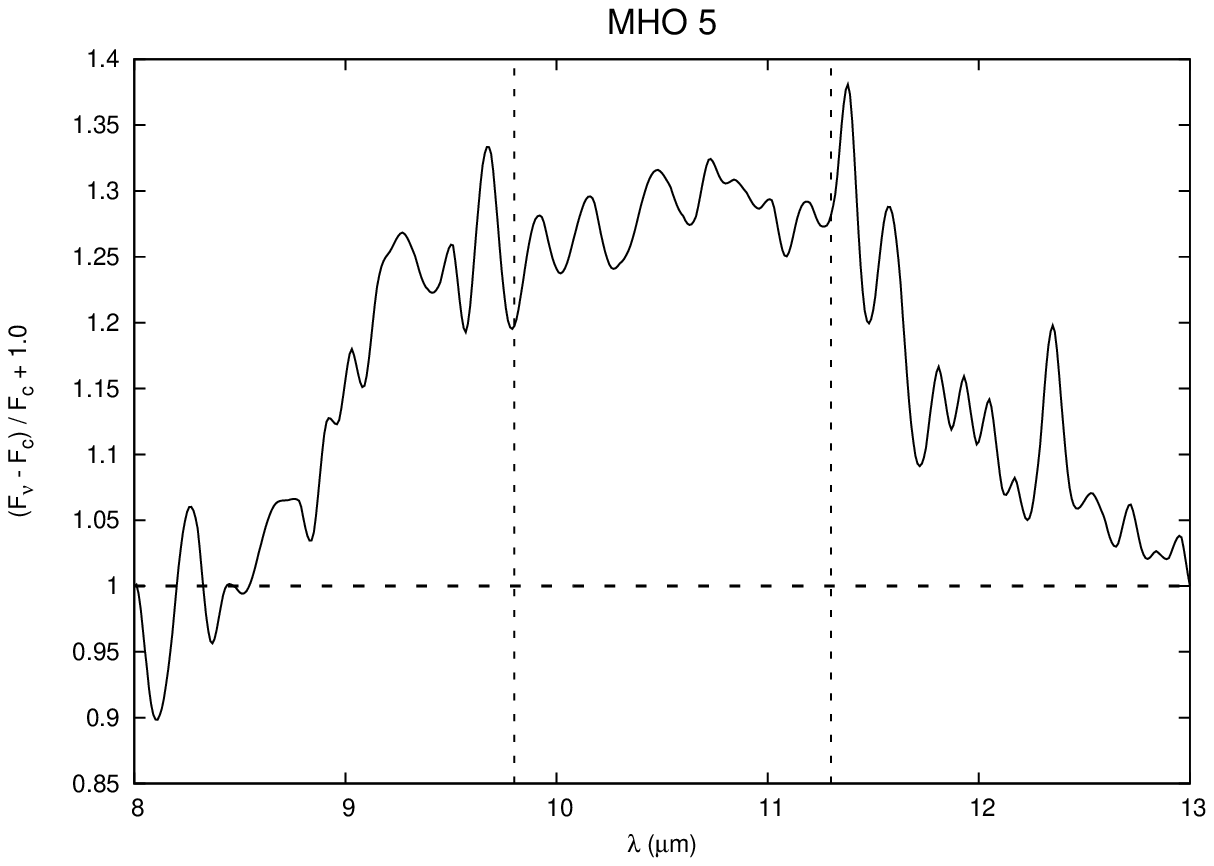}} &
      \resizebox{50mm}{!}{\includegraphics[angle=0]{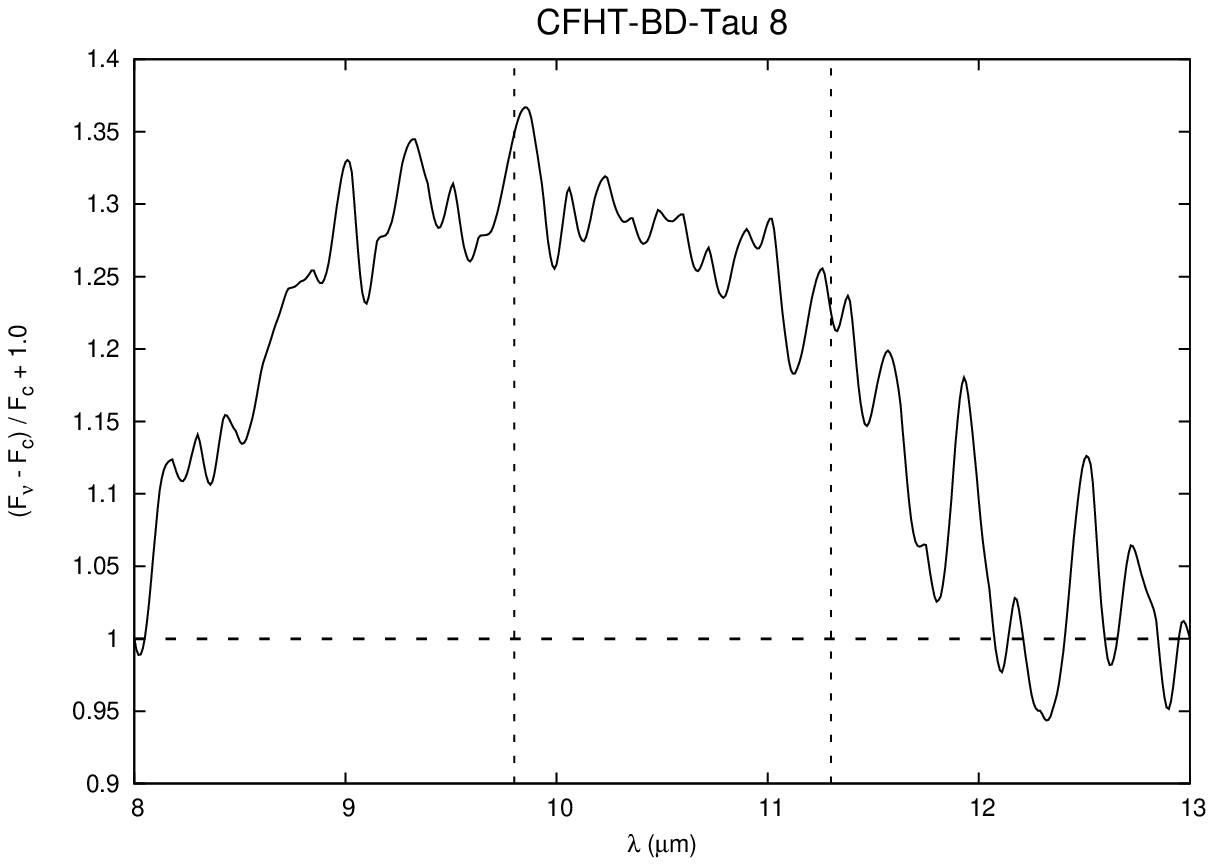}}  \\   
      \resizebox{50mm}{!}{\includegraphics[angle=0]{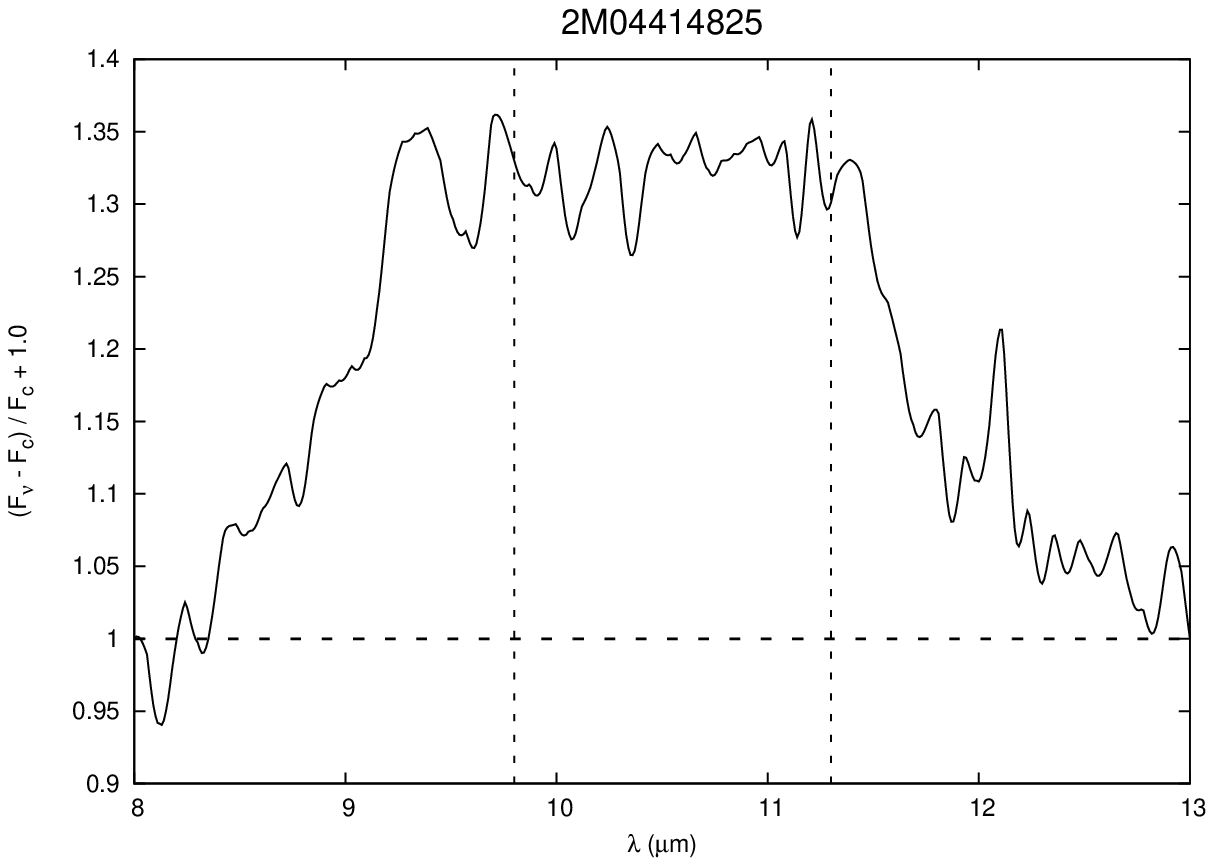}} &
      \resizebox{50mm}{!}{\includegraphics[angle=0]{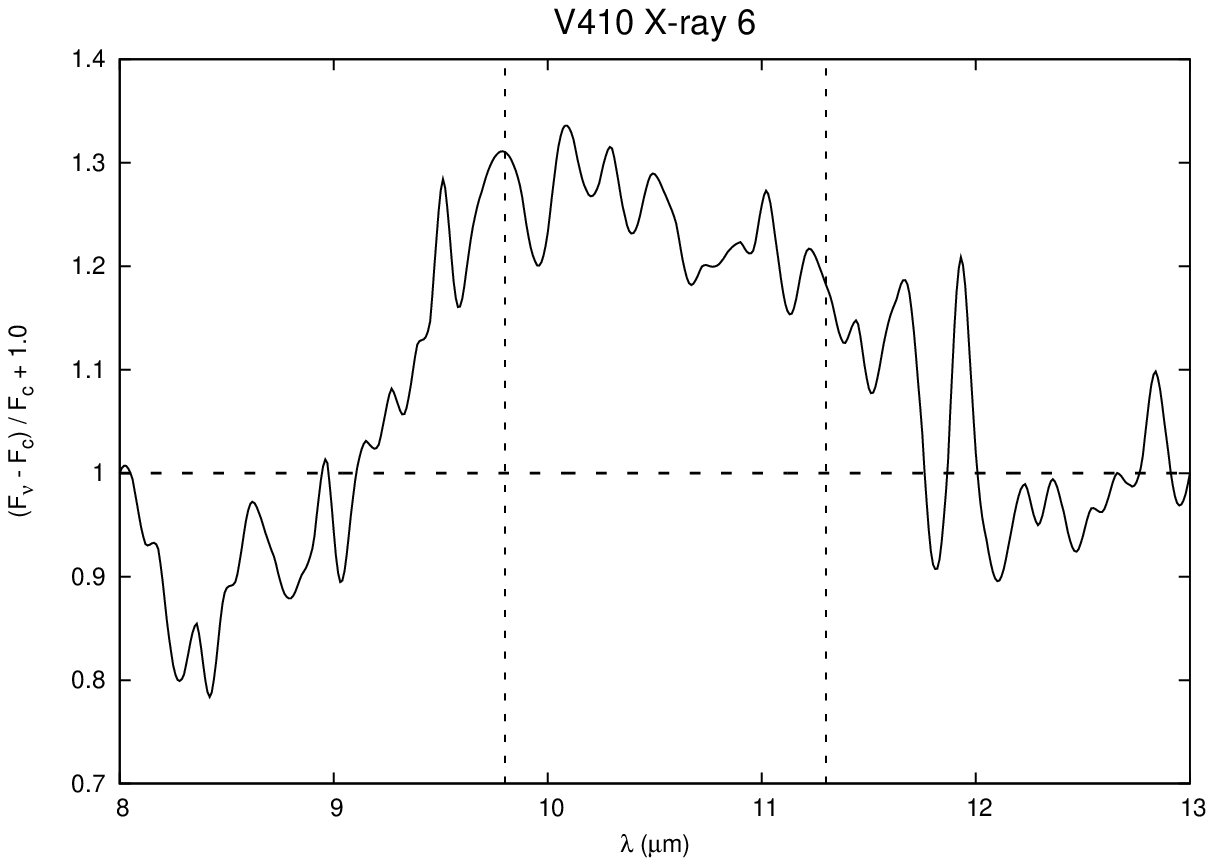}} &    
      \resizebox{50mm}{!}{\includegraphics[angle=0]{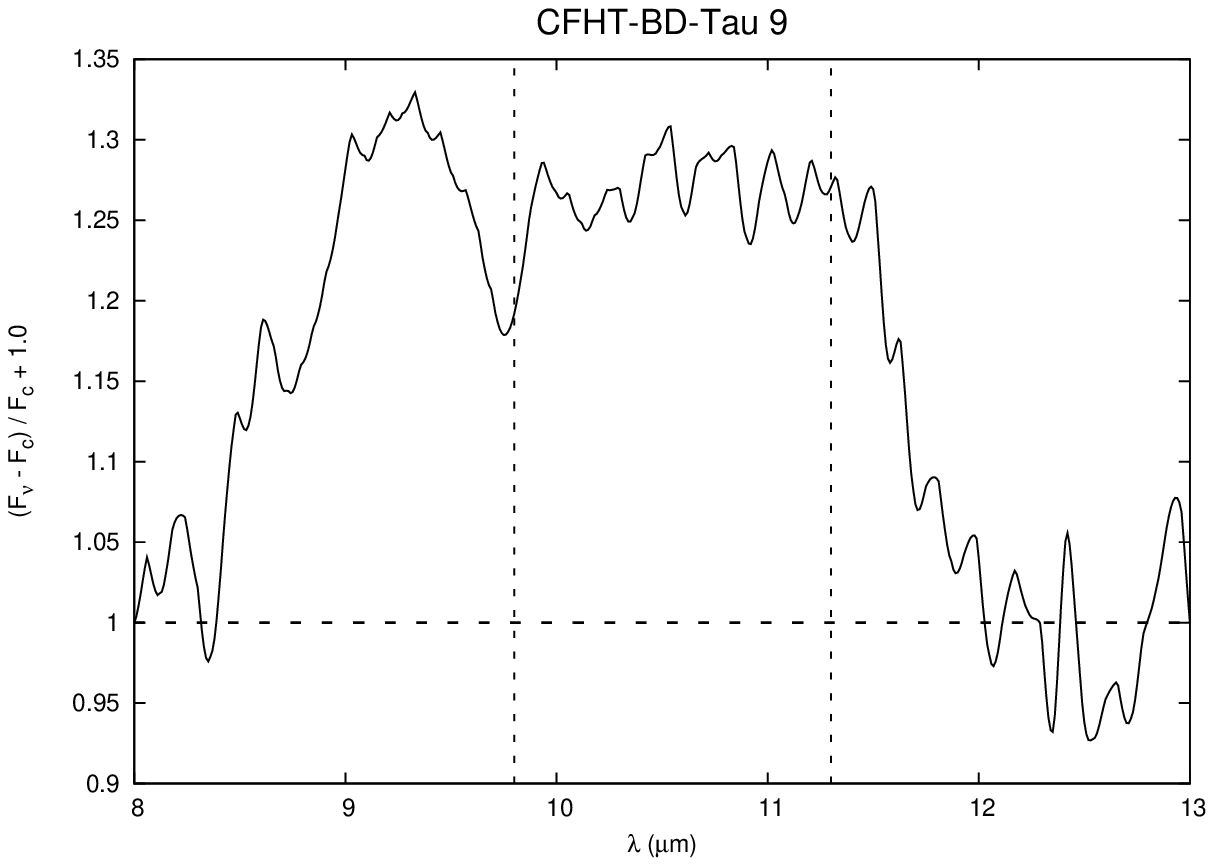}}   \\   
      \resizebox{50mm}{!}{\includegraphics[angle=0]{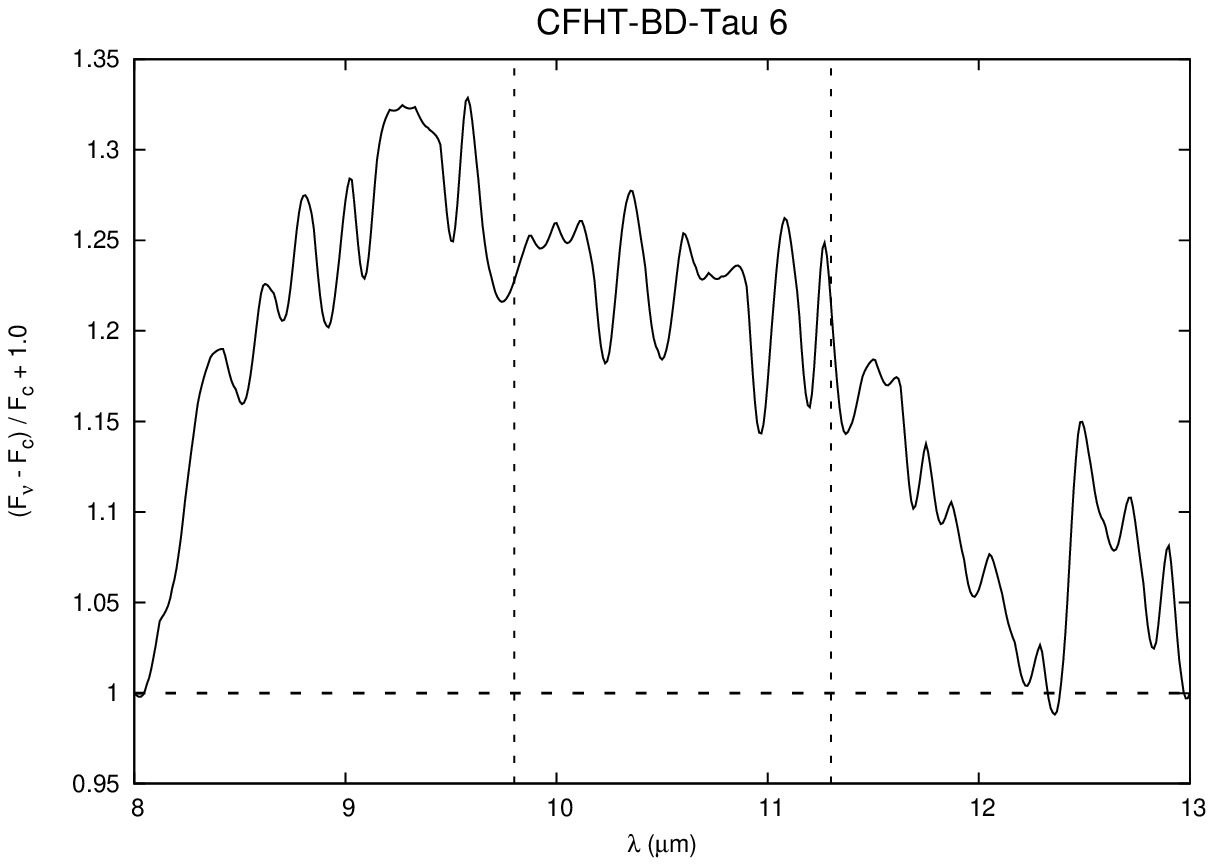}}  &  
      \resizebox{50mm}{!}{\includegraphics[angle=0]{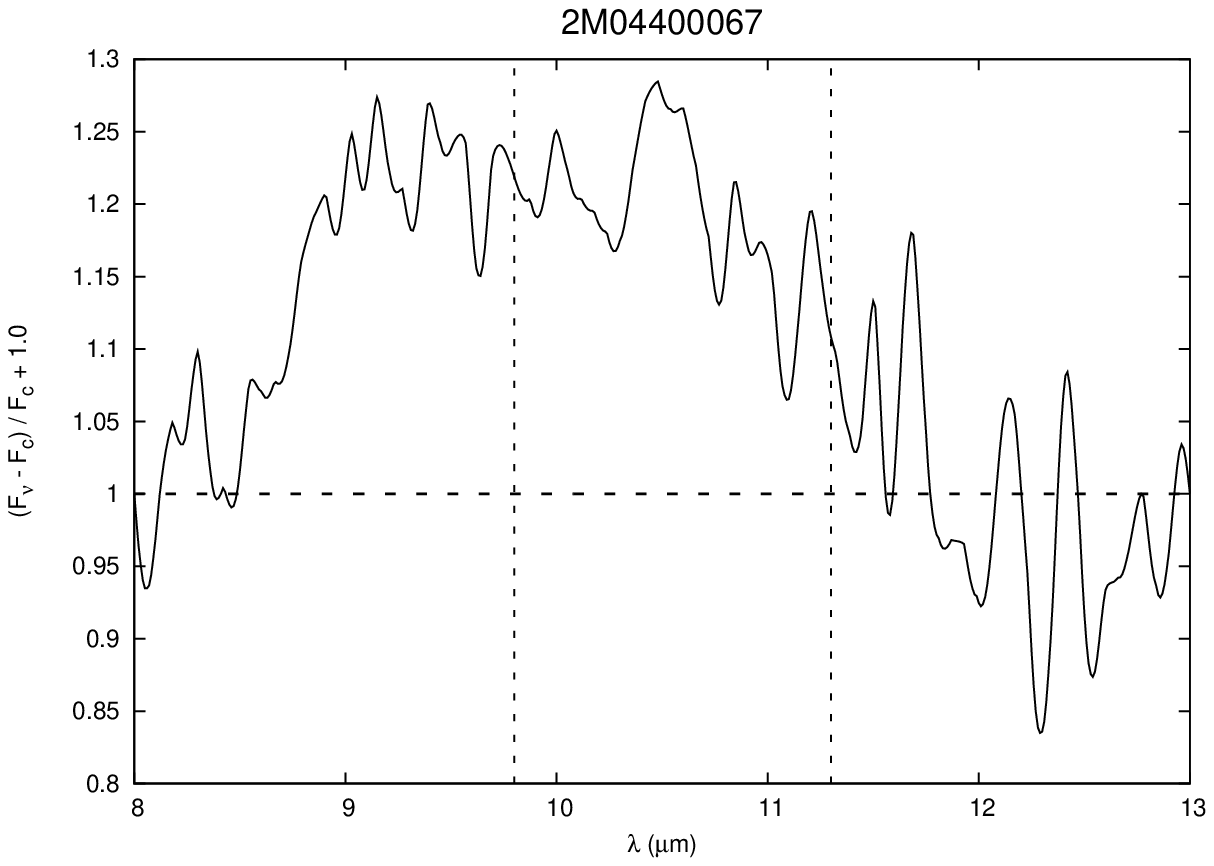}} &                    
      \resizebox{50mm}{!}{\includegraphics[angle=0]{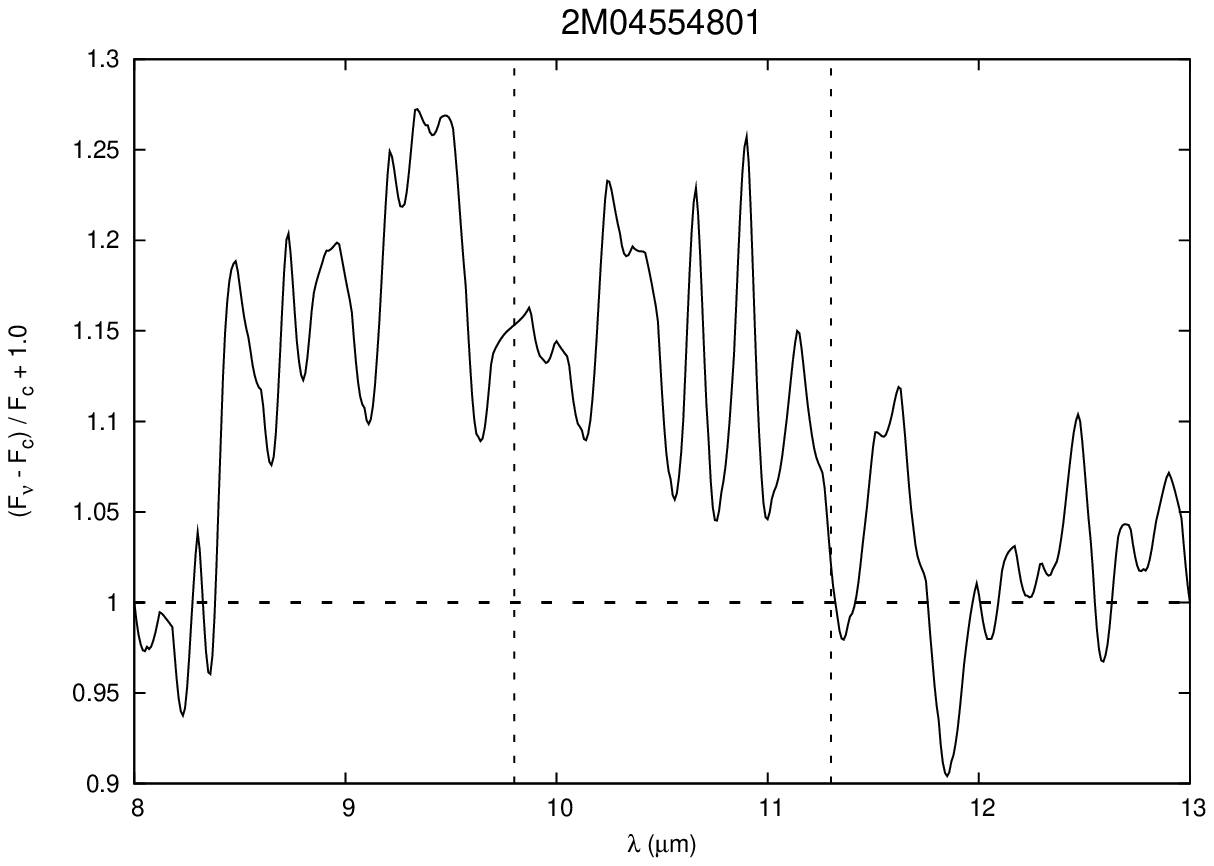}} \\    
    \end{tabular}
    \caption{The normalized continuum-subtracted spectra in units of ($F_{\nu} - F_{c})/F_{c}$. The dashed horizontal line represents the continuum. The vertical lines mark the amorphous olivine and crystalline forsterite peaks at 9.8 and 11.3 $\micron$, respectively. The spectra have been arranged in the decreasing order of $F_{peak}$ values.}
    \label{cont-subt}
  \end{center}
 \end{figure}

\begin{figure}
\addtocounter{figure}{-1}
 \begin{center}
    \begin{tabular}{ccc}   
       \resizebox{50mm}{!}{\includegraphics[angle=0]{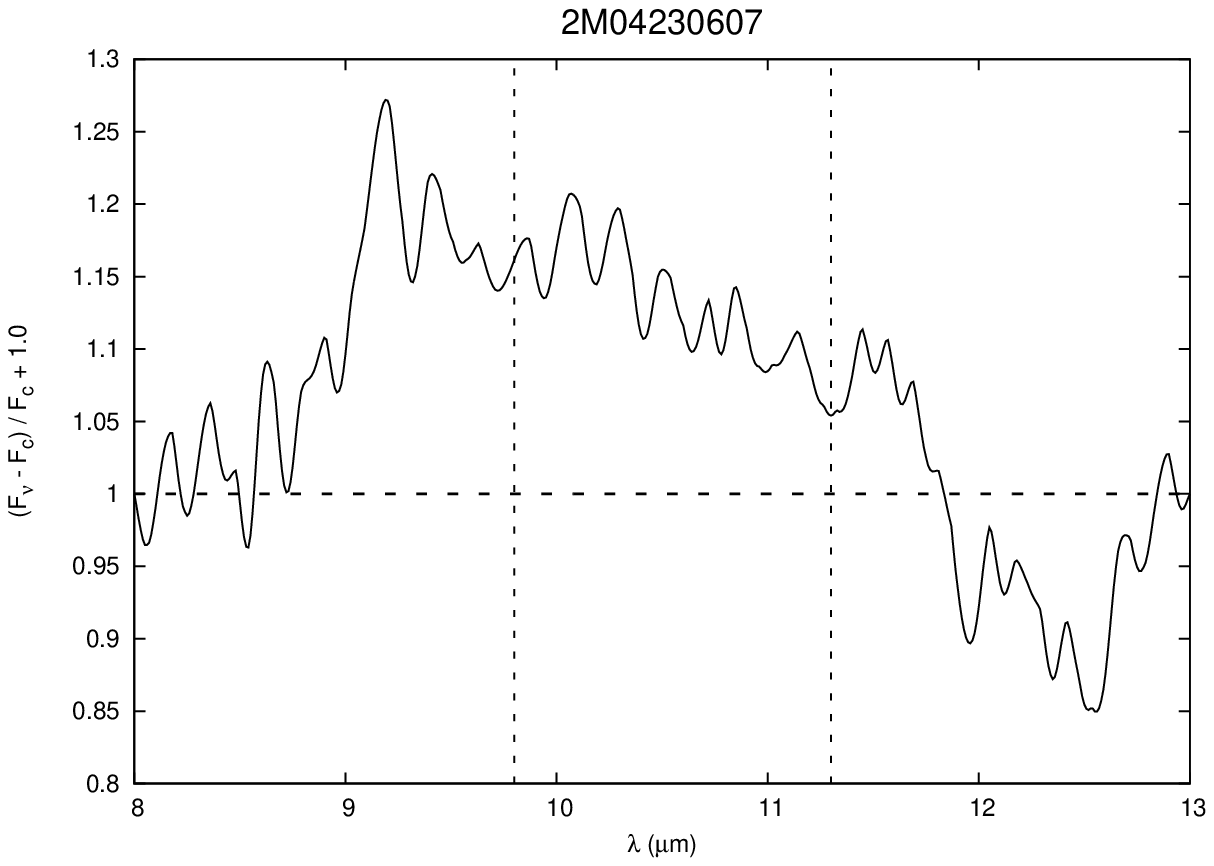}} &     
      \resizebox{50mm}{!}{\includegraphics[angle=0]{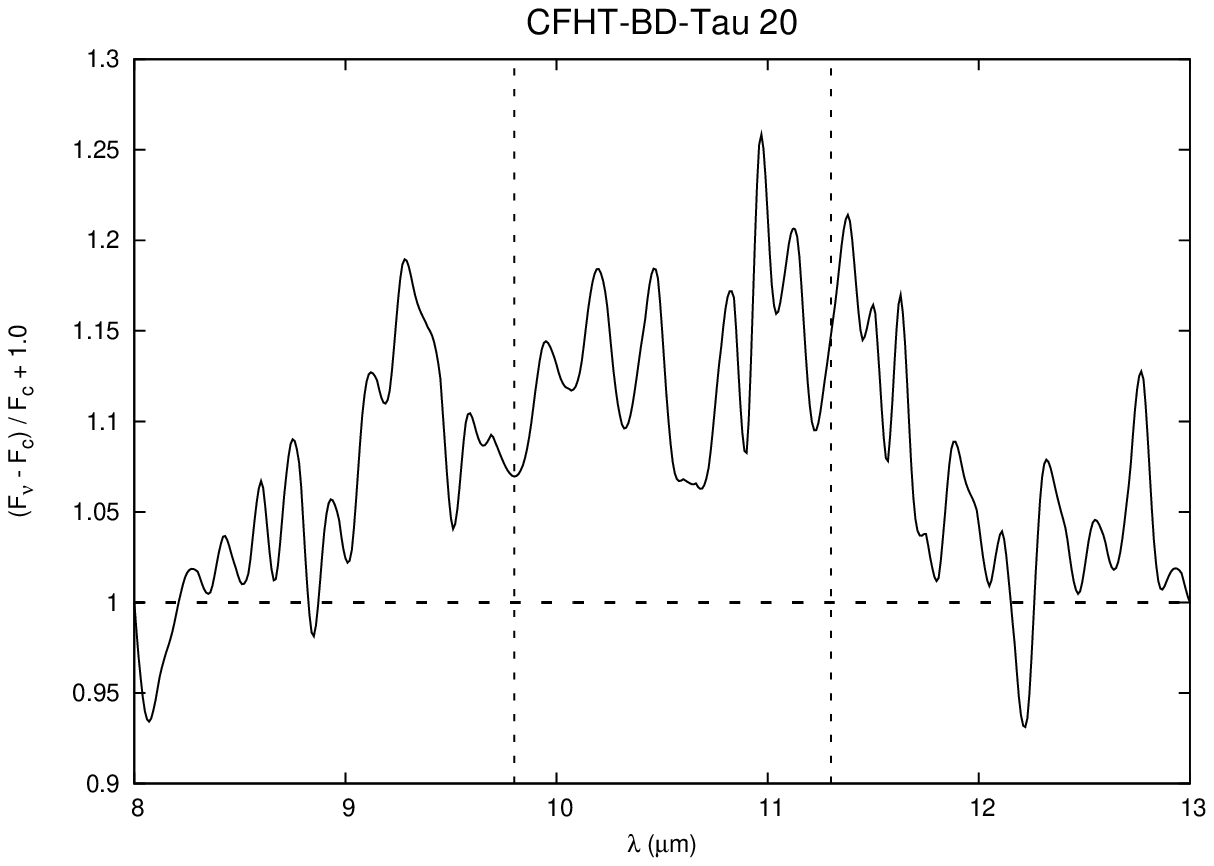}} &
      \resizebox{50mm}{!}{\includegraphics[angle=0]{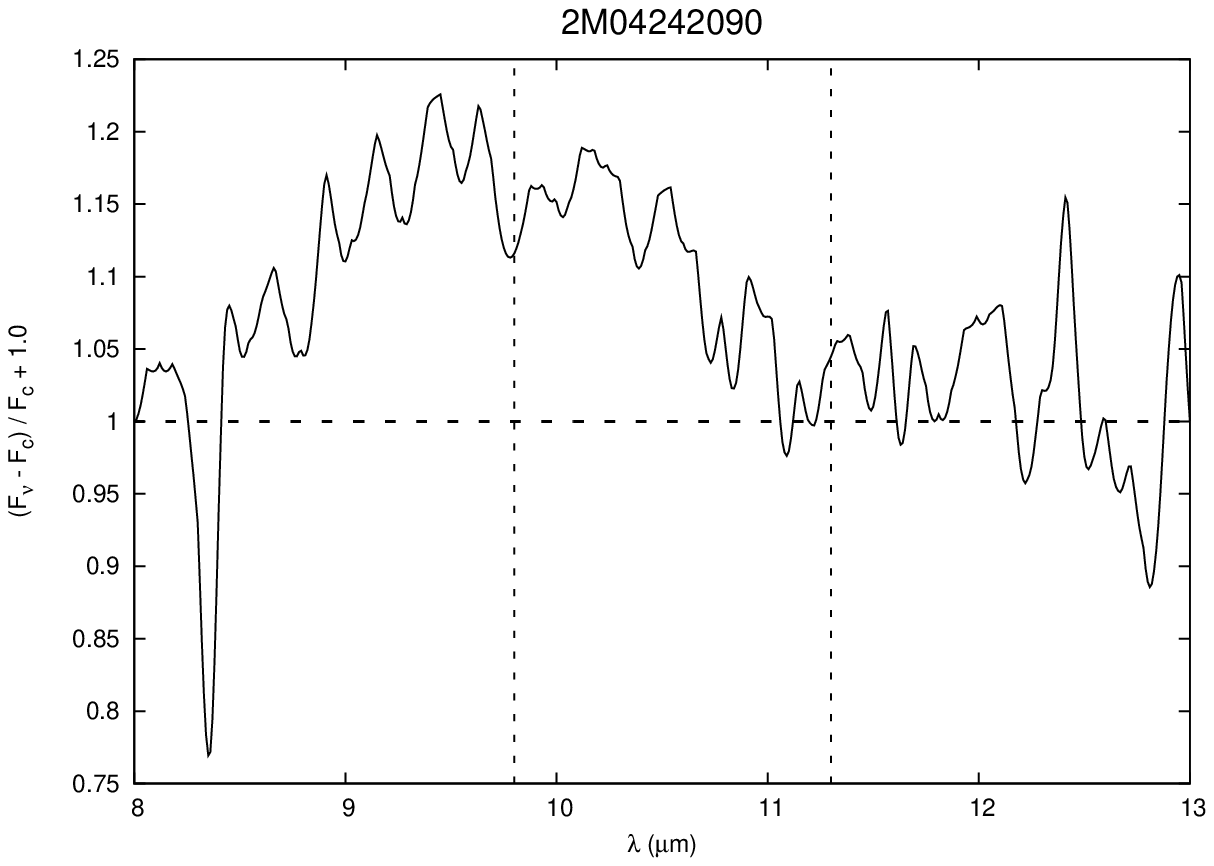}} \\   
      \resizebox{50mm}{!}{\includegraphics[angle=0]{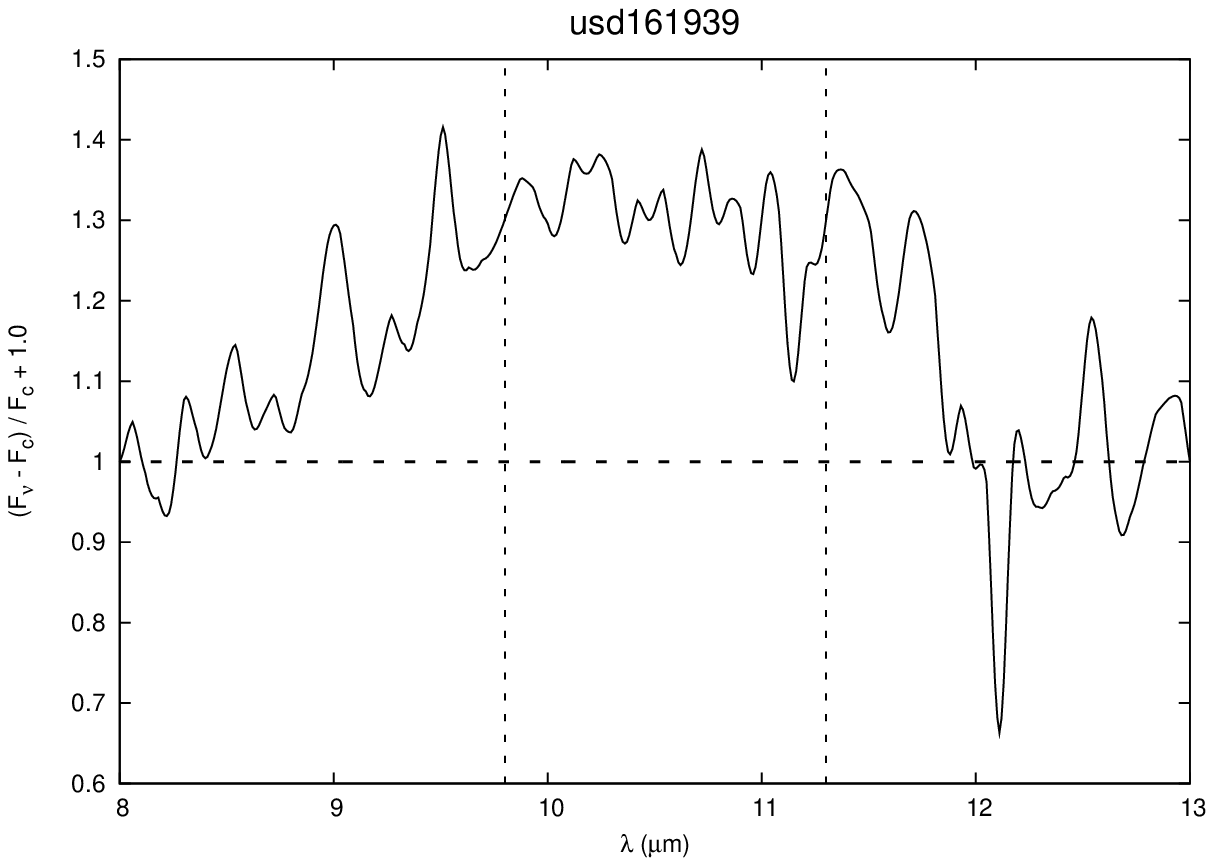}} & 
      \resizebox{50mm}{!}{\includegraphics[angle=0]{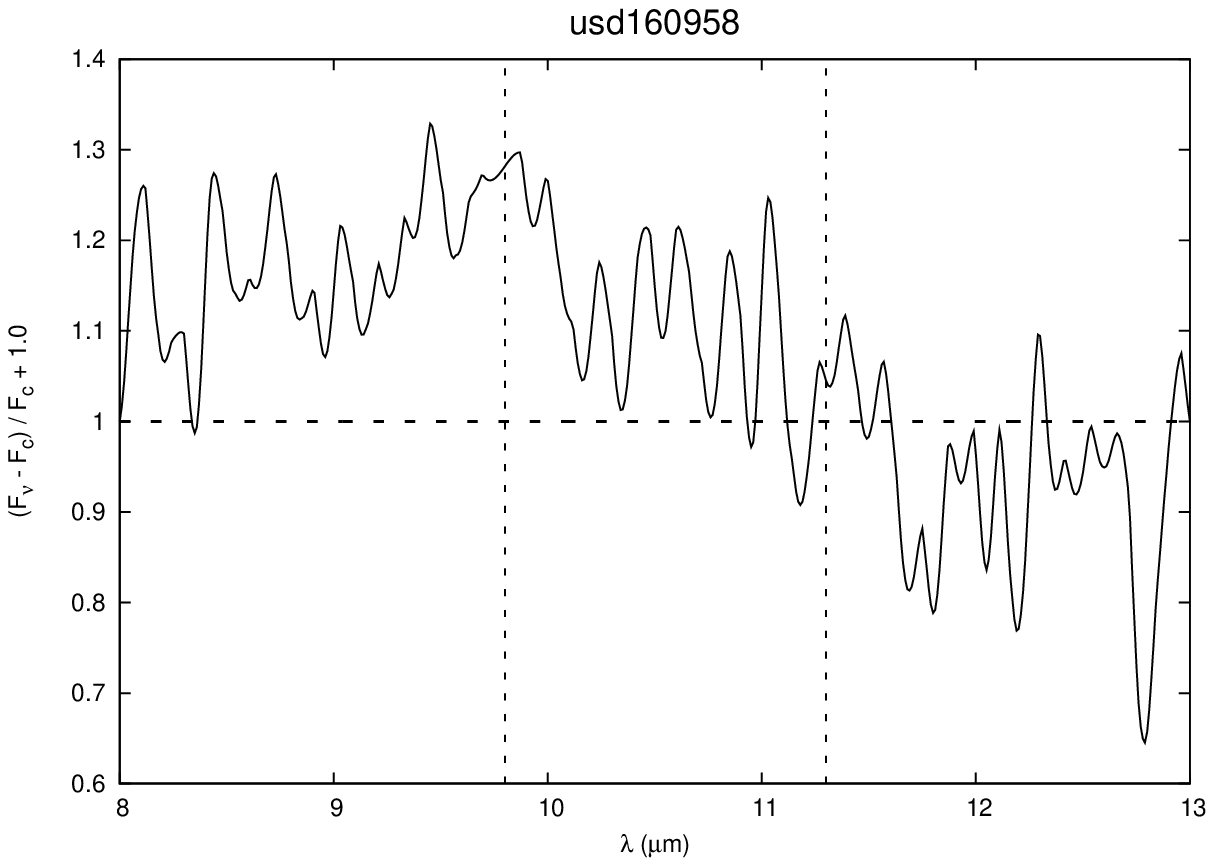}} &
      \resizebox{50mm}{!}{\includegraphics[angle=0]{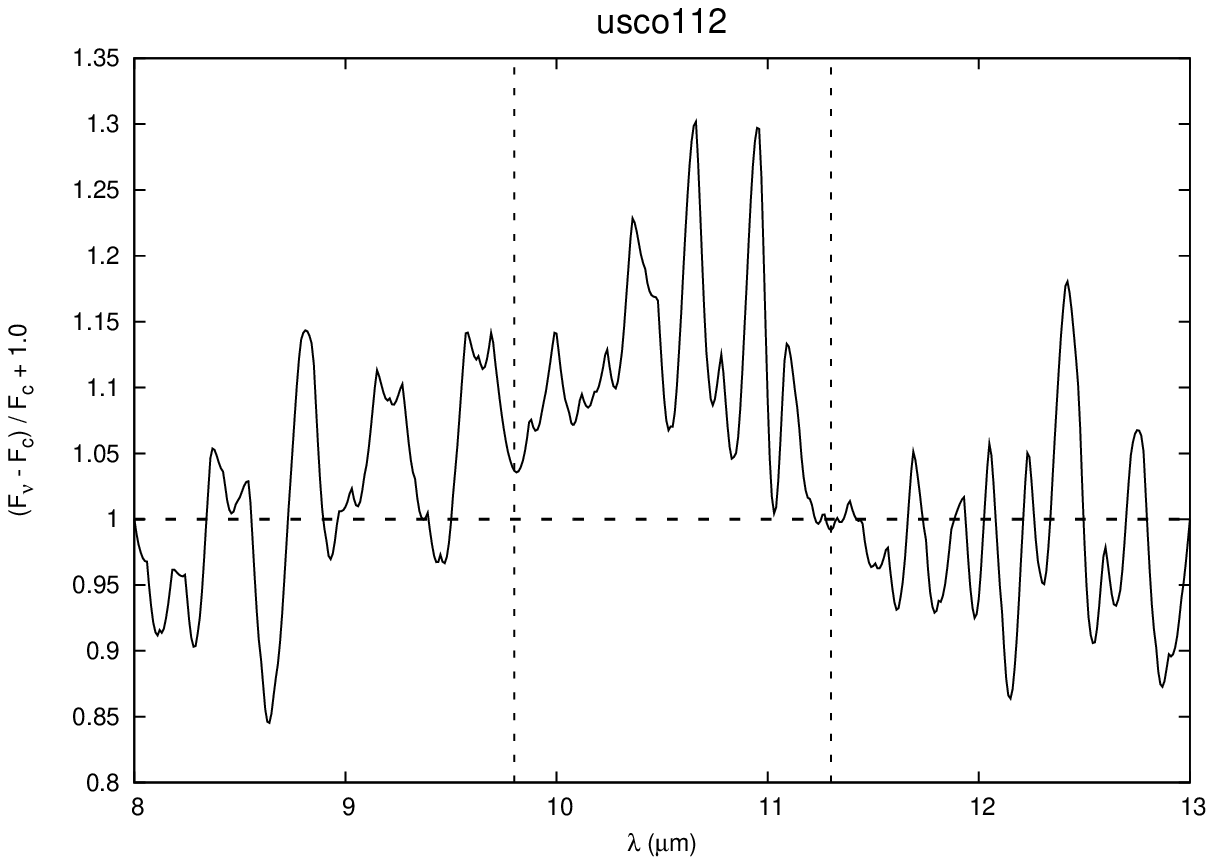}} \\  
      \resizebox{50mm}{!}{\includegraphics[angle=0]{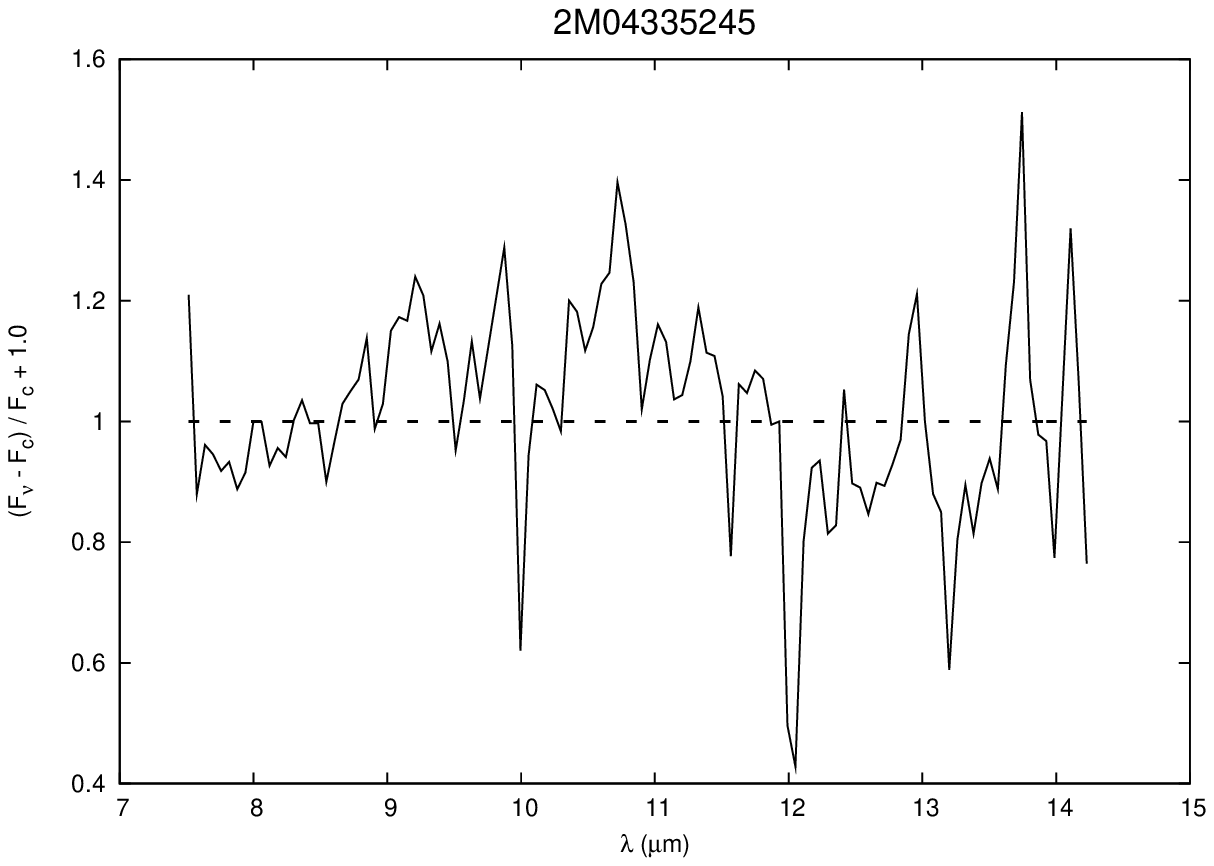}} & 
      \resizebox{50mm}{!}{\includegraphics[angle=0]{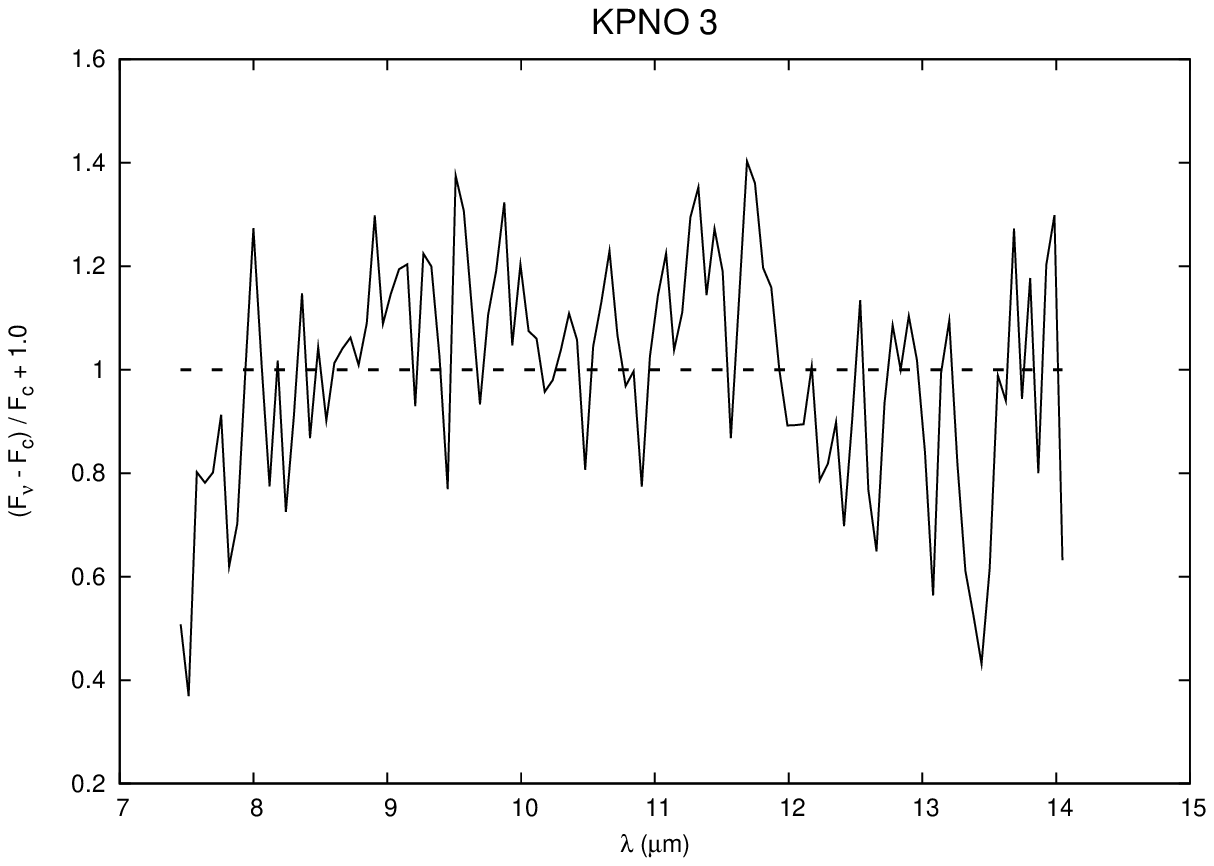}} &
      \resizebox{50mm}{!}{\includegraphics[angle=0]{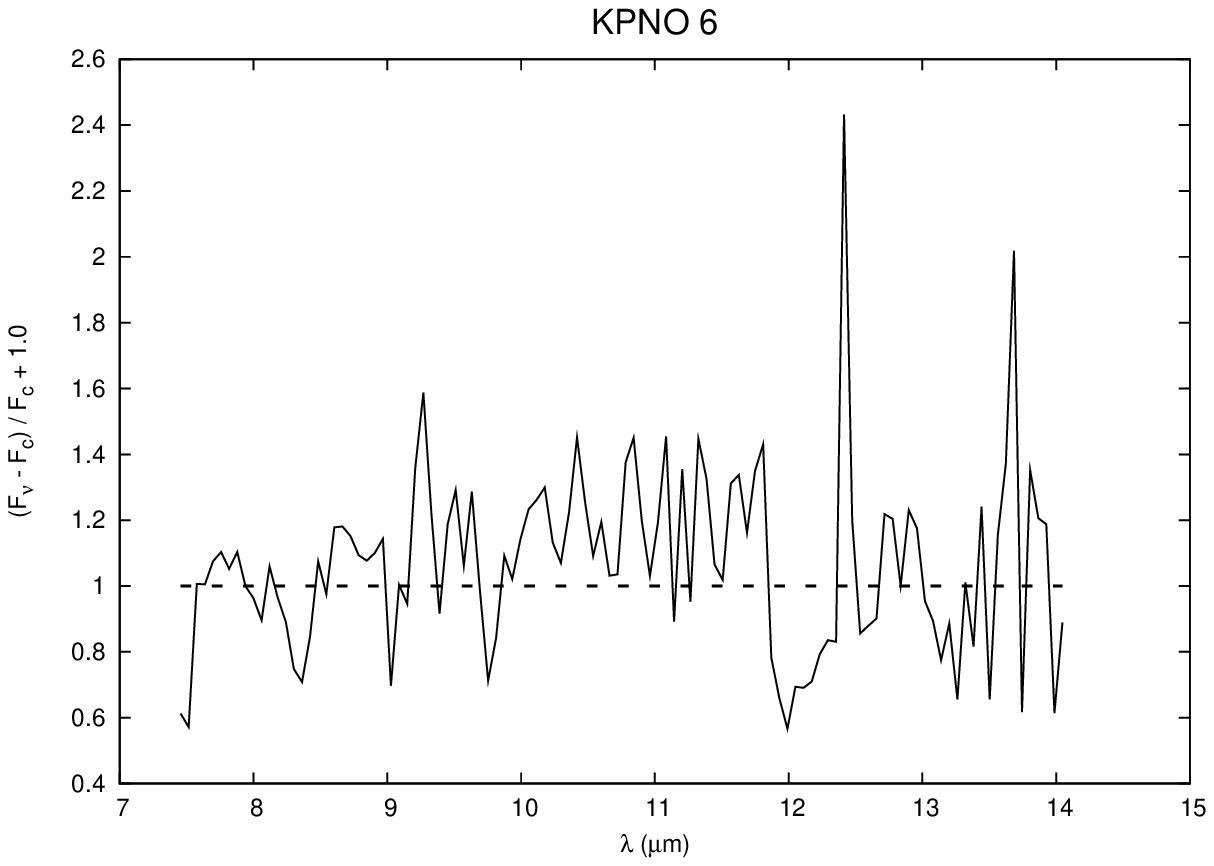}} \\     
      \resizebox{50mm}{!}{\includegraphics[angle=0]{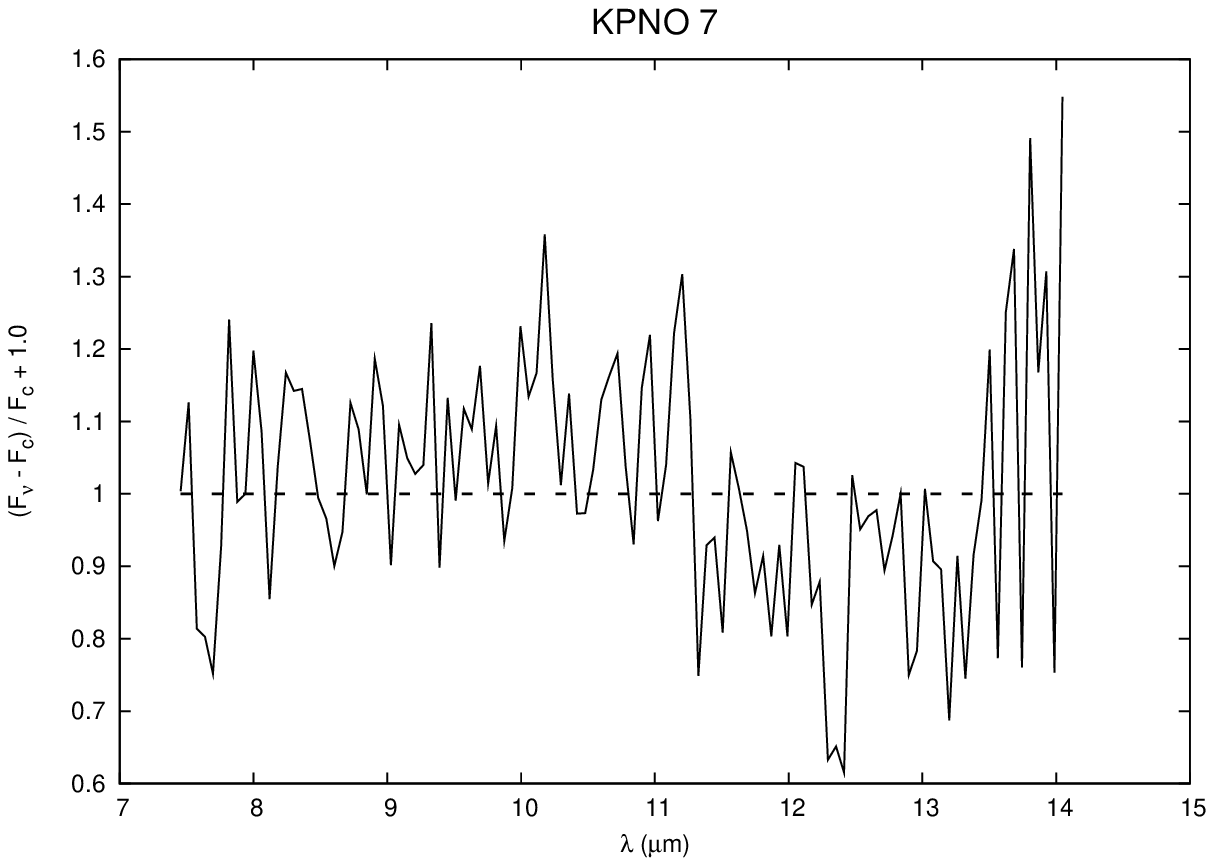}} &
      \resizebox{50mm}{!}{\includegraphics[angle=0]{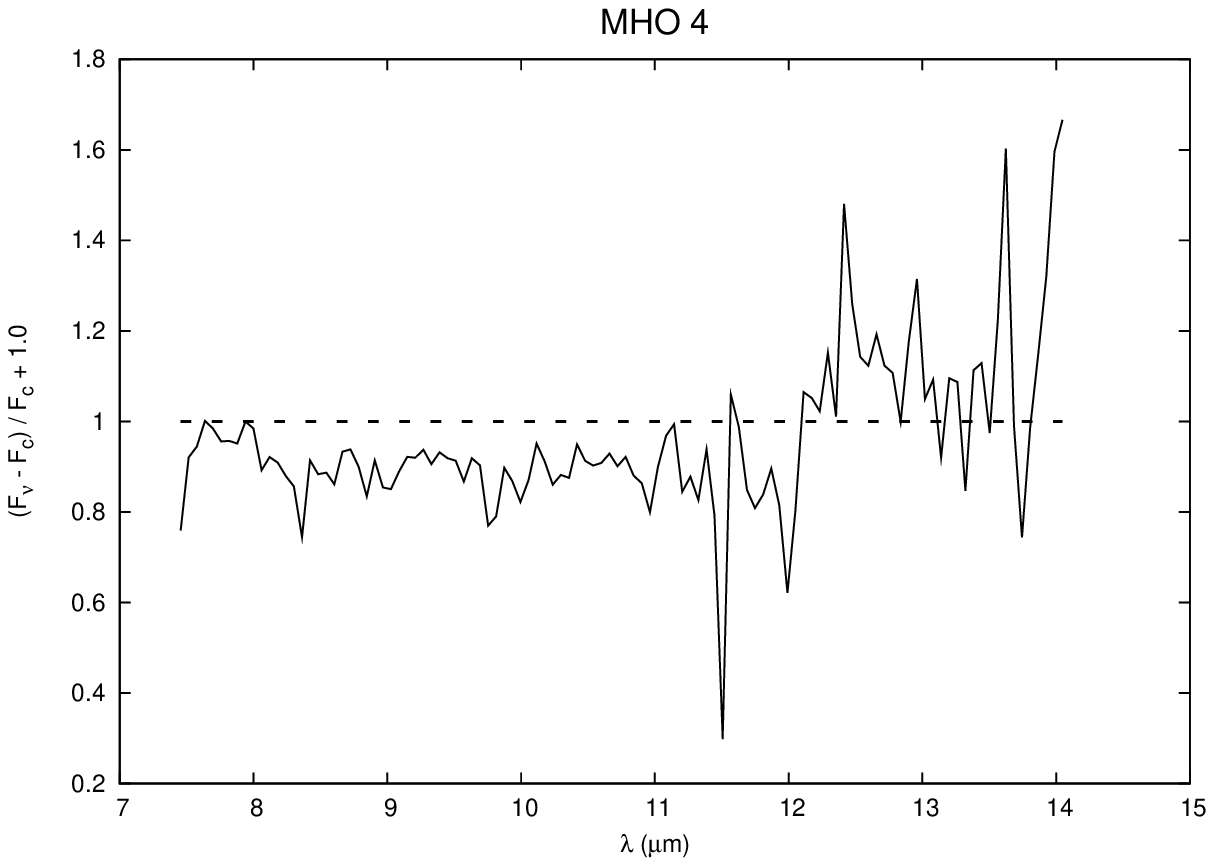}} \\                
    \end{tabular}
    \caption{{\it Continued}. The second panel shows the UppSco objects. Bottom two panels show the spectra for the 5 Taurus objects that do not show any silicate emission.}
  \end{center}
 \end{figure}

  \begin{figure}
 \begin{center}
    \begin{tabular}{cc}                 
     \resizebox{100mm}{!}{\includegraphics[angle=0]{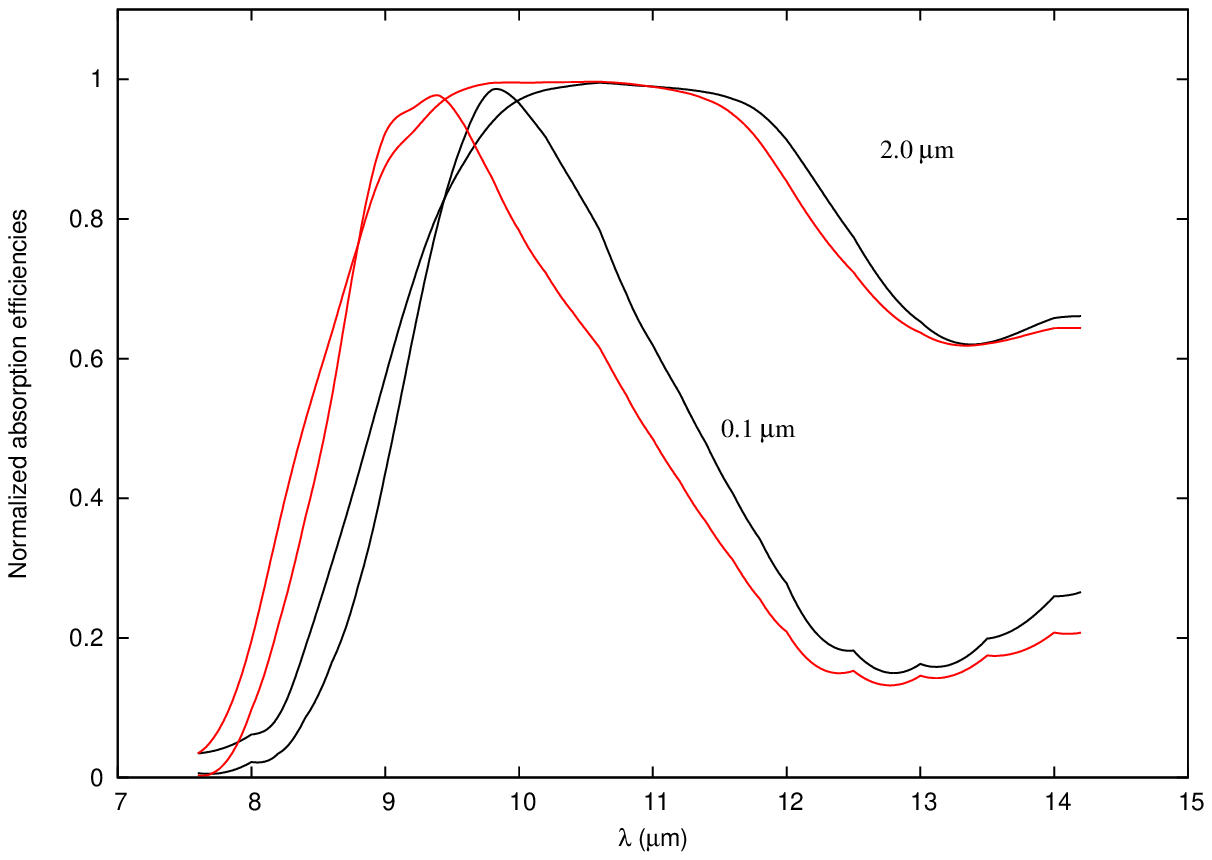}} \\
     \resizebox{100mm}{!}{\includegraphics[angle=0]{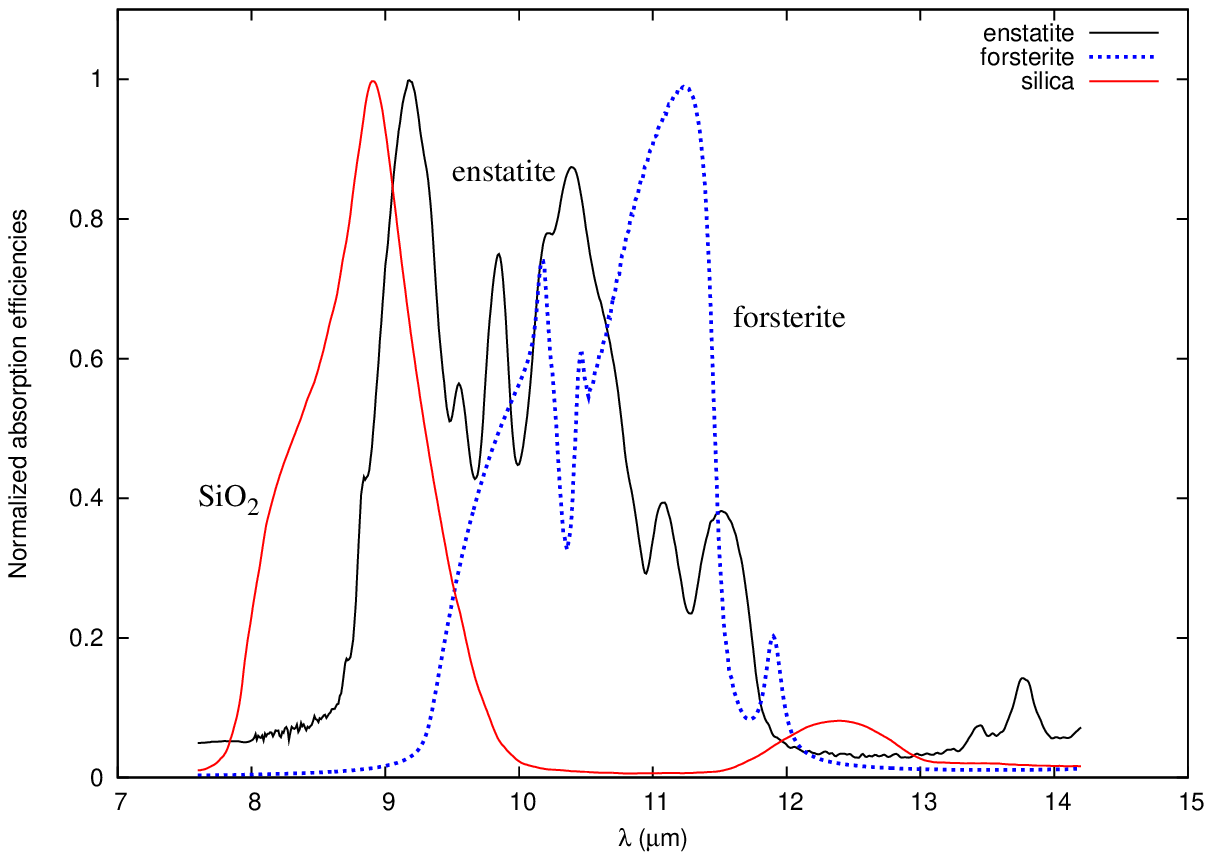}}   \\       
    \end{tabular}
    \caption{Normalized spectral profiles for the five dust species considered for modeling. Upper panel shows the profiles for small and large amorphous olivine ({\it black}) and pyroxene ({\it red}) silicates. Bottom panel shows the profiles for the crystalline silicates.}
    \label{species}
  \end{center}
 \end{figure}  
 
  \begin{figure}
 \begin{center}
    \begin{tabular}{cc}                 
     \resizebox{100mm}{!}{\includegraphics[angle=0]{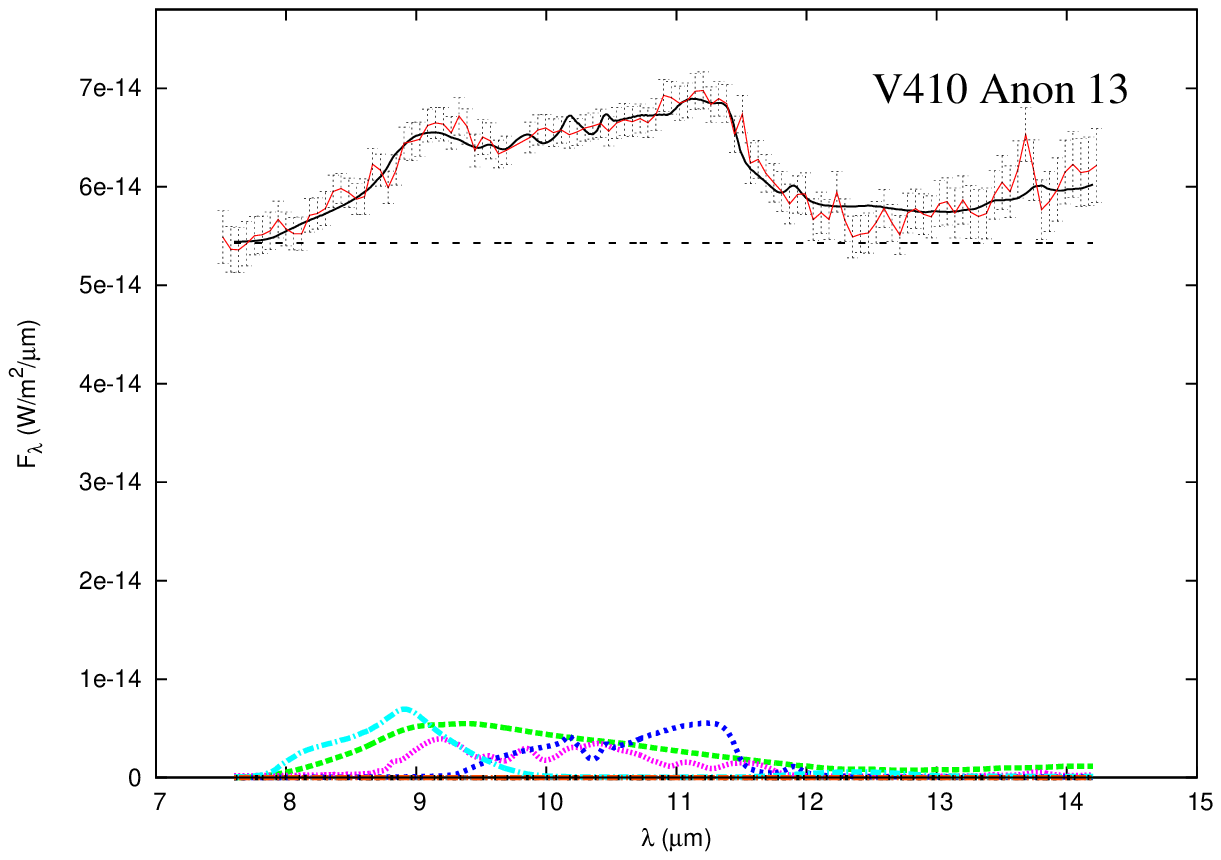}} \\
     \resizebox{100mm}{!}{\includegraphics[angle=0]{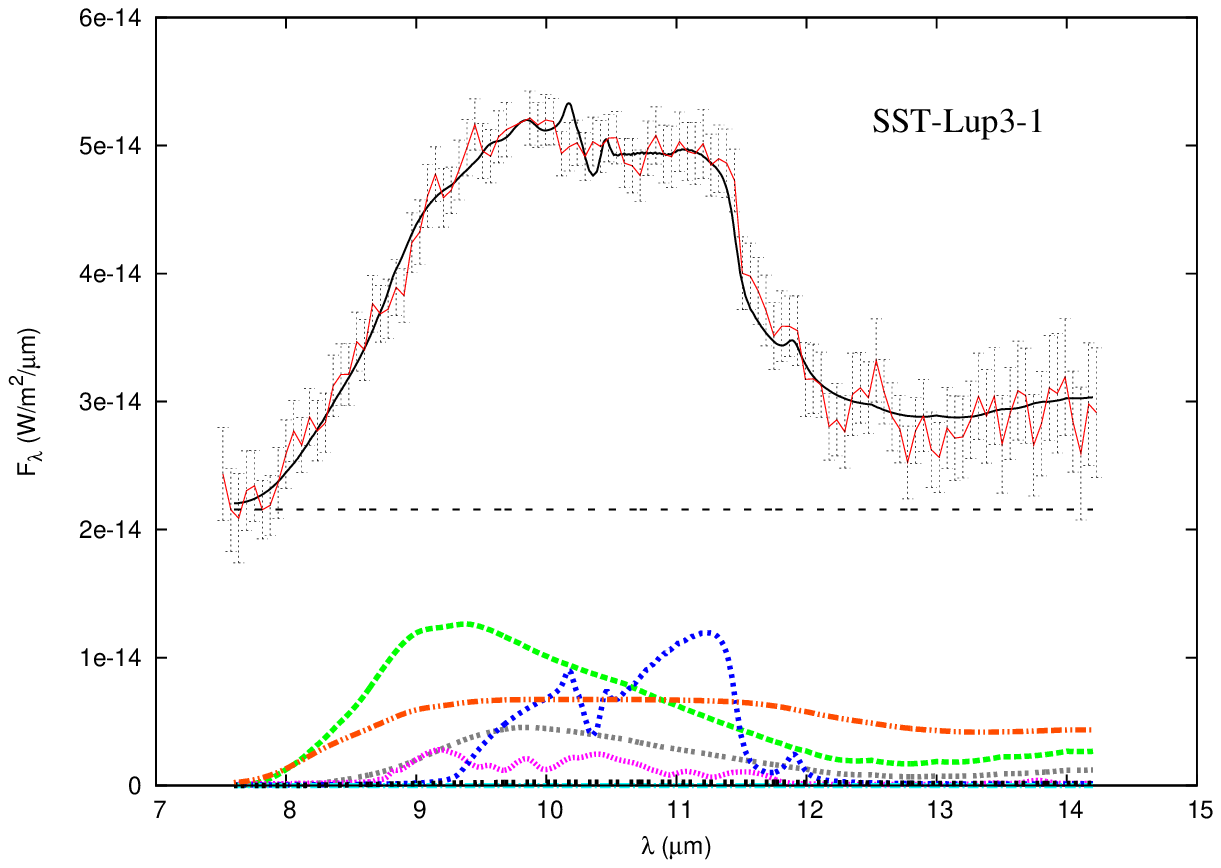}}   \\         
    \end{tabular}
    \caption{Model-fits for V410 Anon 13 ({\it top}) and SST-Lup3-1 ({\it bottom}). }
    \label{validate}
  \end{center}
 \end{figure} 

\begin{figure}
 \begin{center}
    \begin{tabular}{ccc}      
      \resizebox{50mm}{!}{\includegraphics[angle=0]{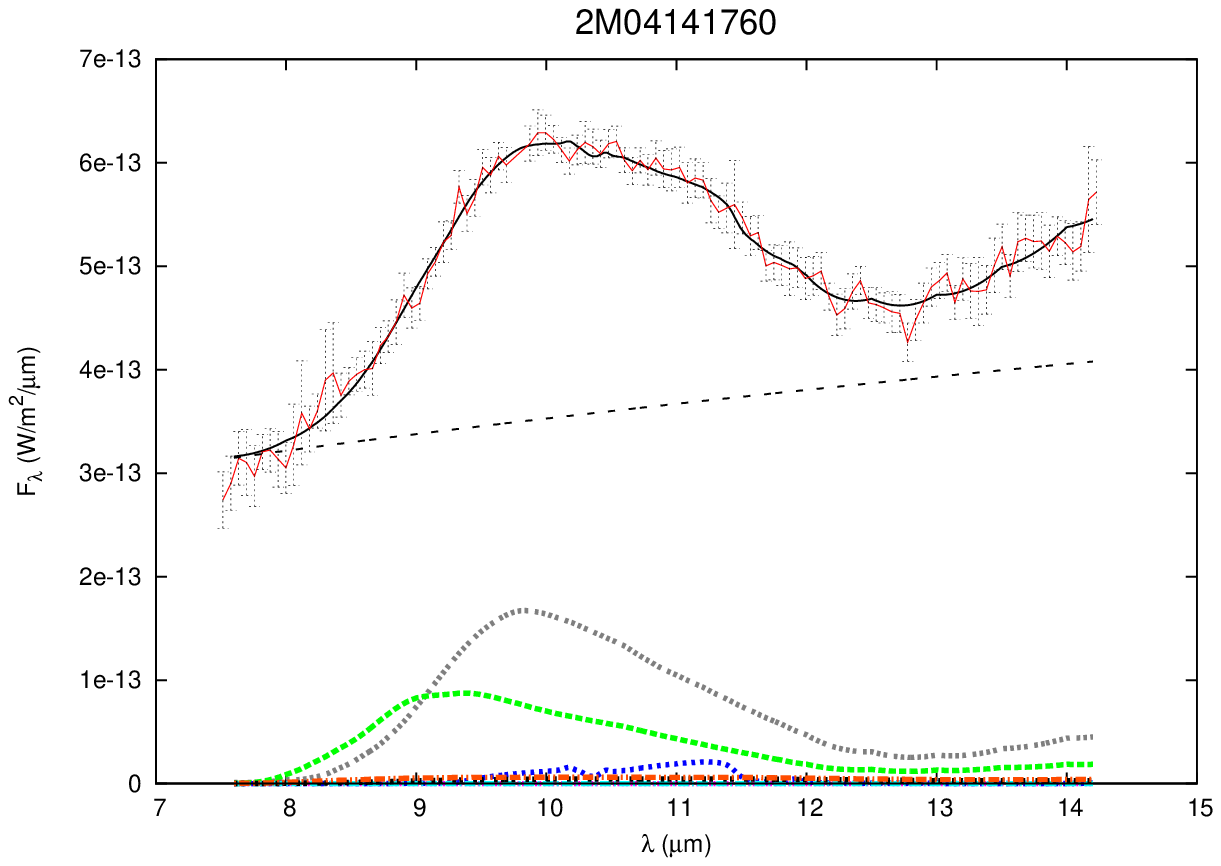}} &    
      \resizebox{50mm}{!}{\includegraphics[angle=0]{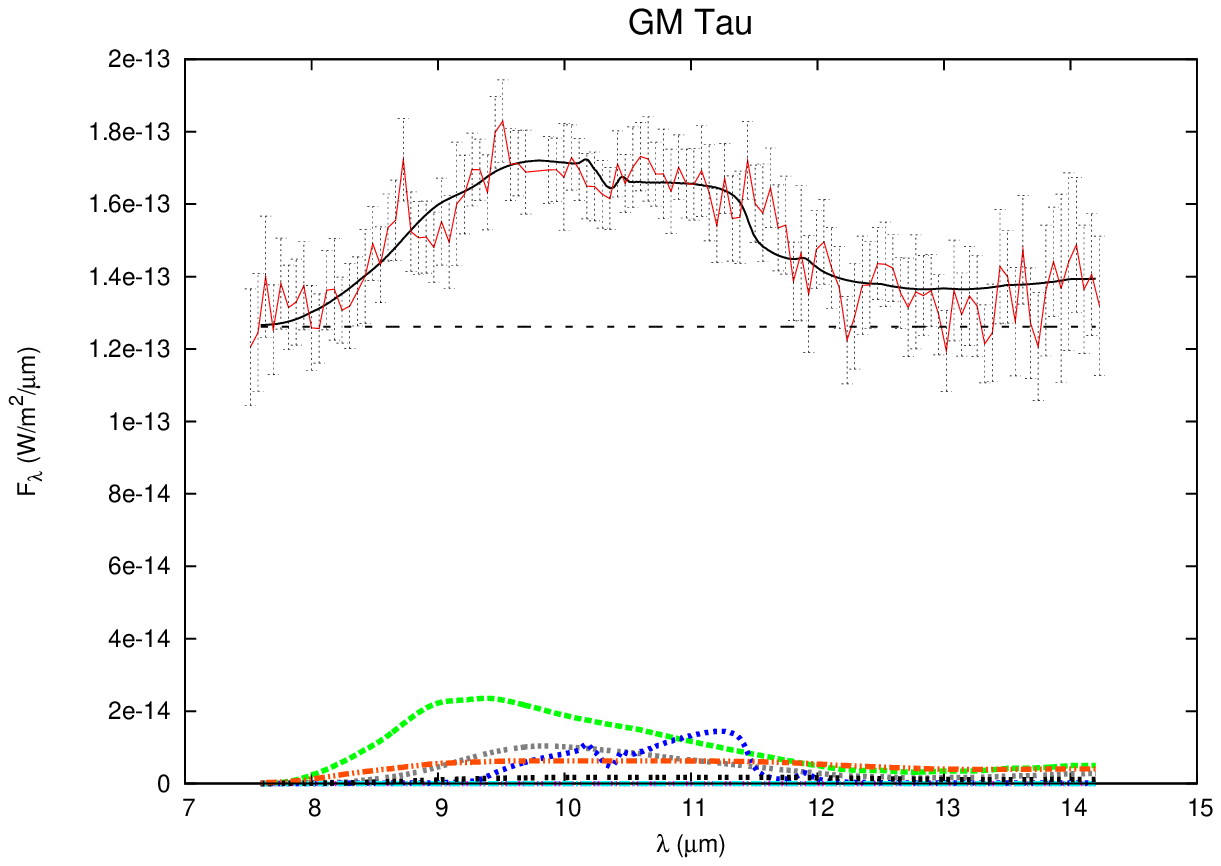}}  &   
      \resizebox{50mm}{!}{\includegraphics[angle=0]{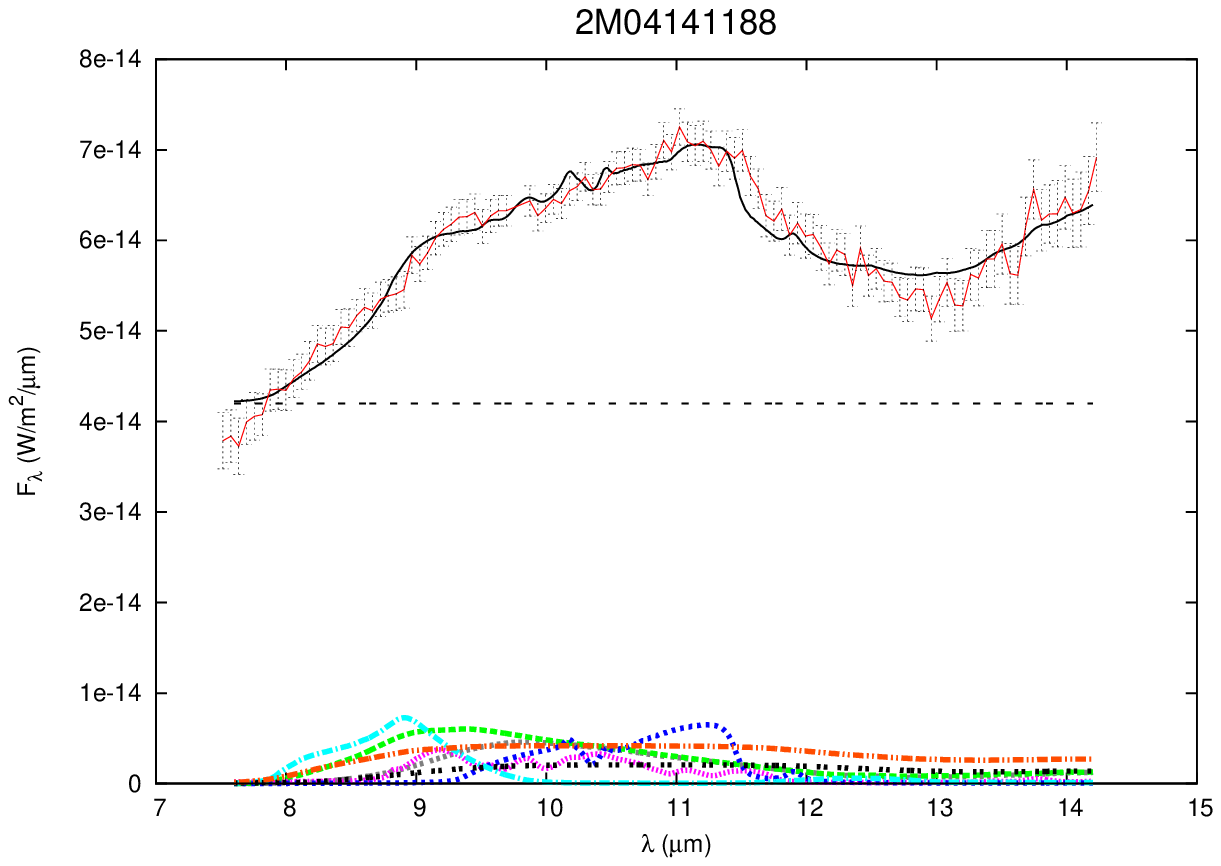}} \\
      \resizebox{50mm}{!}{\includegraphics[angle=0]{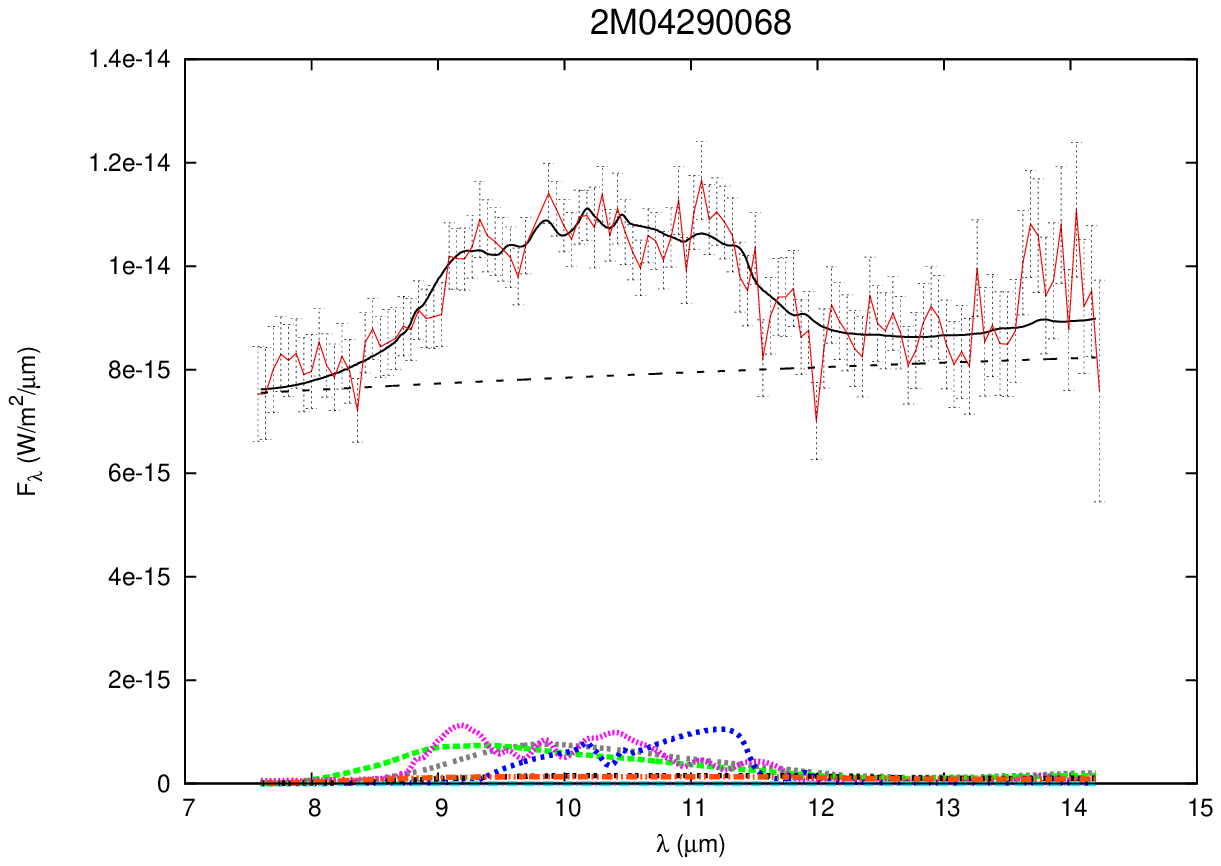}} &
      \resizebox{50mm}{!}{\includegraphics[angle=0]{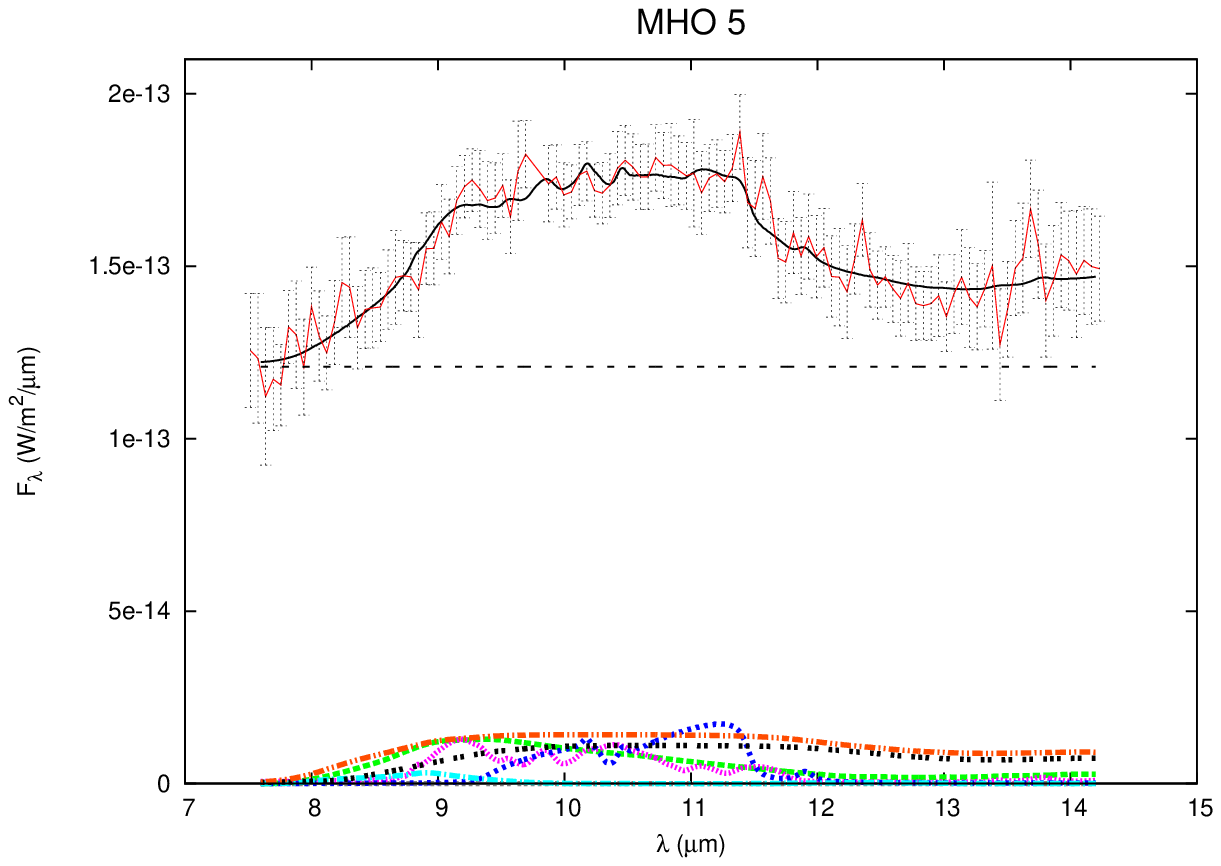}} &
      \resizebox{50mm}{!}{\includegraphics[angle=0]{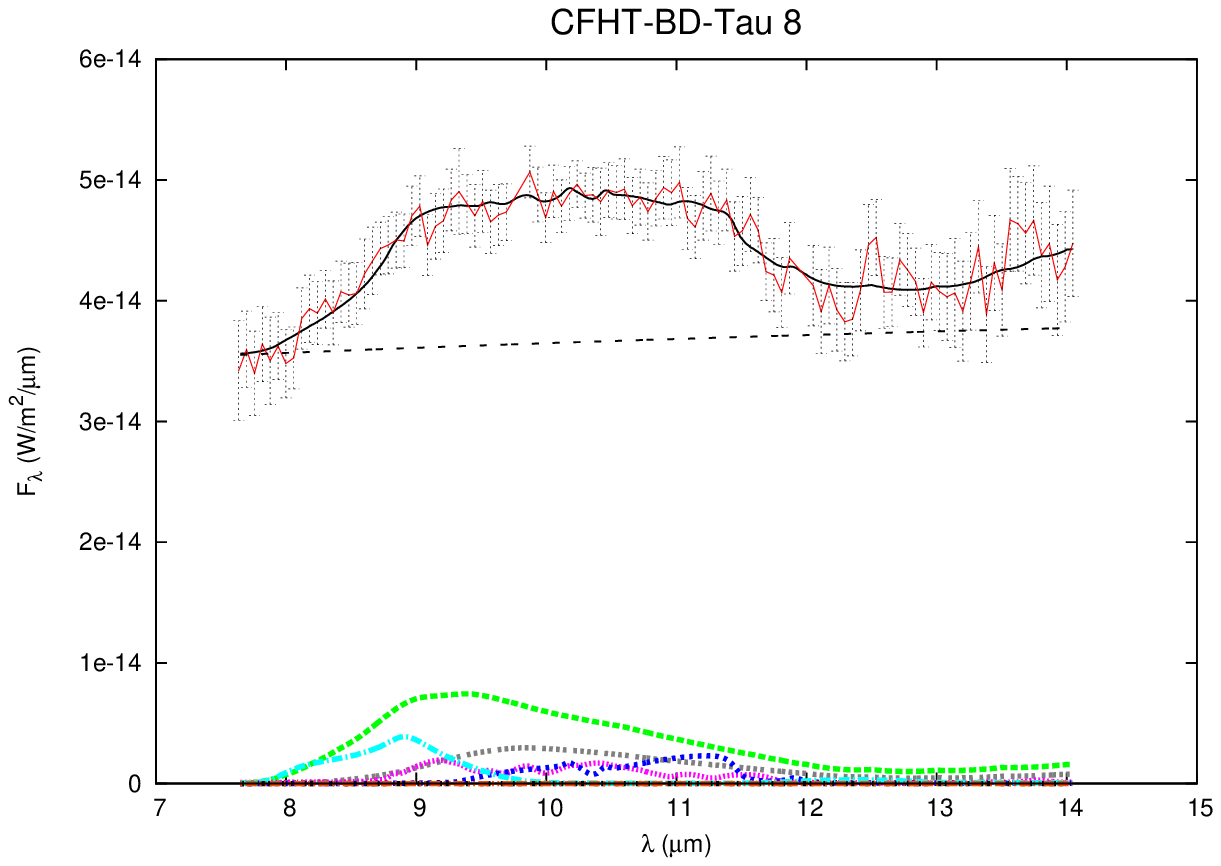}}  \\   
      \resizebox{50mm}{!}{\includegraphics[angle=0]{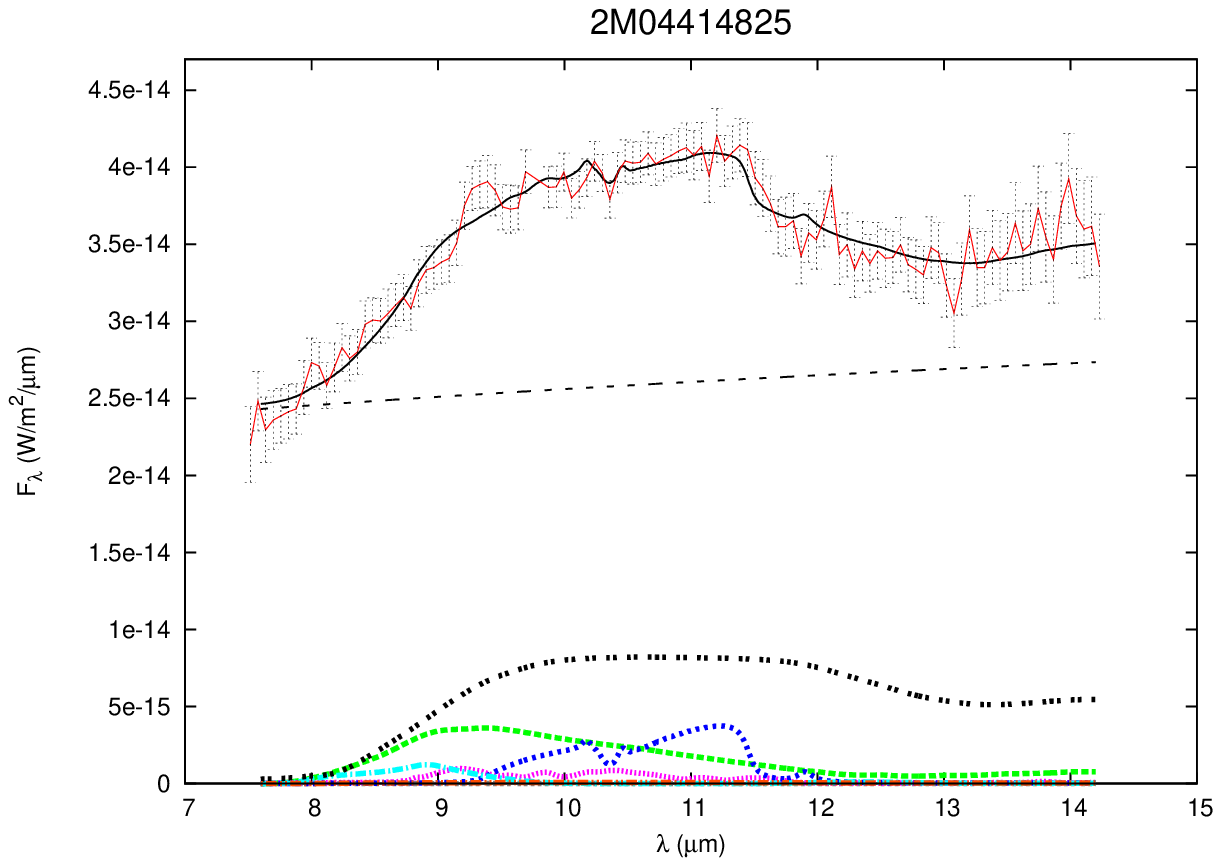}} &
      \resizebox{50mm}{!}{\includegraphics[angle=0]{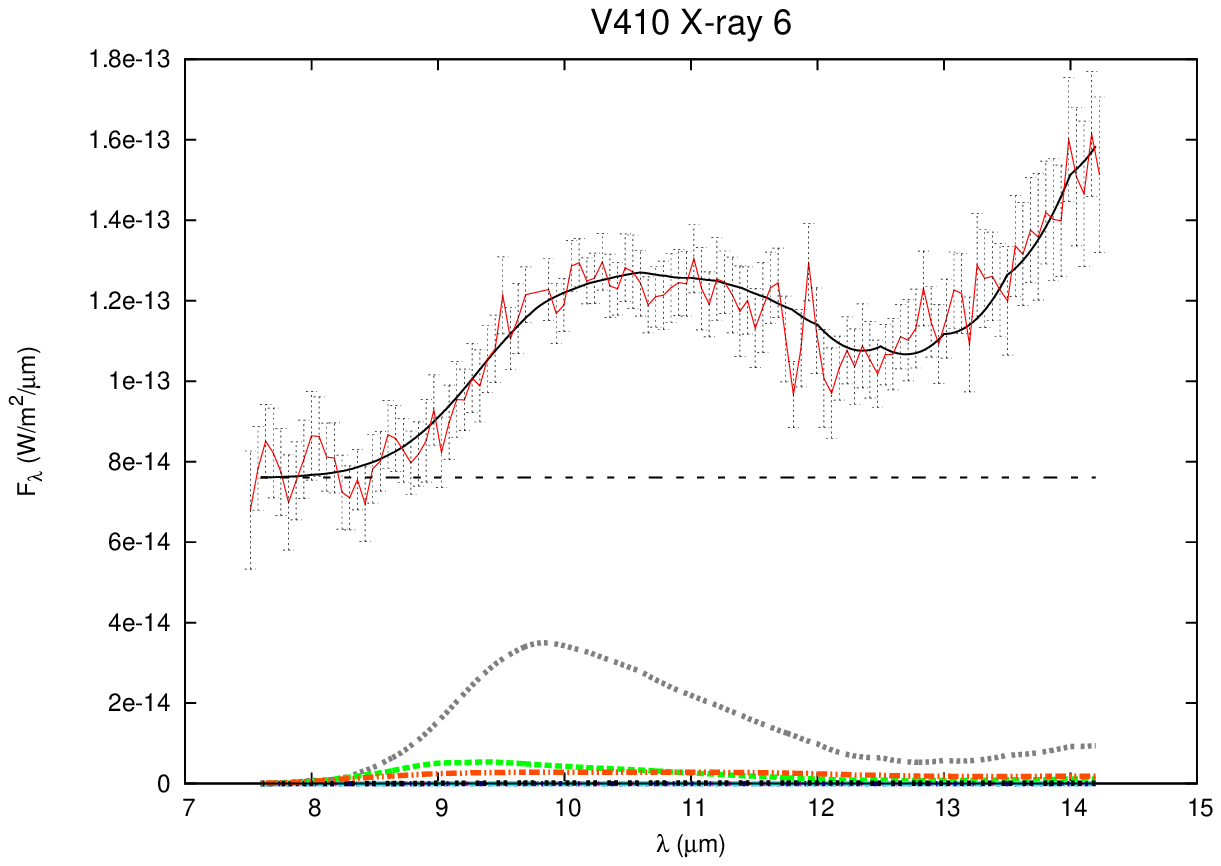}} &    
      \resizebox{50mm}{!}{\includegraphics[angle=0]{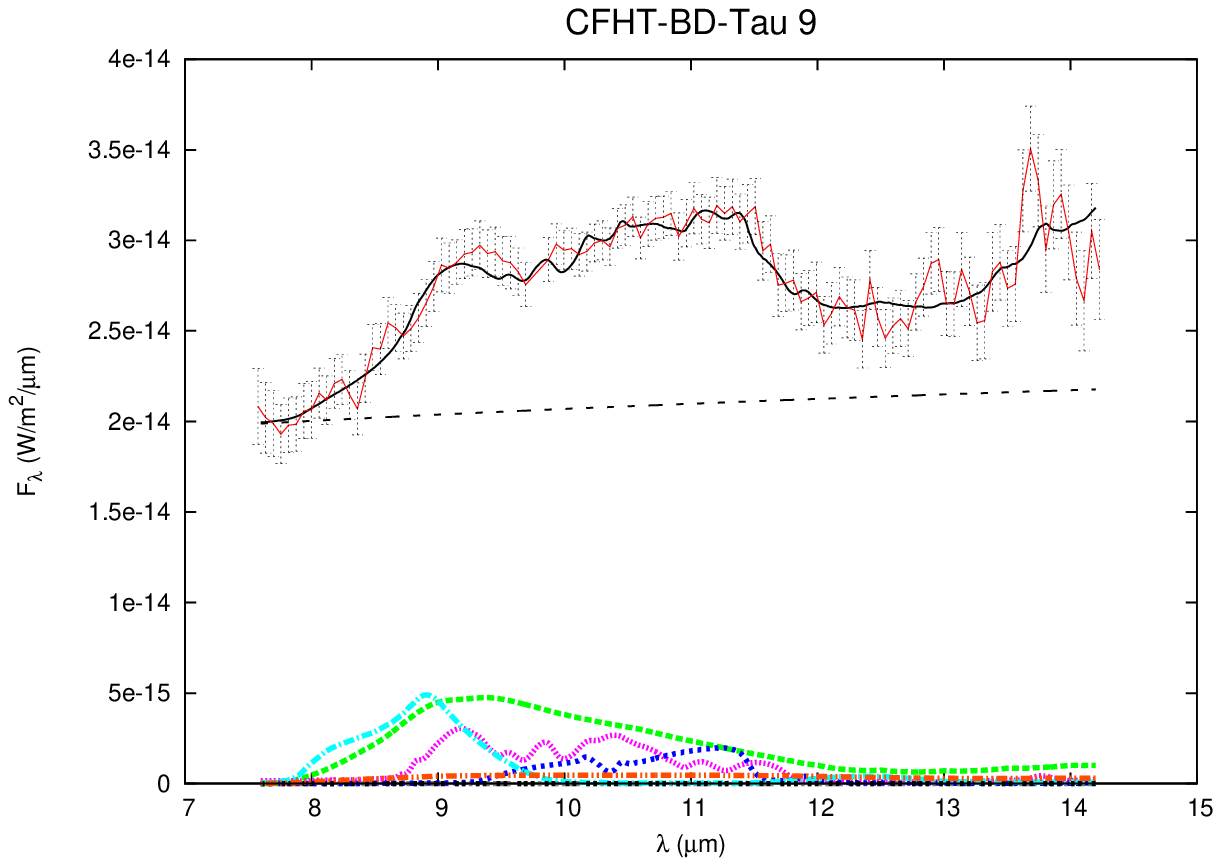}}   \\       
    \end{tabular}
    \caption{Model-fit to the 10$\micron$ silicate feature for Taurus brown dwarfs. Colors represent the following: red--observed spectrum; black--model fit, grey--small amorphous olivine; green--small amorphous pyroxene; cyan--silica; blue--forsterite; pink--enstatite; black dashed--large amorphous olivine; orange--large amorphous pyroxene. Thin dashed line represents the continuum. The spectra have been arranged in the same order as Fig.~\ref{cont-subt}.}
    \label{taurus}
  \end{center}
 \end{figure}

\begin{figure}
\addtocounter{figure}{-1}
 \begin{center}
    \begin{tabular}{ccc}   
      \resizebox{50mm}{!}{\includegraphics[angle=0]{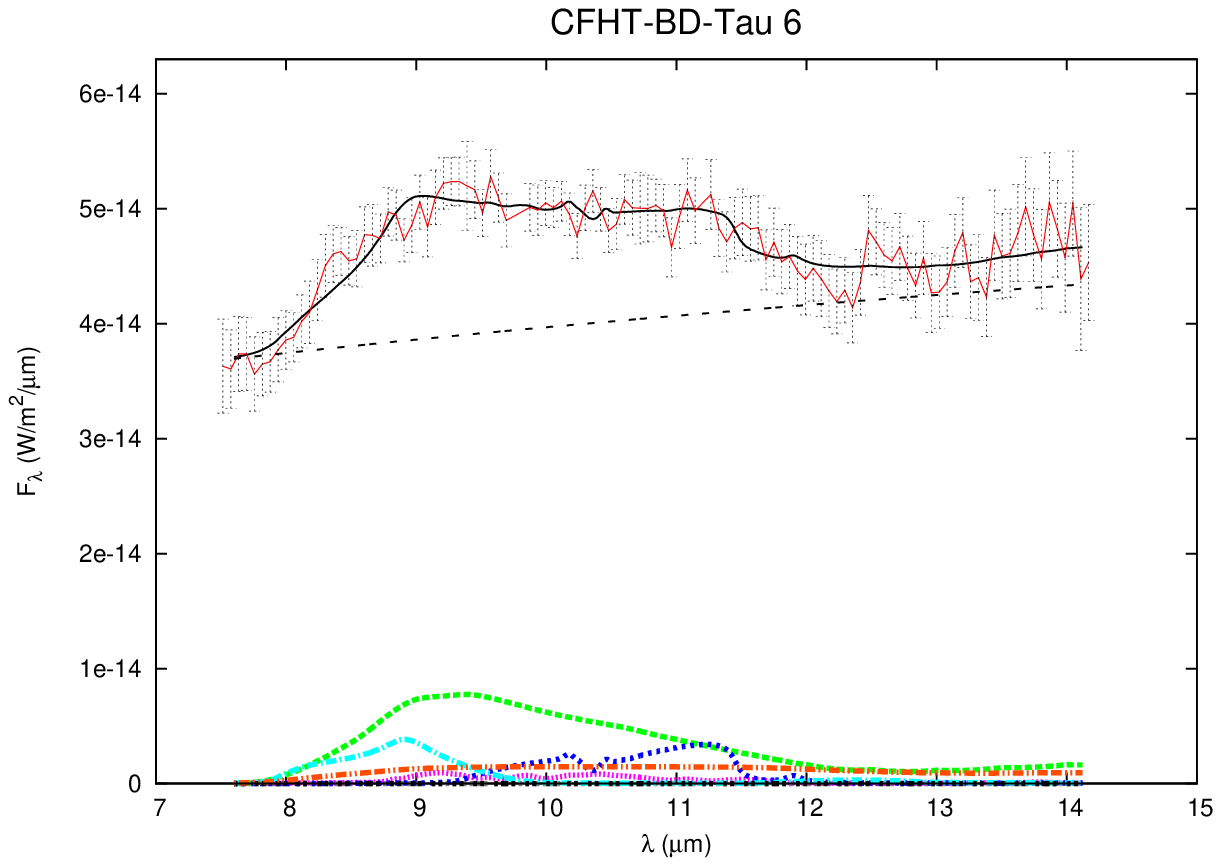}}  &  
      \resizebox{50mm}{!}{\includegraphics[angle=0]{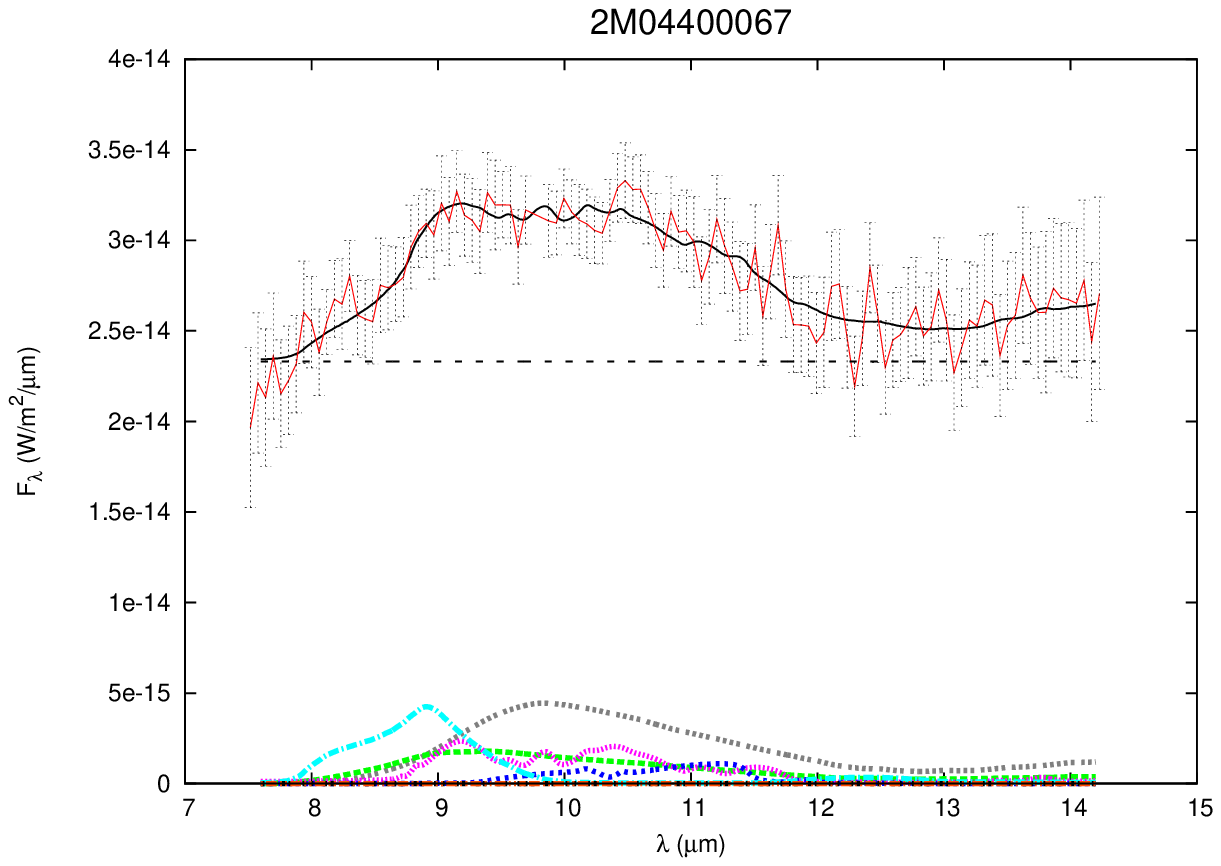}} &                    
      \resizebox{50mm}{!}{\includegraphics[angle=0]{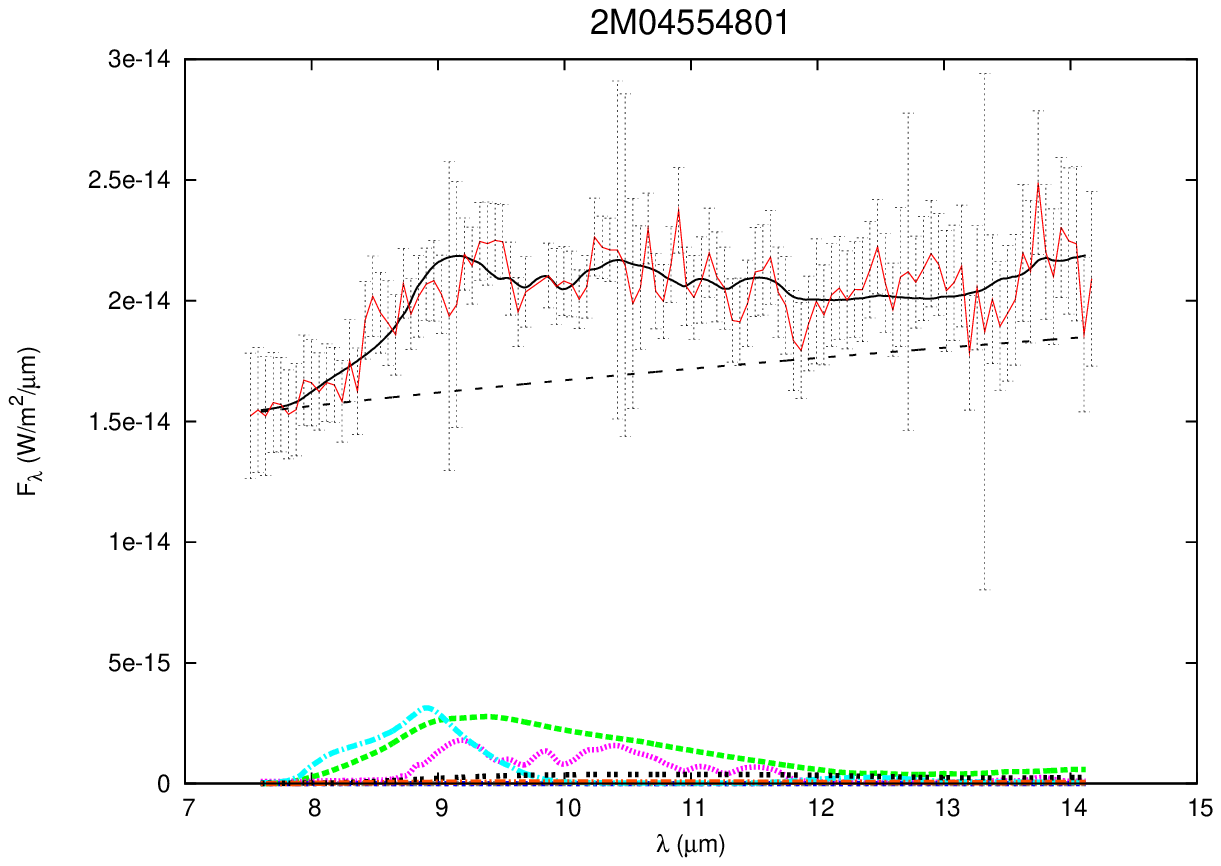}} \\
       \resizebox{50mm}{!}{\includegraphics[angle=0]{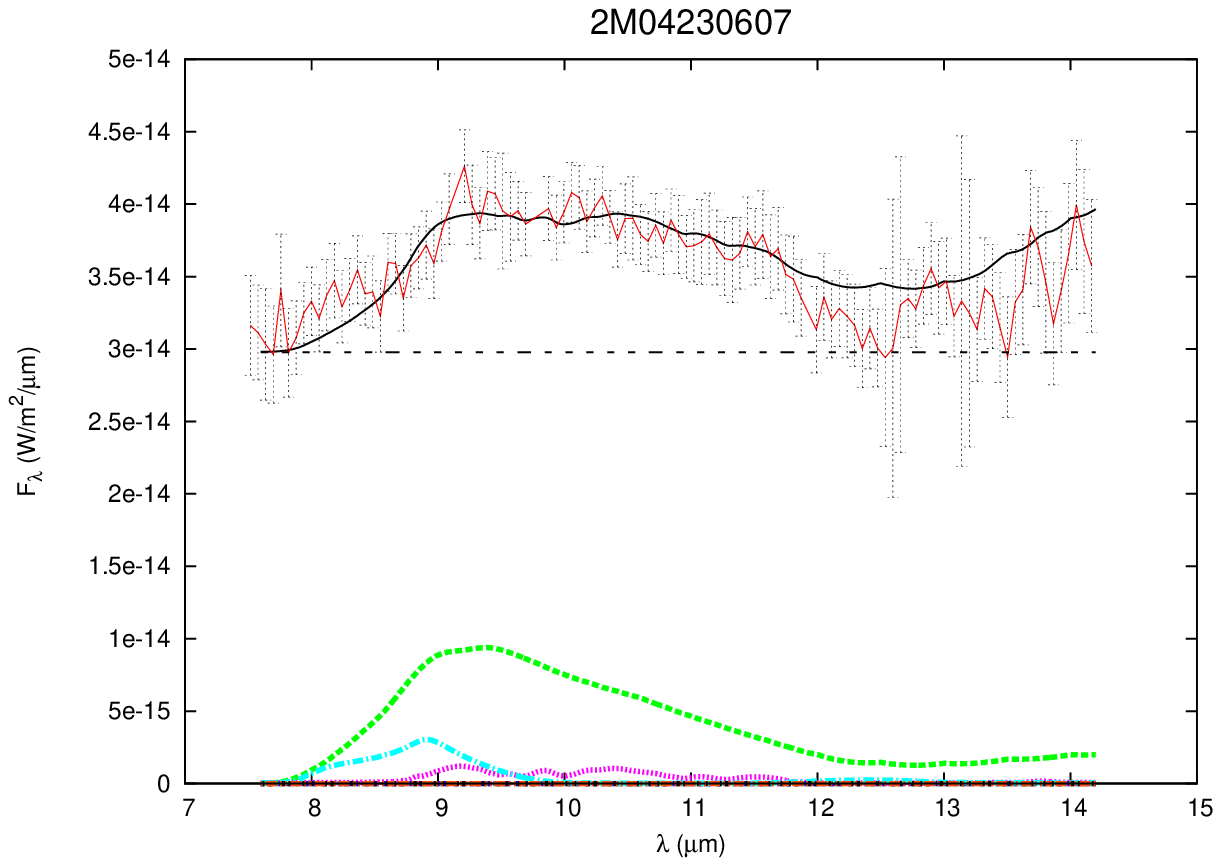}} &     
      \resizebox{50mm}{!}{\includegraphics[angle=0]{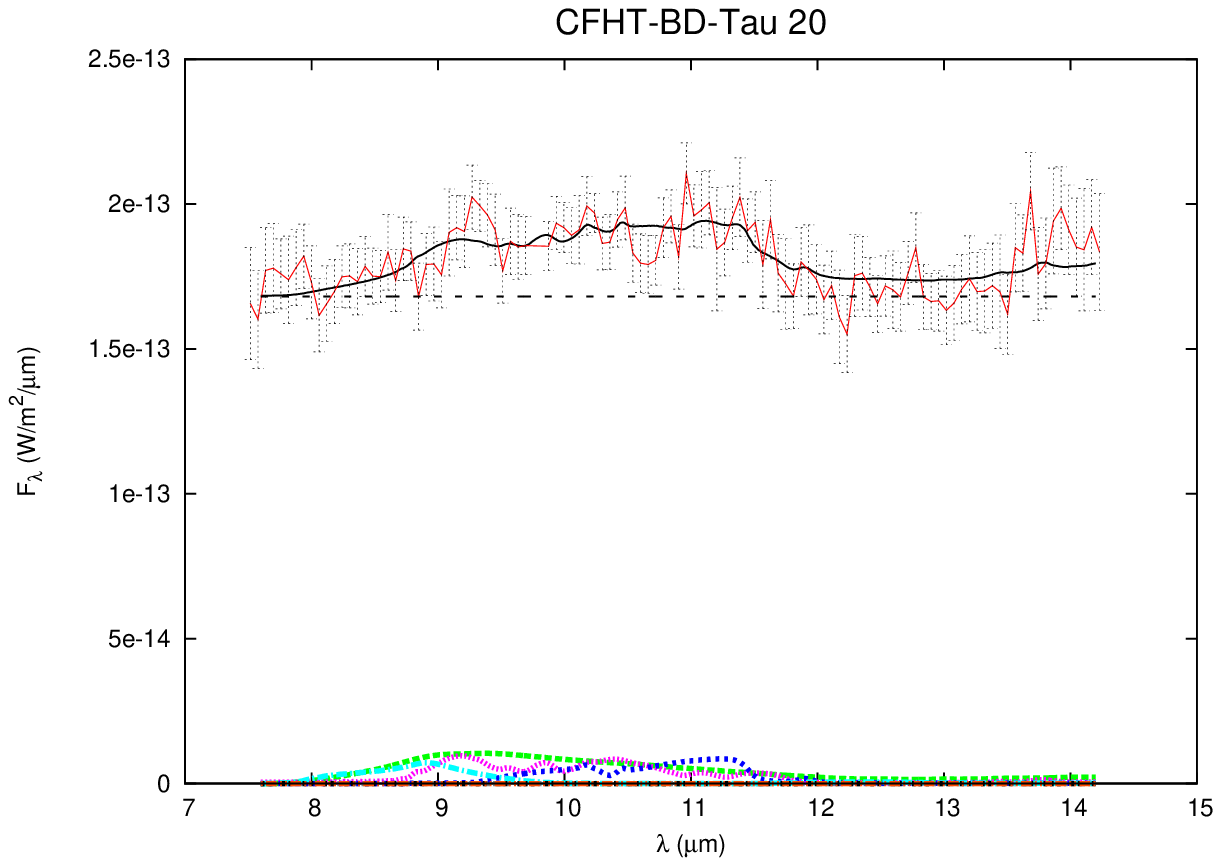}} &
      \resizebox{50mm}{!}{\includegraphics[angle=0]{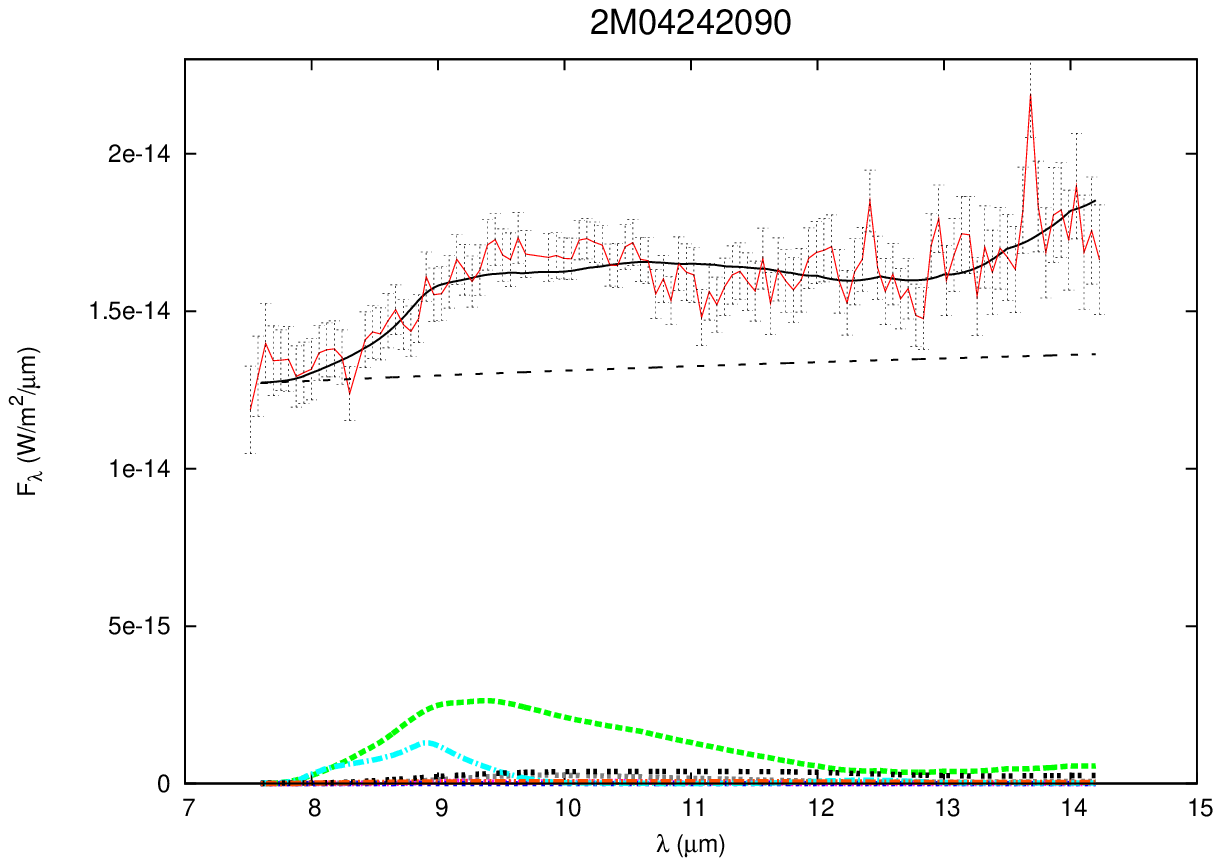}} \\   
      \resizebox{50mm}{!}{\includegraphics[angle=0]{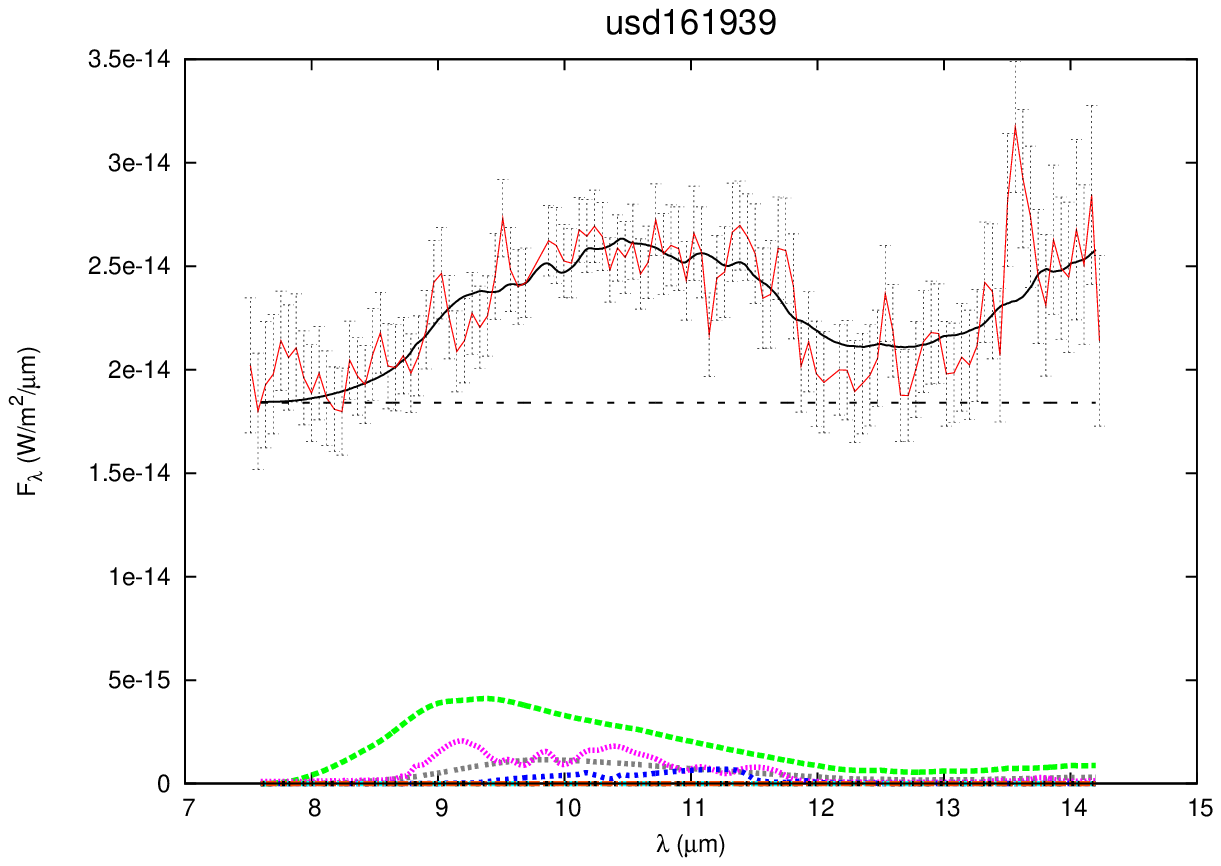}} & 
      \resizebox{50mm}{!}{\includegraphics[angle=0]{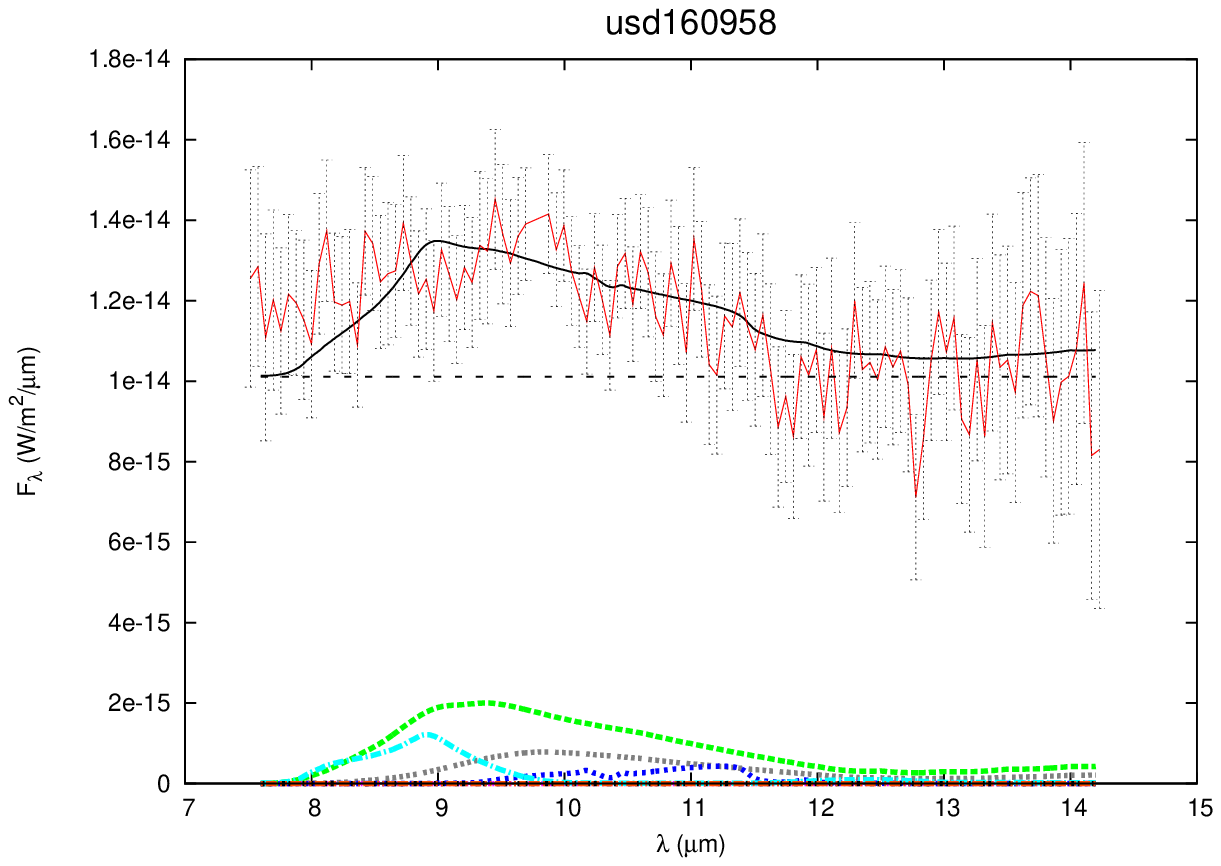}} &
      \resizebox{50mm}{!}{\includegraphics[angle=0]{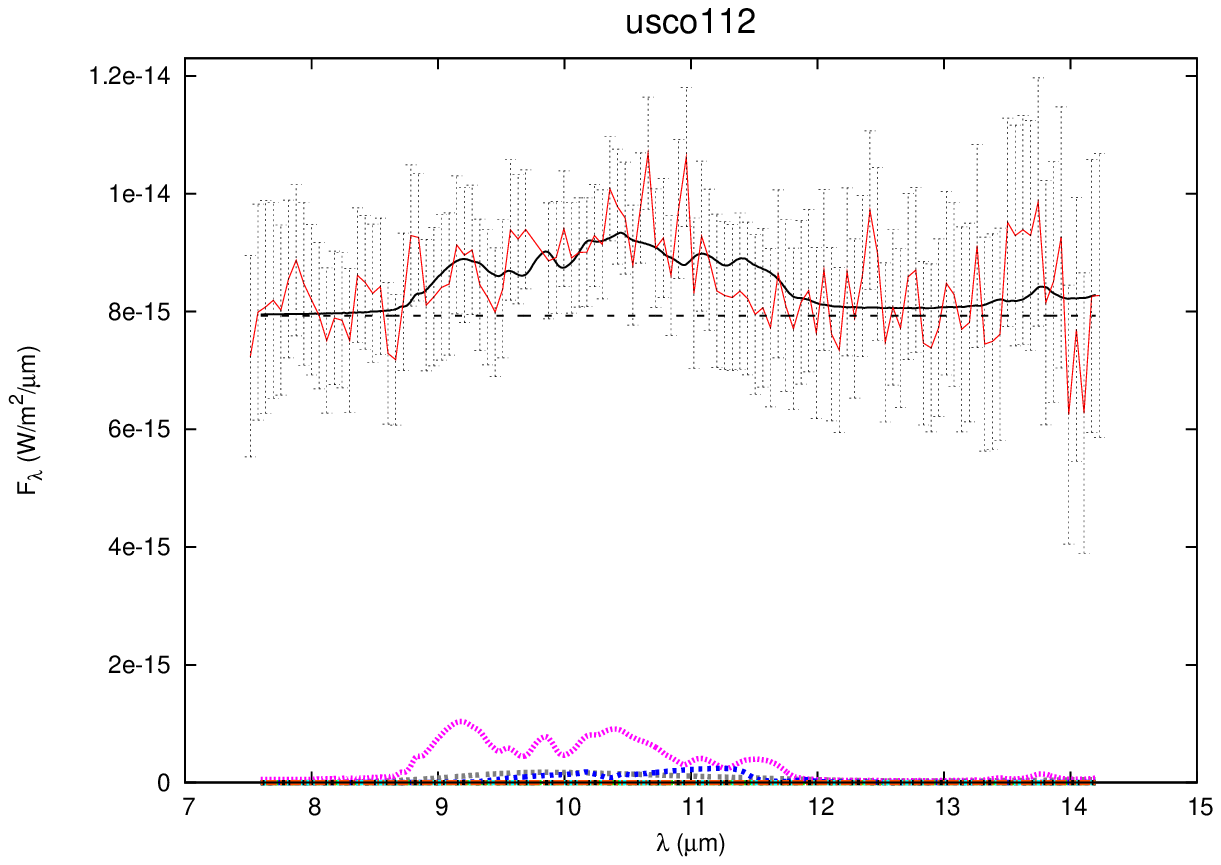}} \\     
    \end{tabular}
    \caption{{\it Continued}. Bottom panel shows the fits for the UppSco brown dwarfs.}
  \end{center}
 \end{figure}

\begin{figure}
 \begin{center}
    \begin{tabular}{ccc}      
      \resizebox{55mm}{!}{\includegraphics[angle=0]{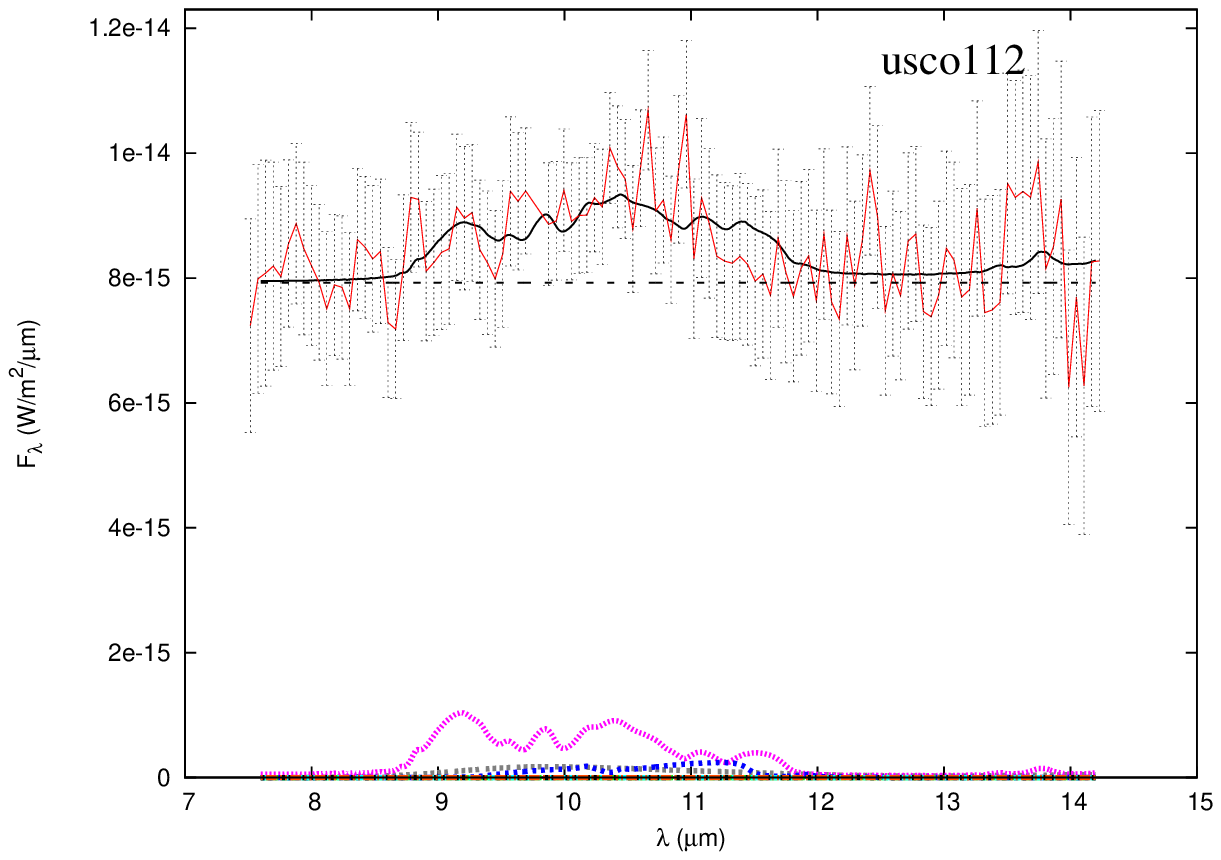}} &    
      \resizebox{55mm}{!}{\includegraphics[angle=0]{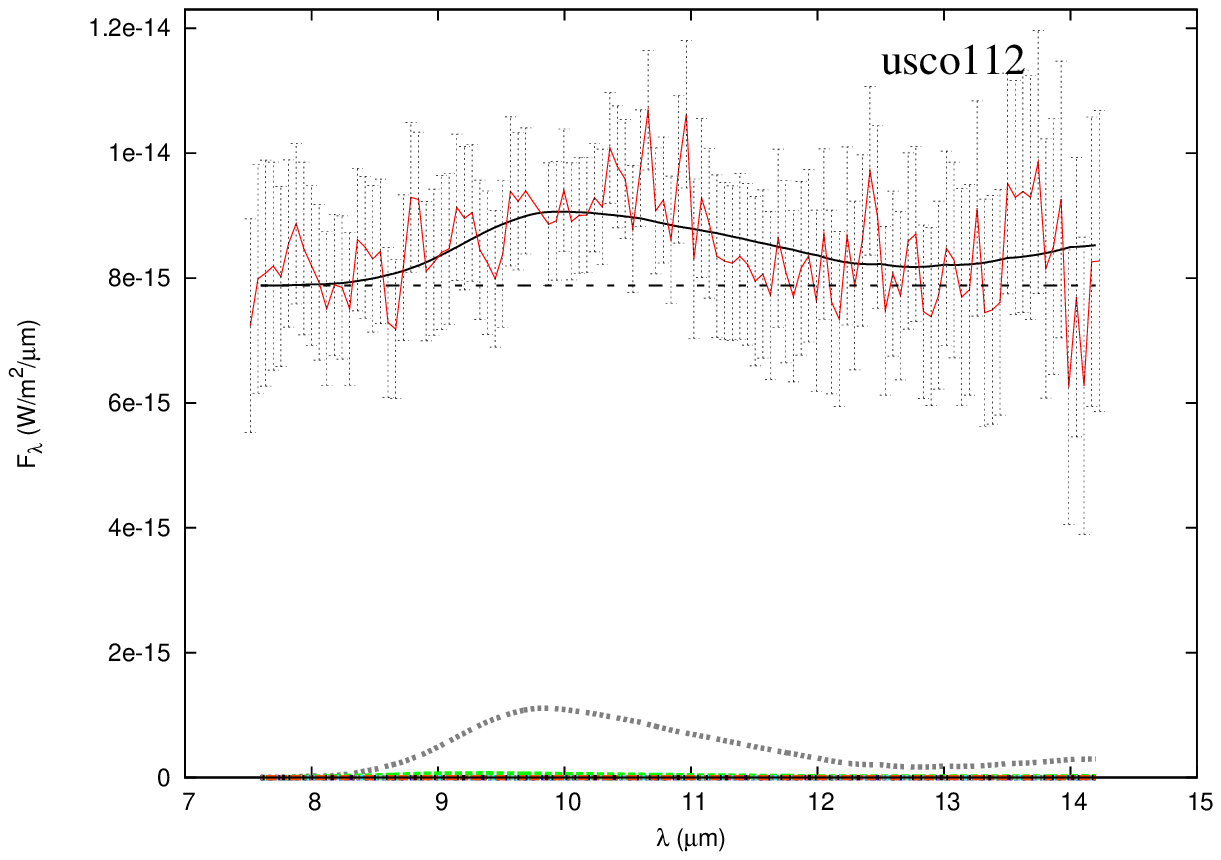}} & 
      \resizebox{55mm}{!}{\includegraphics[angle=0]{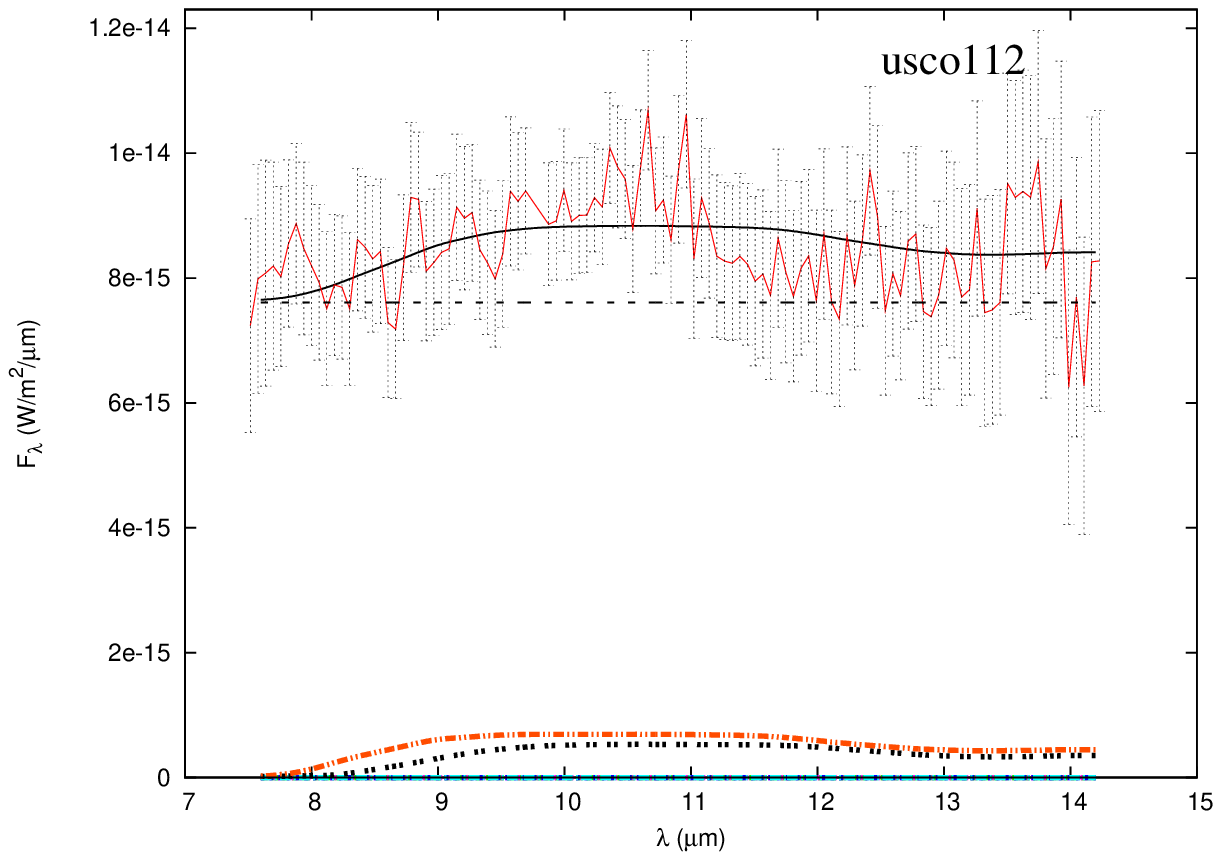}} \\
      \resizebox{55mm}{!}{\includegraphics[angle=0]{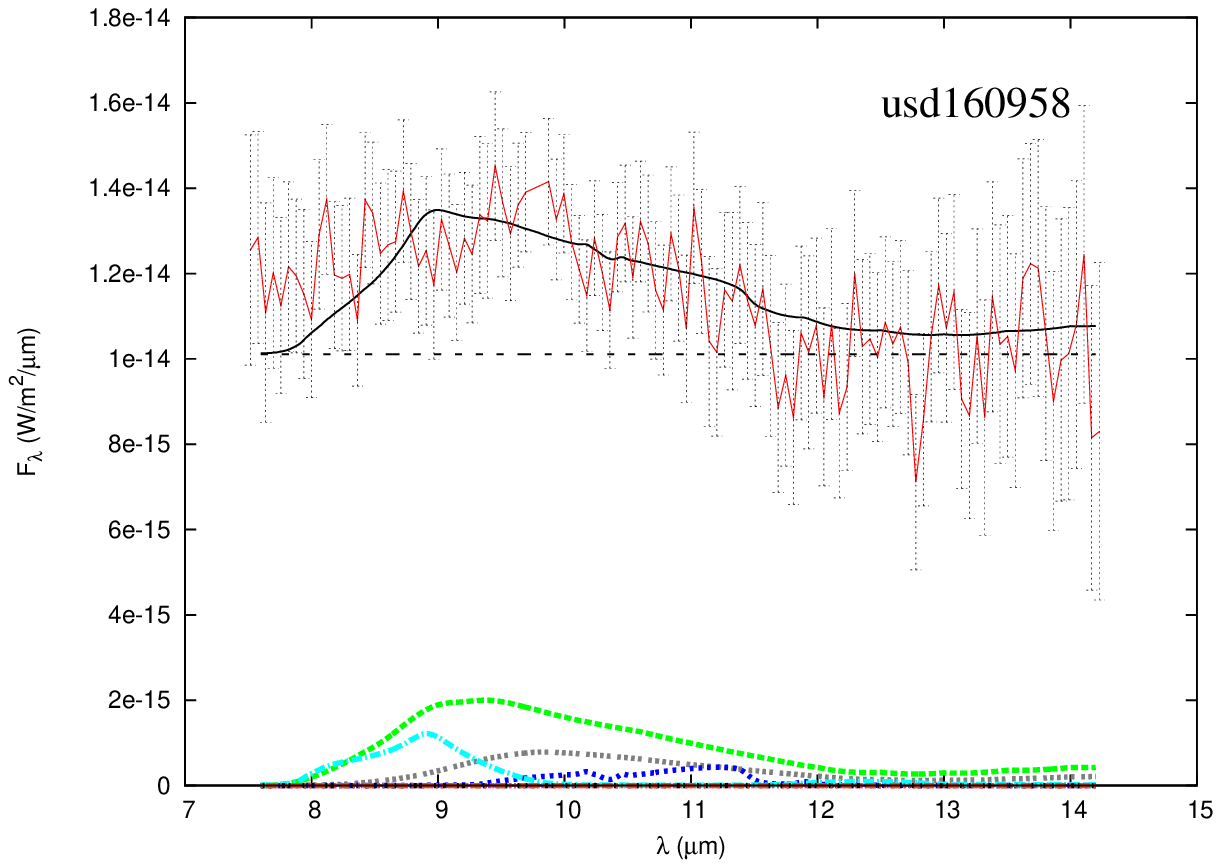}} &
      \resizebox{55mm}{!}{\includegraphics[angle=0]{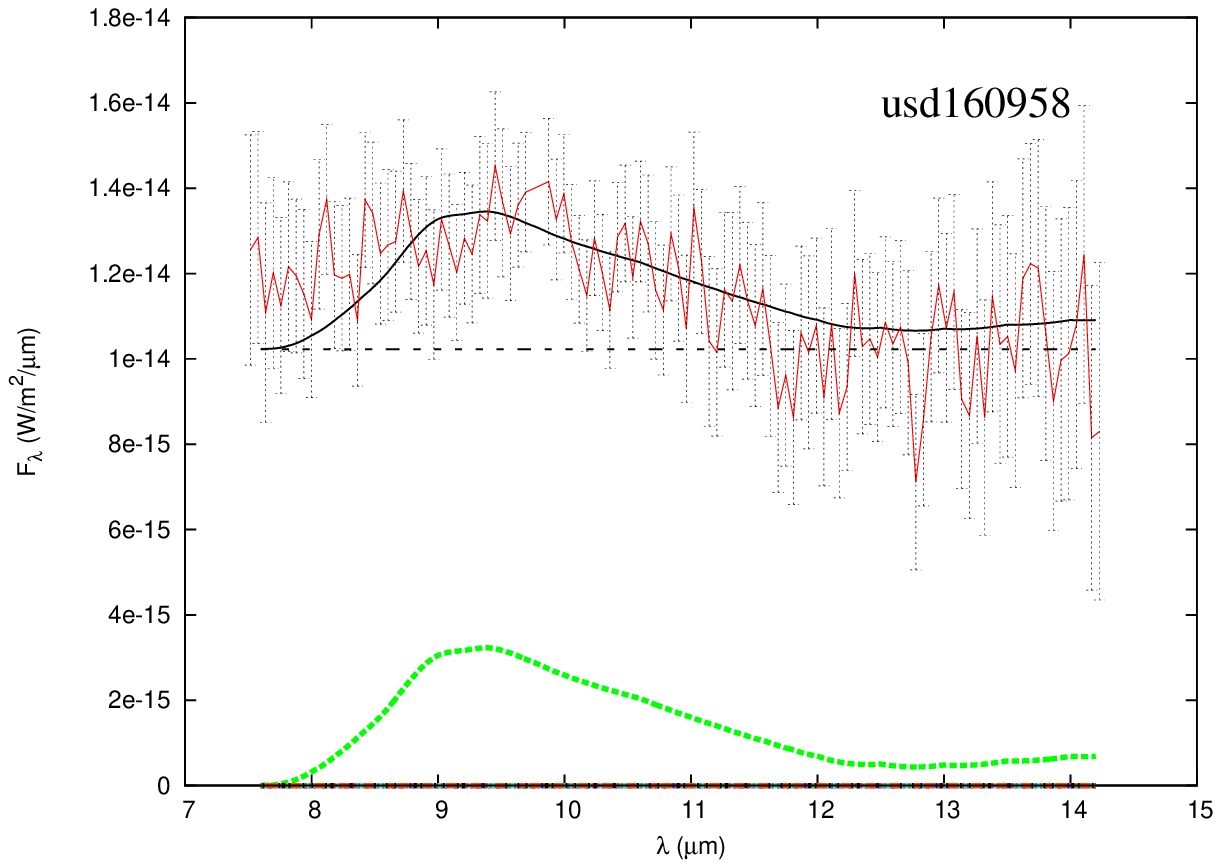}} &
      \resizebox{55mm}{!}{\includegraphics[angle=0]{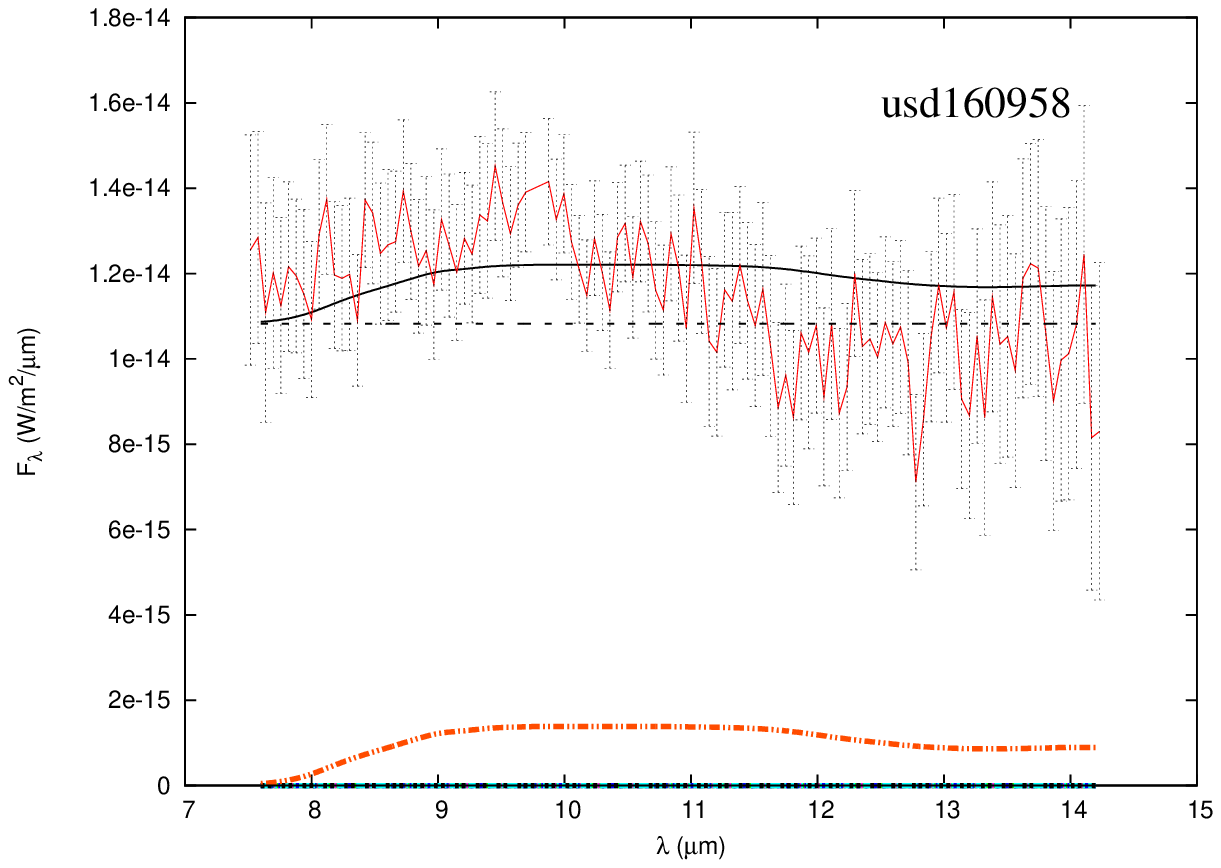}}  \\   
      \resizebox{55mm}{!}{\includegraphics[angle=0]{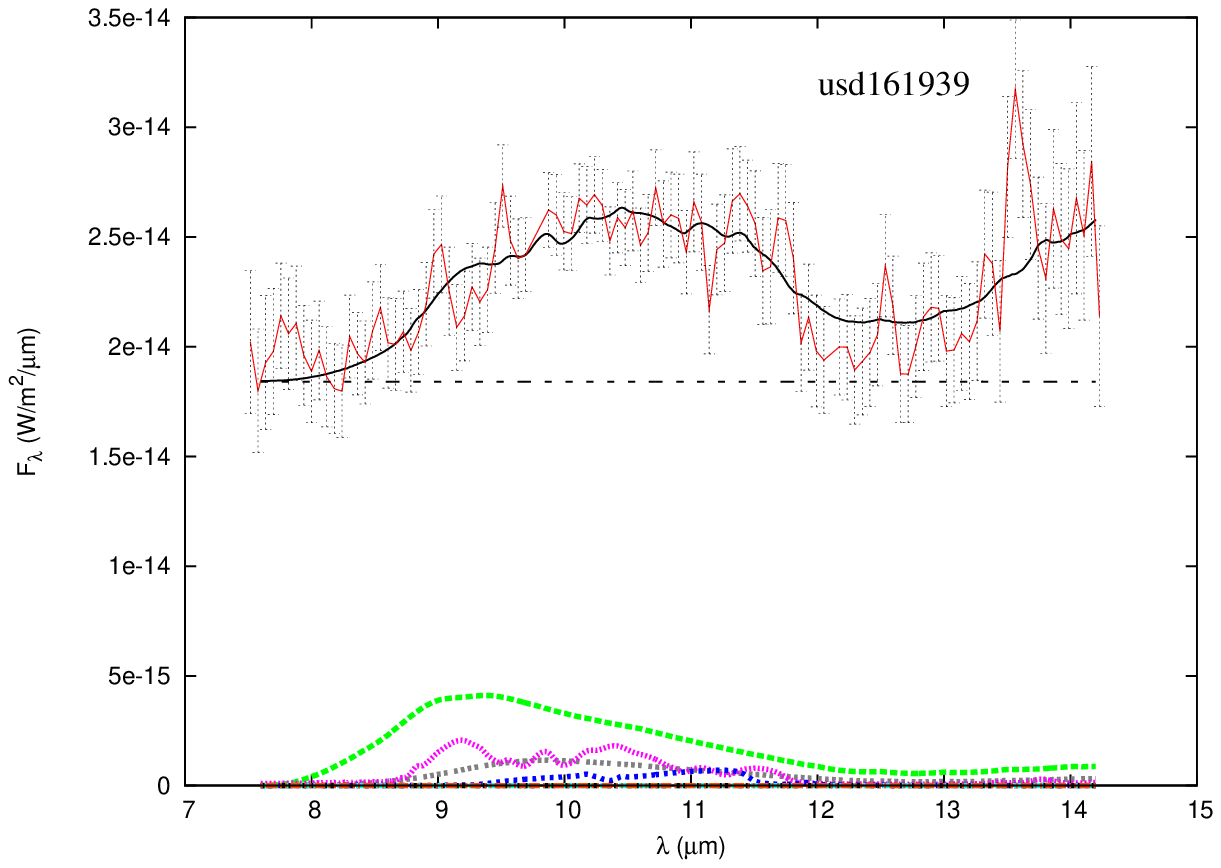}} &
      \resizebox{55mm}{!}{\includegraphics[angle=0]{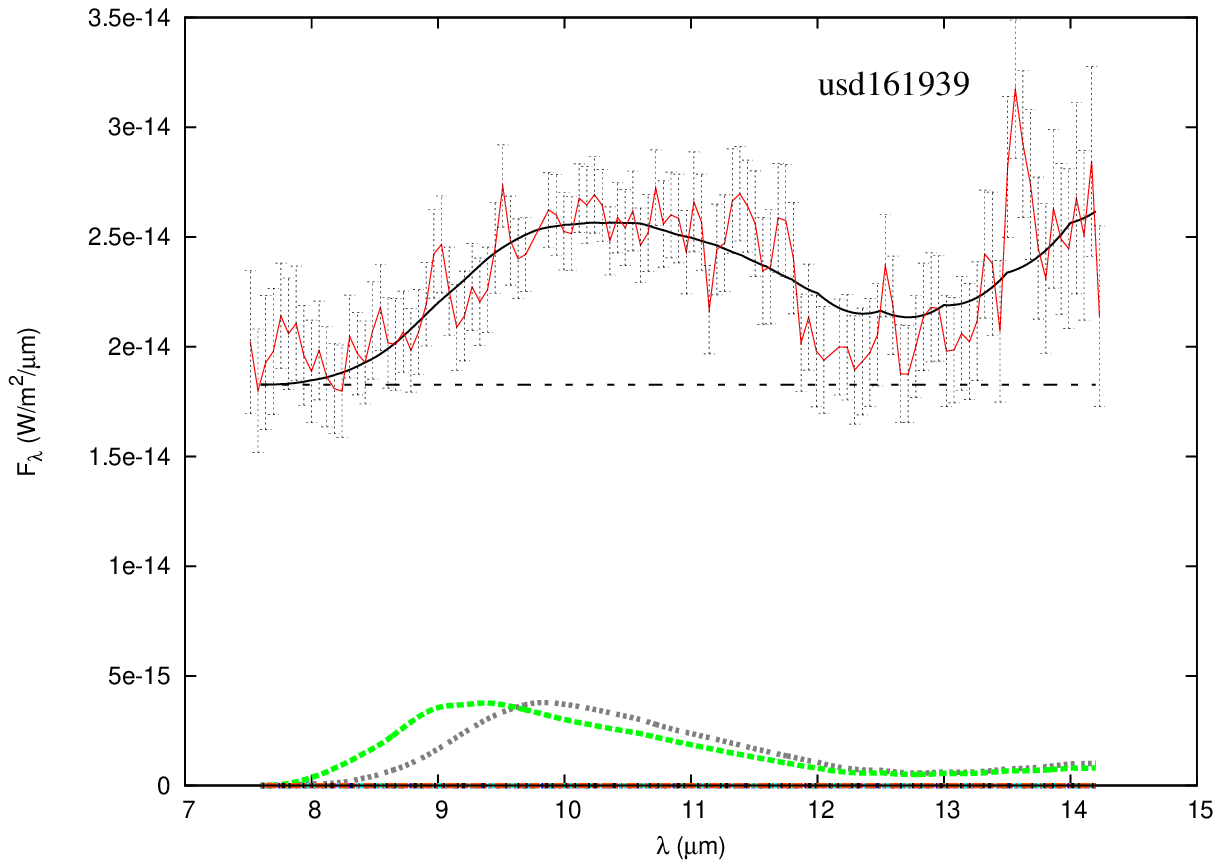}} &      
      \resizebox{55mm}{!}{\includegraphics[angle=0]{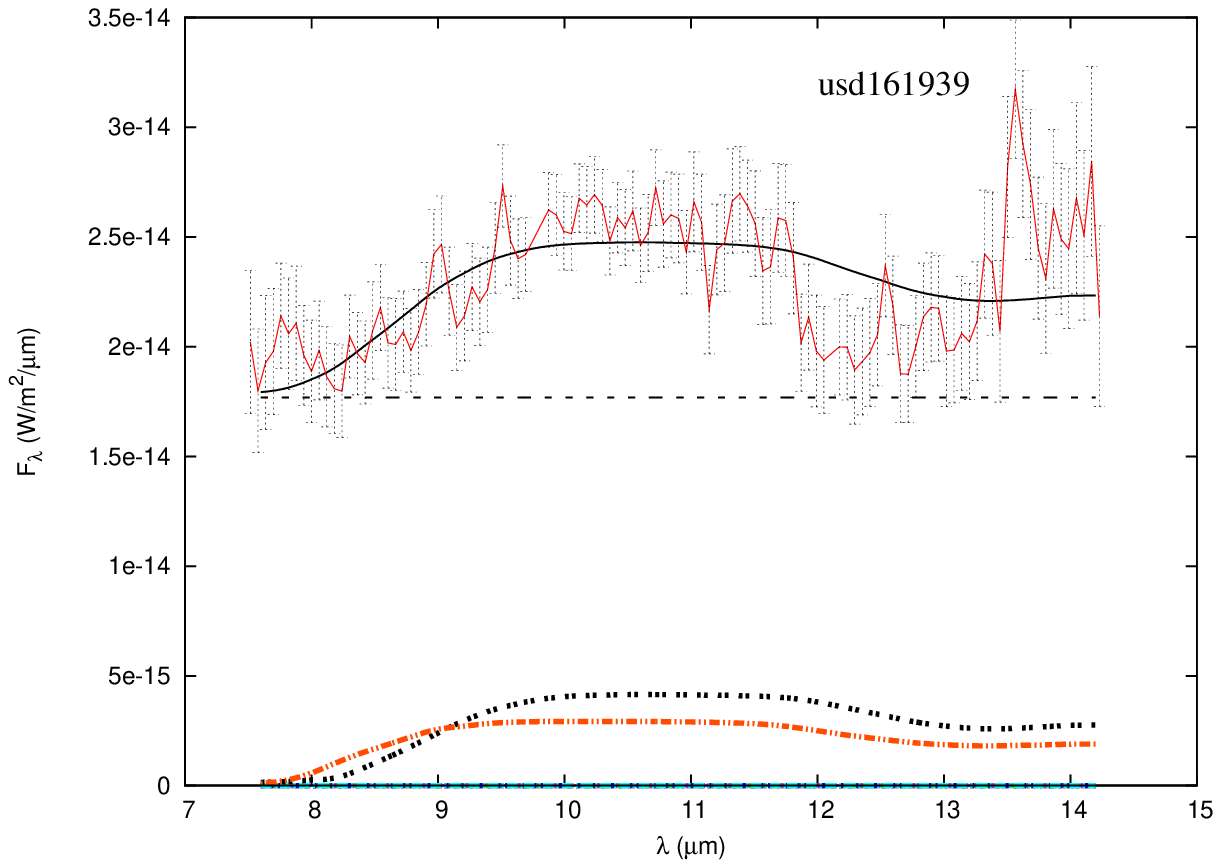}}  \\
      \resizebox{55mm}{!}{\includegraphics[angle=0]{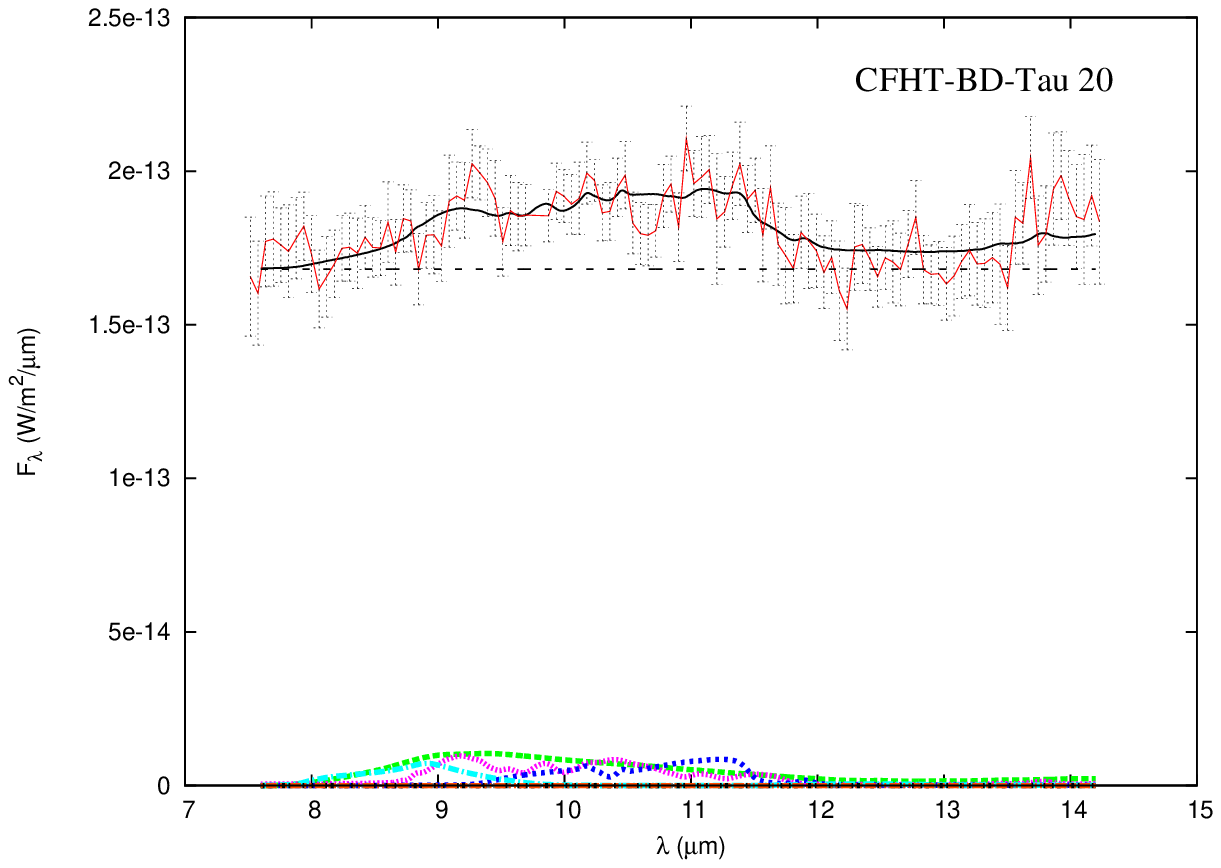}}  & 
      \resizebox{55mm}{!}{\includegraphics[angle=0]{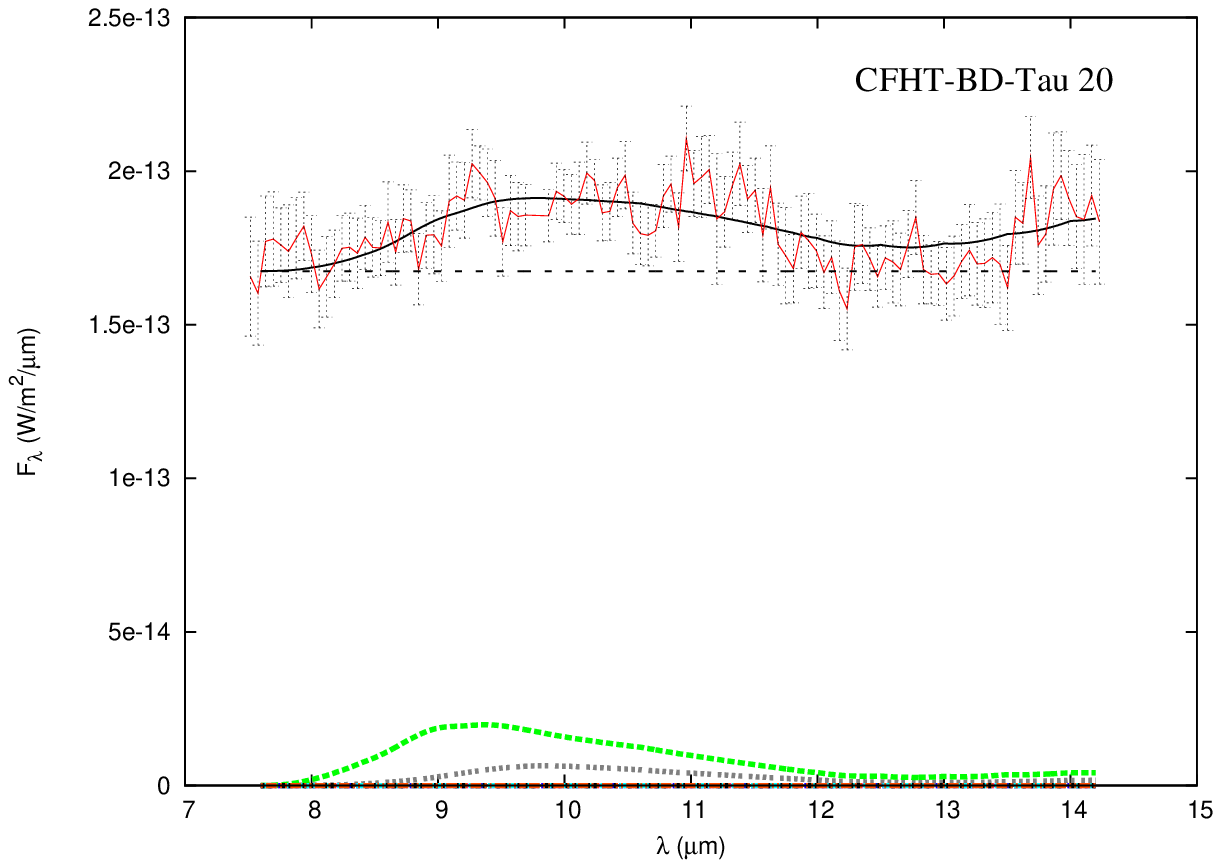}} &                    
      \resizebox{55mm}{!}{\includegraphics[angle=0]{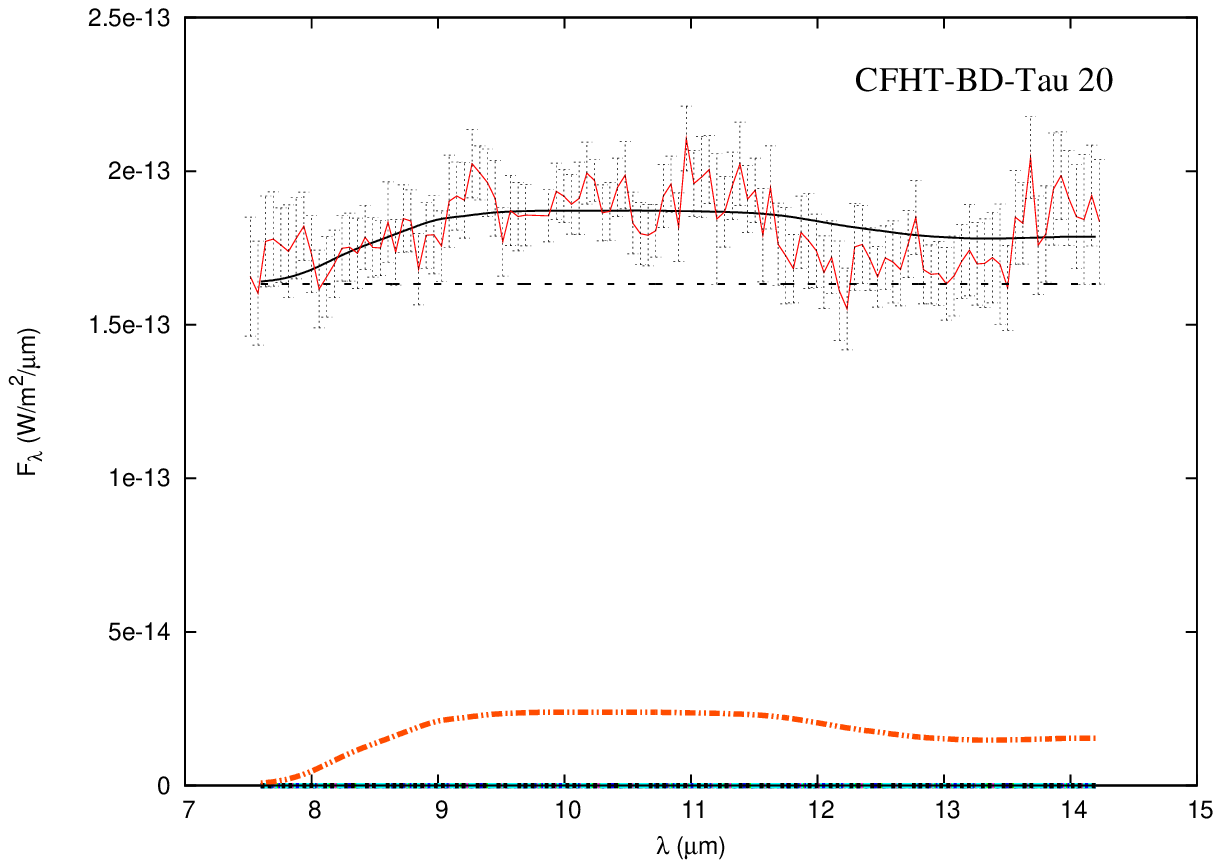}} \\
    \end{tabular}
    \caption{(a): Model-fits for objects with negligible large grain mass fractions. Left panel shows the best-fit with the lowest reduced-$\chi^{2}$ value, obtained using all five dust species. Middle panel shows the fits obtained without using any crystalline silicates. Right panel shows the fits obtained using only large amorphous olivine and pyroxene grains.}
    \label{nolarge}
  \end{center}
 \end{figure}

\begin{figure}
\addtocounter{figure}{-1}
 \begin{center}
    \begin{tabular}{ccc}   
     \resizebox{80mm}{!}{\includegraphics[angle=0]{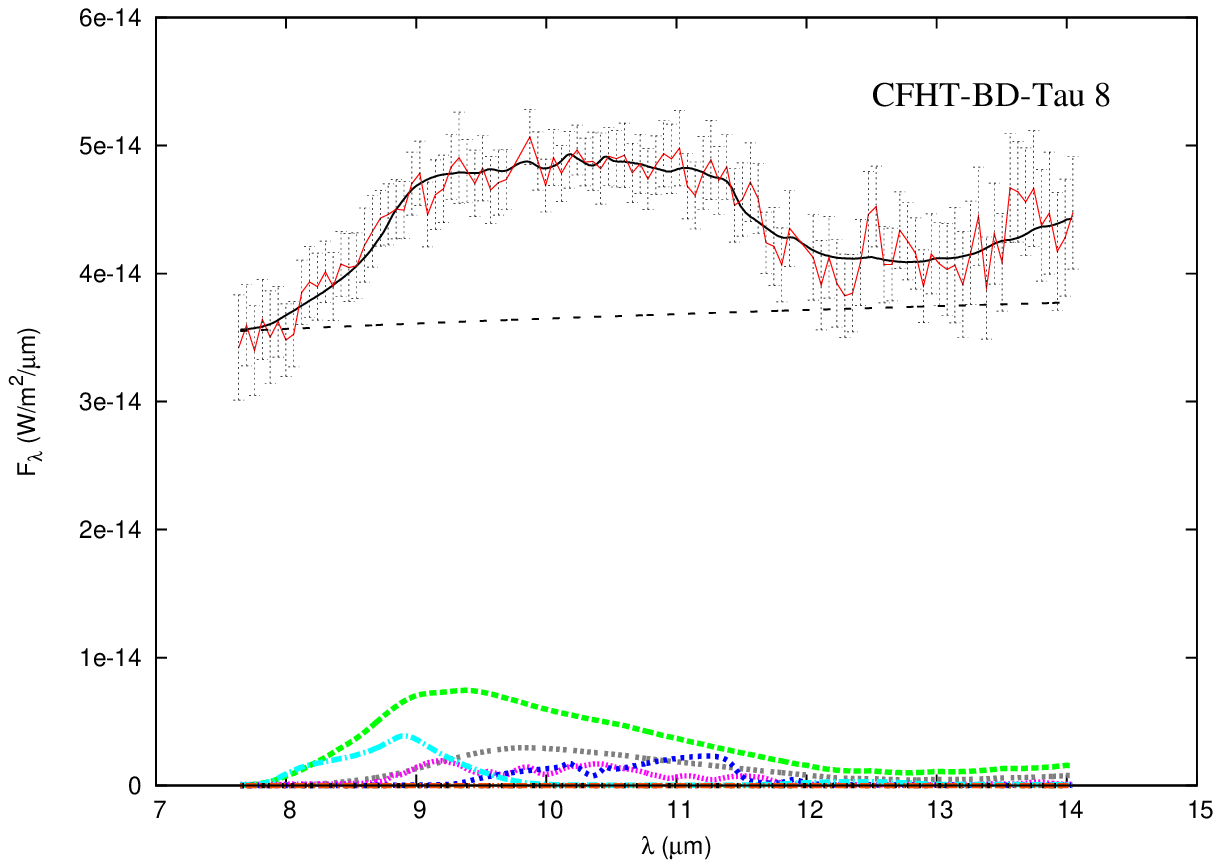}} &     
      \resizebox{80mm}{!}{\includegraphics[angle=0]{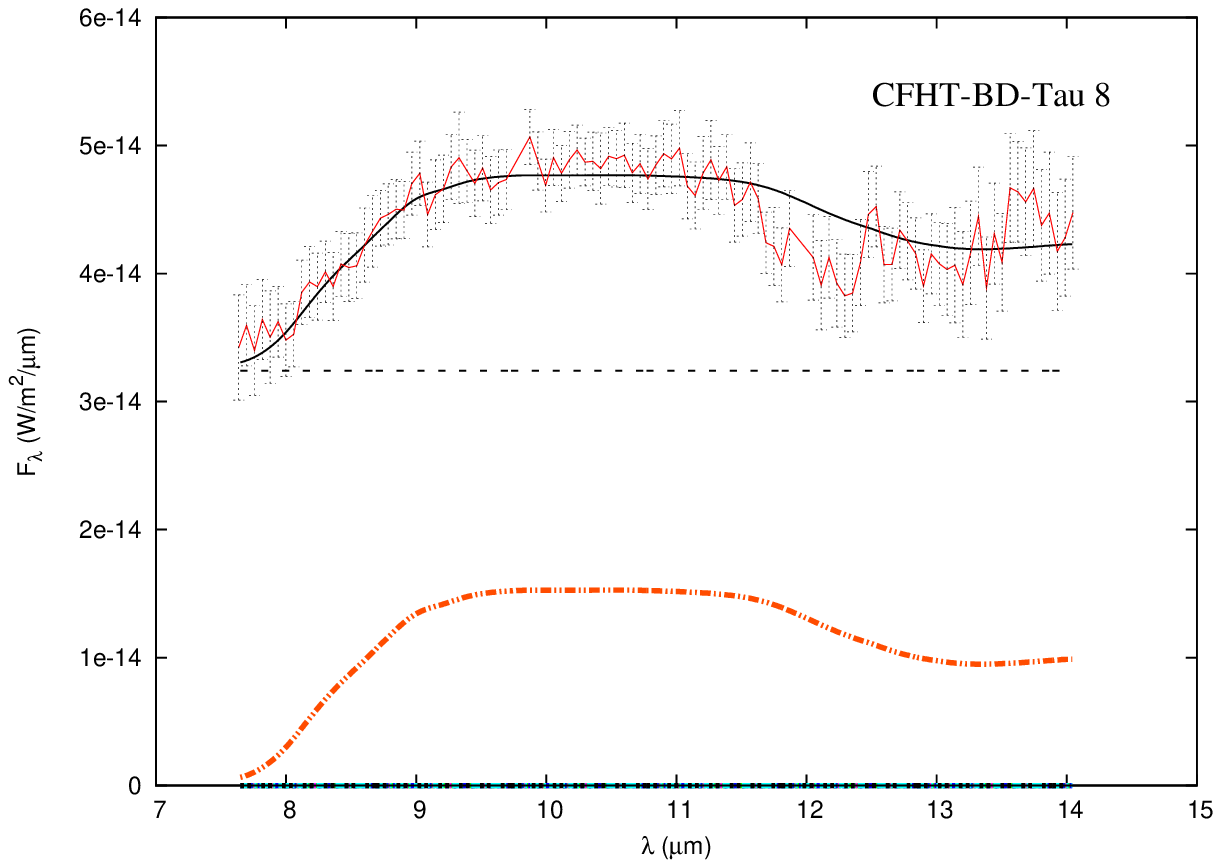}} \\
      \resizebox{80mm}{!}{\includegraphics[angle=0]{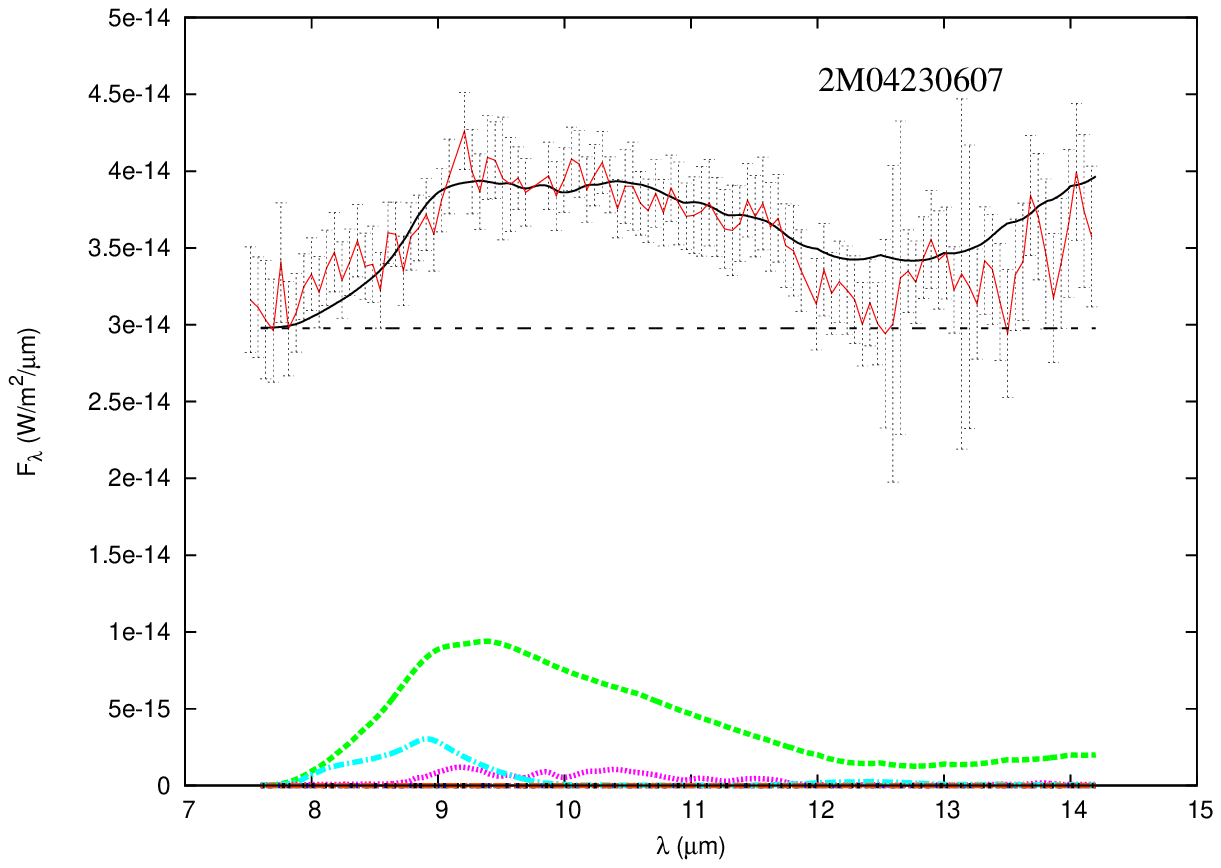}}  &
      \resizebox{80mm}{!}{\includegraphics[angle=0]{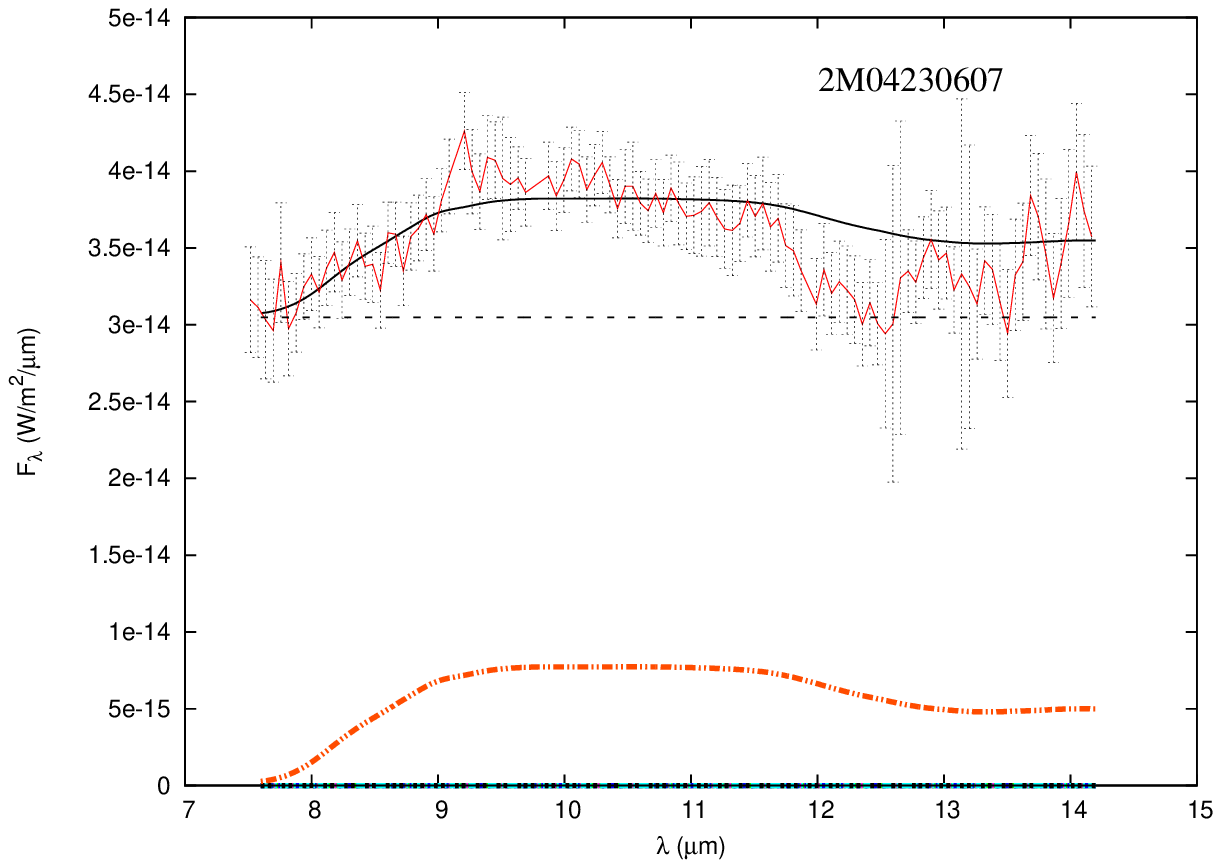}}  \\ 
      \resizebox{80mm}{!}{\includegraphics[angle=0]{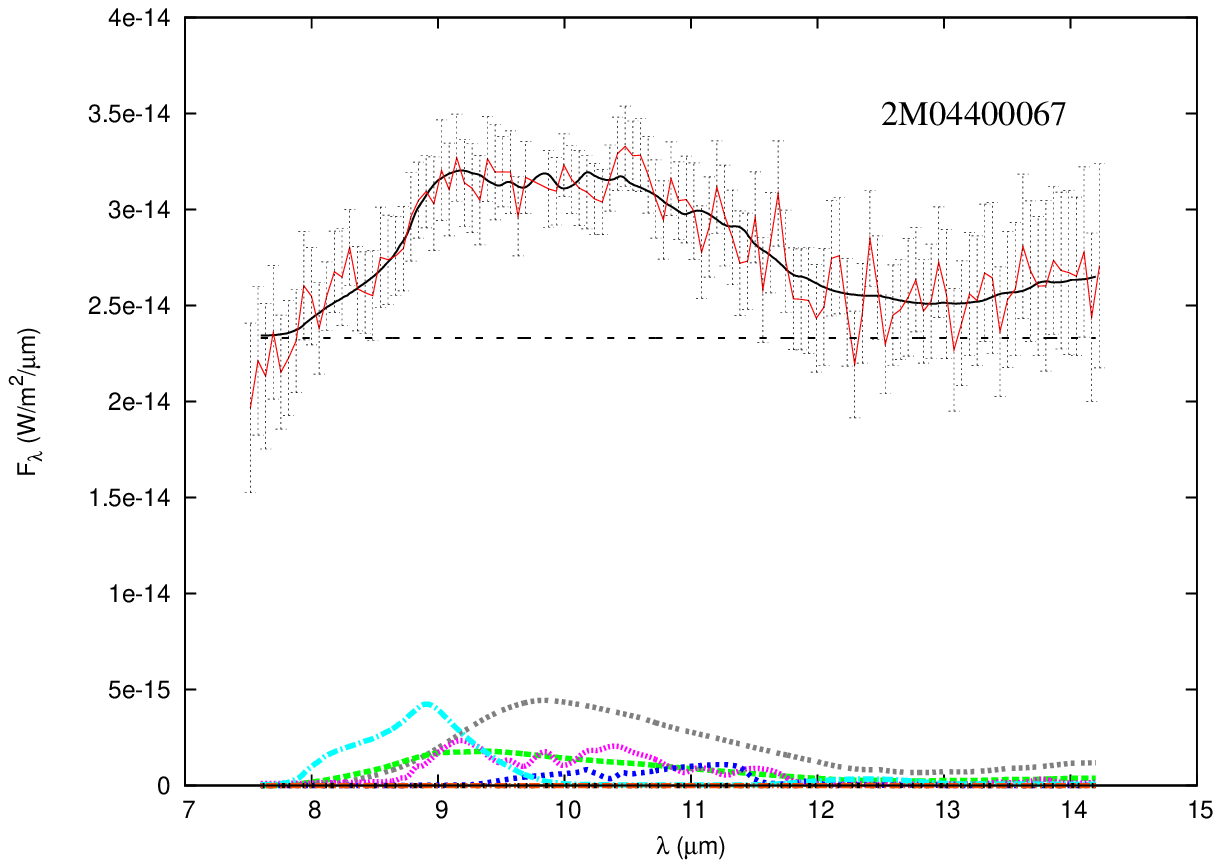}} &                    
      \resizebox{80mm}{!}{\includegraphics[angle=0]{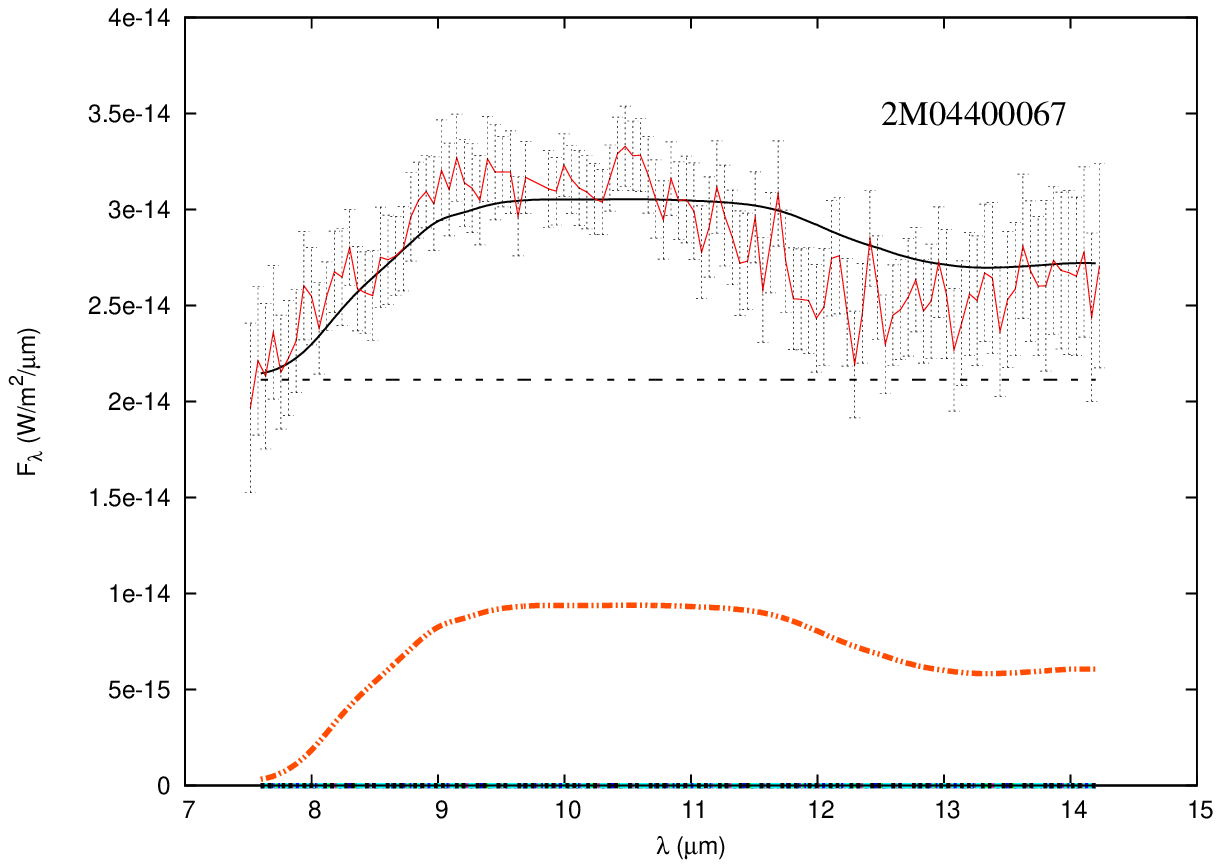}} \\      
    \end{tabular}
    \caption{(b): Model-fits for objects with negligible large grain mass fractions. Left panel shows the best-fit with the lowest reduced-$\chi^{2}$ value, obtained using all five dust species. Right panel shows the fits obtained using only large amorphous olivine and pyroxene grains.}
  \end{center}
 \end{figure}
 
 \begin{figure}
 \begin{center}
    \begin{tabular}{cc}     
      \resizebox{80mm}{!}{\includegraphics[angle=0]{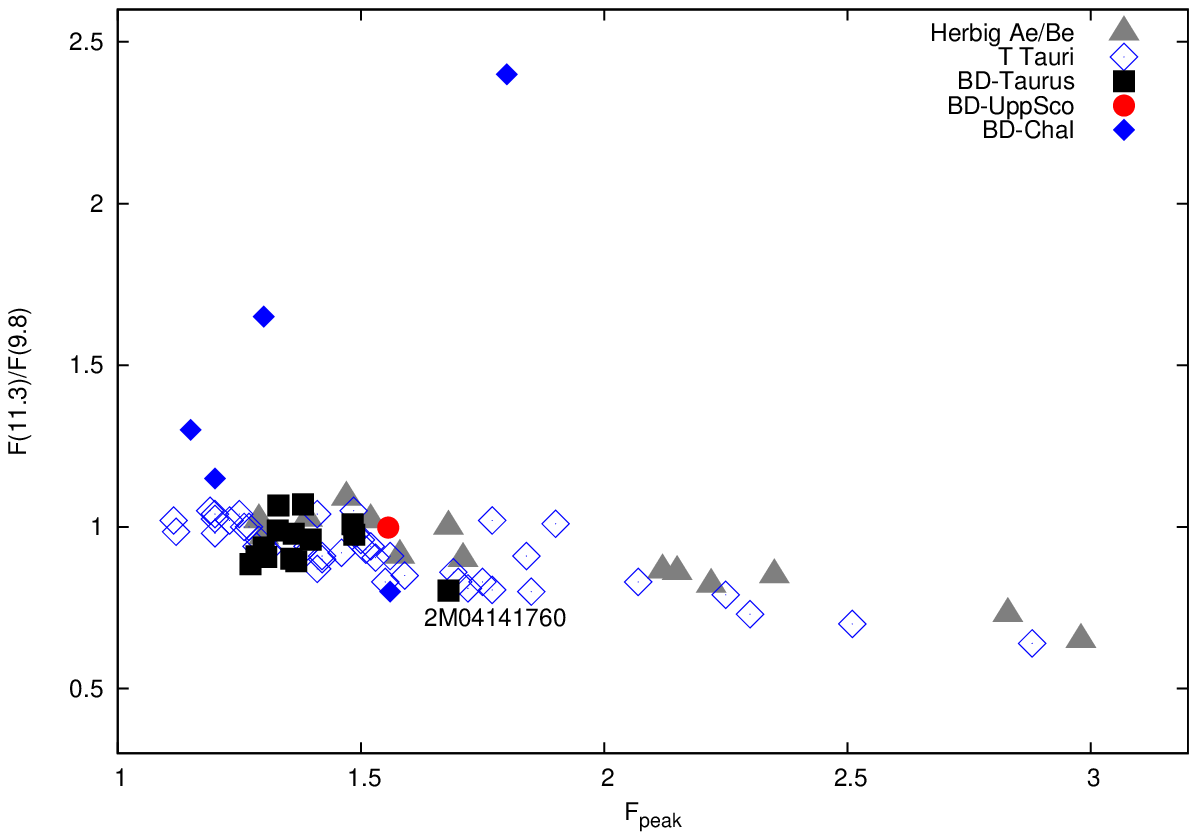}} \\   
      \resizebox{80mm}{!}{\includegraphics[angle=0]{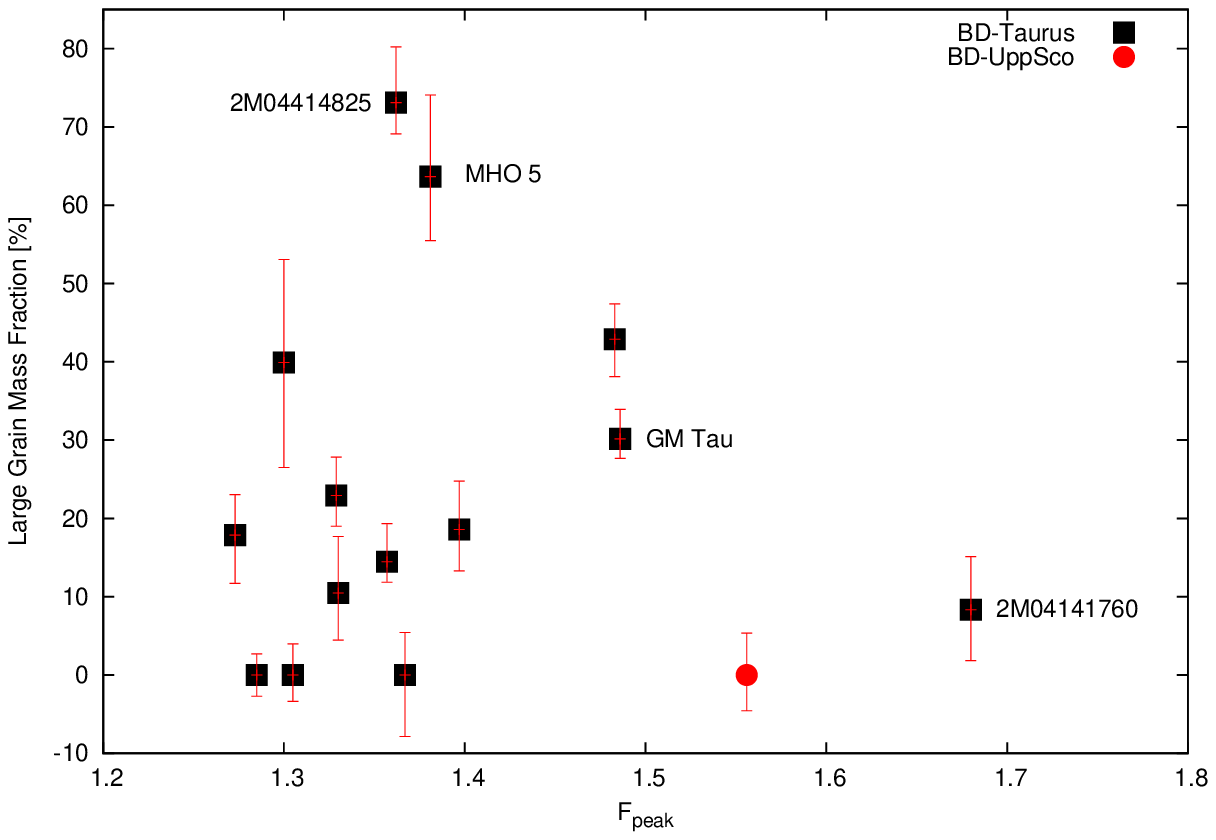}} &
      \resizebox{80mm}{!}{\includegraphics[angle=0]{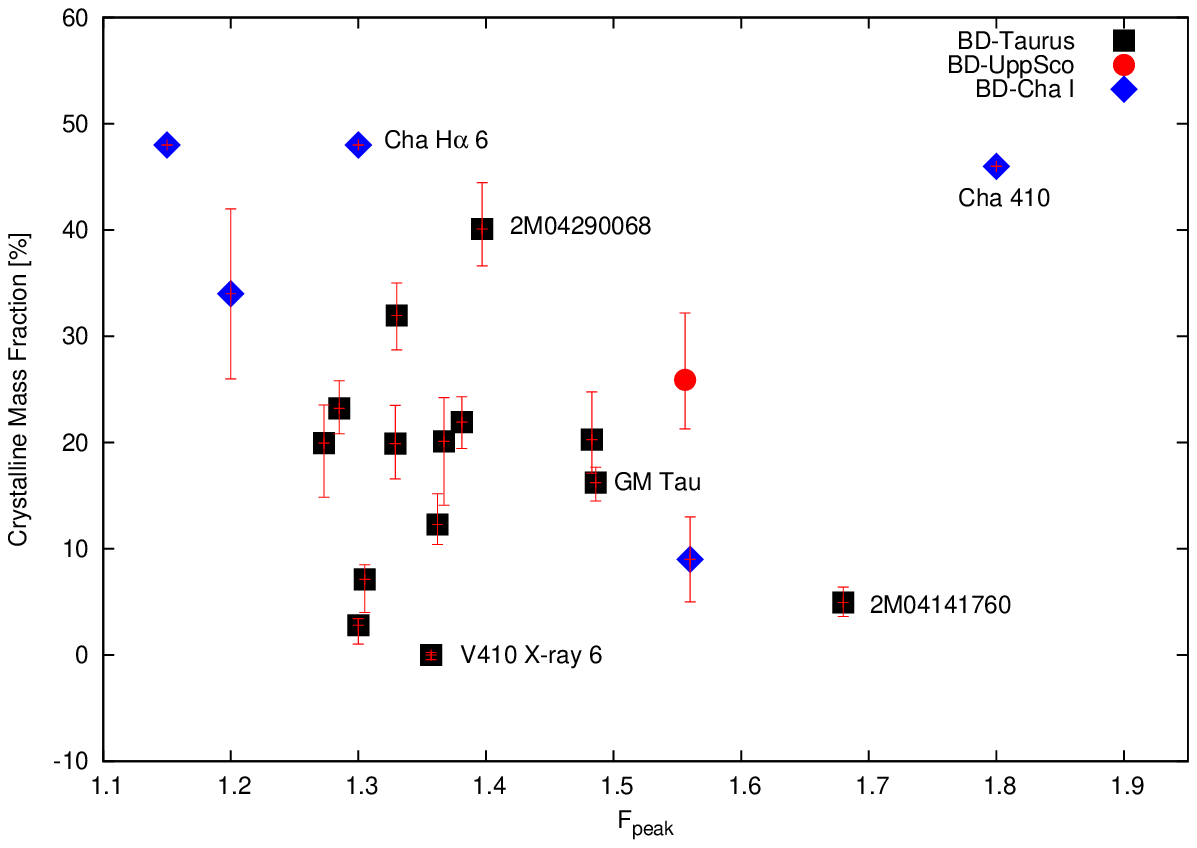}} \\         
      \resizebox{80mm}{!}{\includegraphics[angle=0]{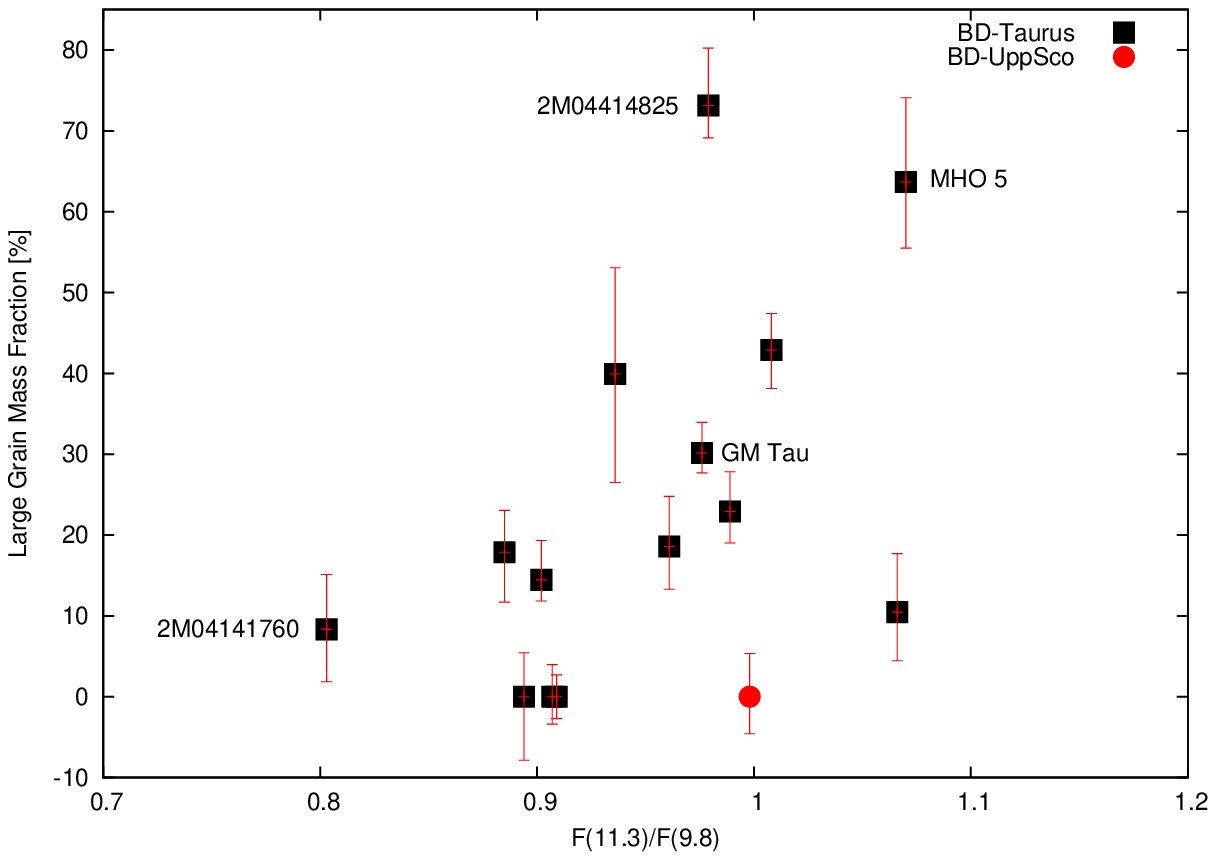}}  &
      \resizebox{80mm}{!}{\includegraphics[angle=0]{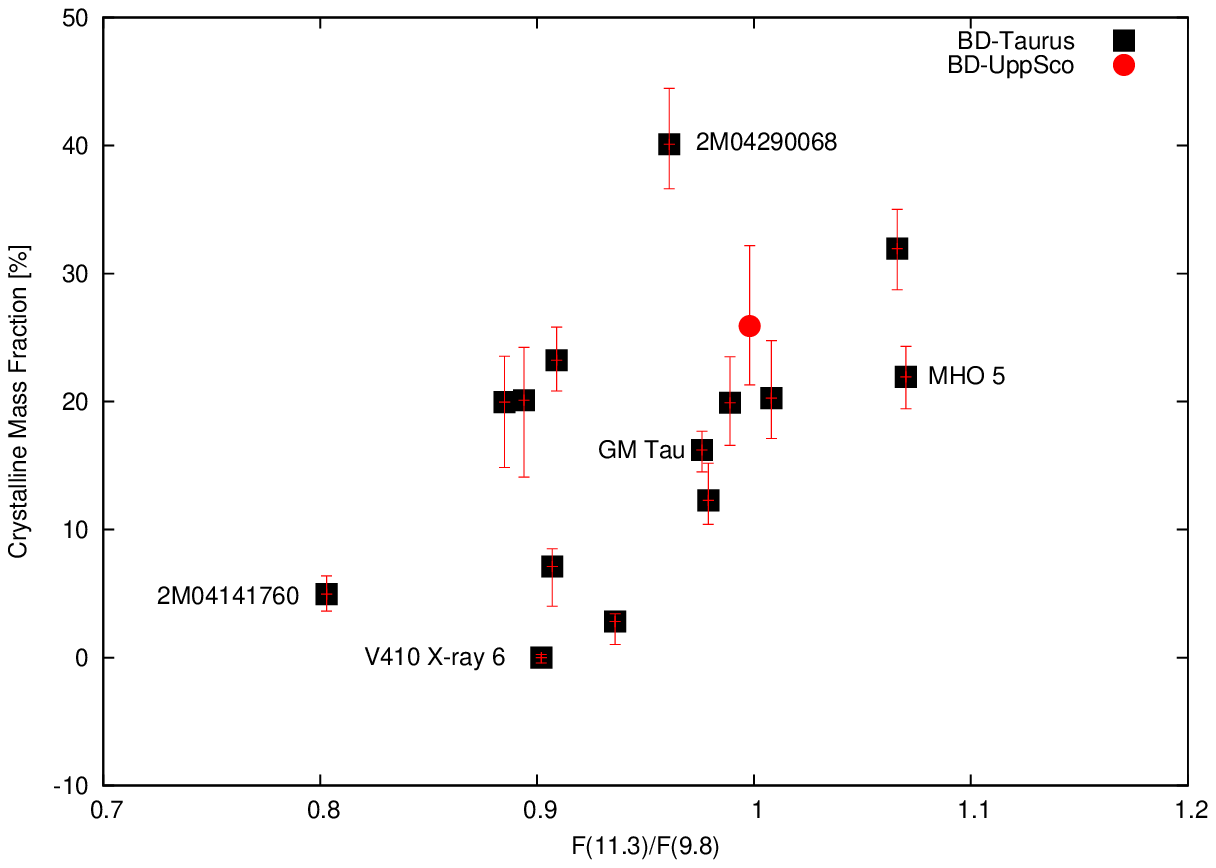}} \\         
    \end{tabular}
    \caption{{\it Top}: (a) The shape vs. strength of the silicate emission features. Taurus brown dwarfs are indicated by filled black squares, UppSco ones by filled red circles. Included for comparison are T Tauri stars from Pascucci et al. (2009) and Herbig Ae/Be stars from van Boekel et al. (2005). {\it Middle, left}: (b) $F_{peak}$ vs. the large grain mass fraction. {\it Middle, right}: (c) $F_{peak}$ vs. the crystalline mass fraction. Blue diamonds indicate Cha I brown dwarfs from Apai et al. (2005). {\it Bottom, left}: (d) $F_{11.3}/F_{9.8}$ vs. the large grain mass fraction. {\it Bottom, right}: (e) $F_{11.3}/F_{9.8}$ vs. the crystalline mass fraction.  }
    \label{plots1}
  \end{center}
 \end{figure}

 \begin{figure}
 \begin{center}
    \begin{tabular}{cc}                 
     \resizebox{80mm}{!}{\includegraphics[angle=0]{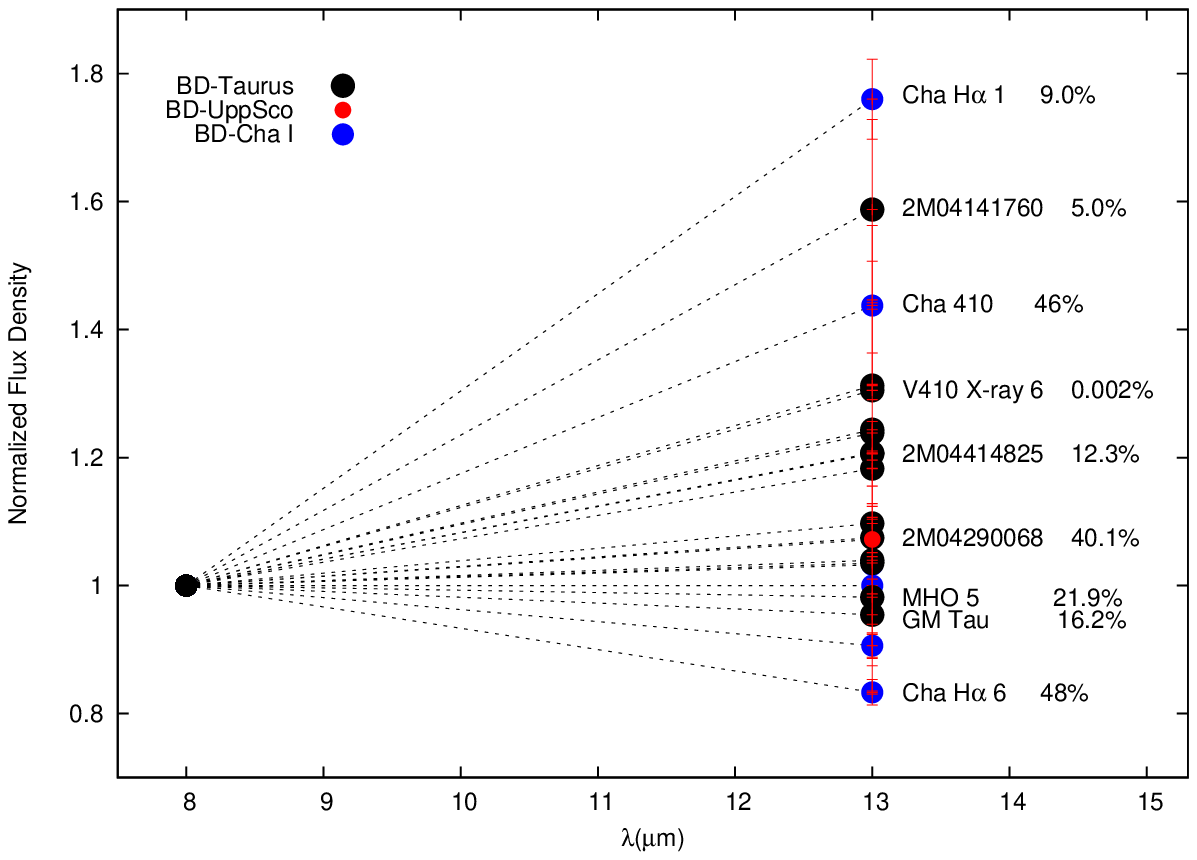}} &
     \resizebox{80mm}{!}{\includegraphics[angle=0]{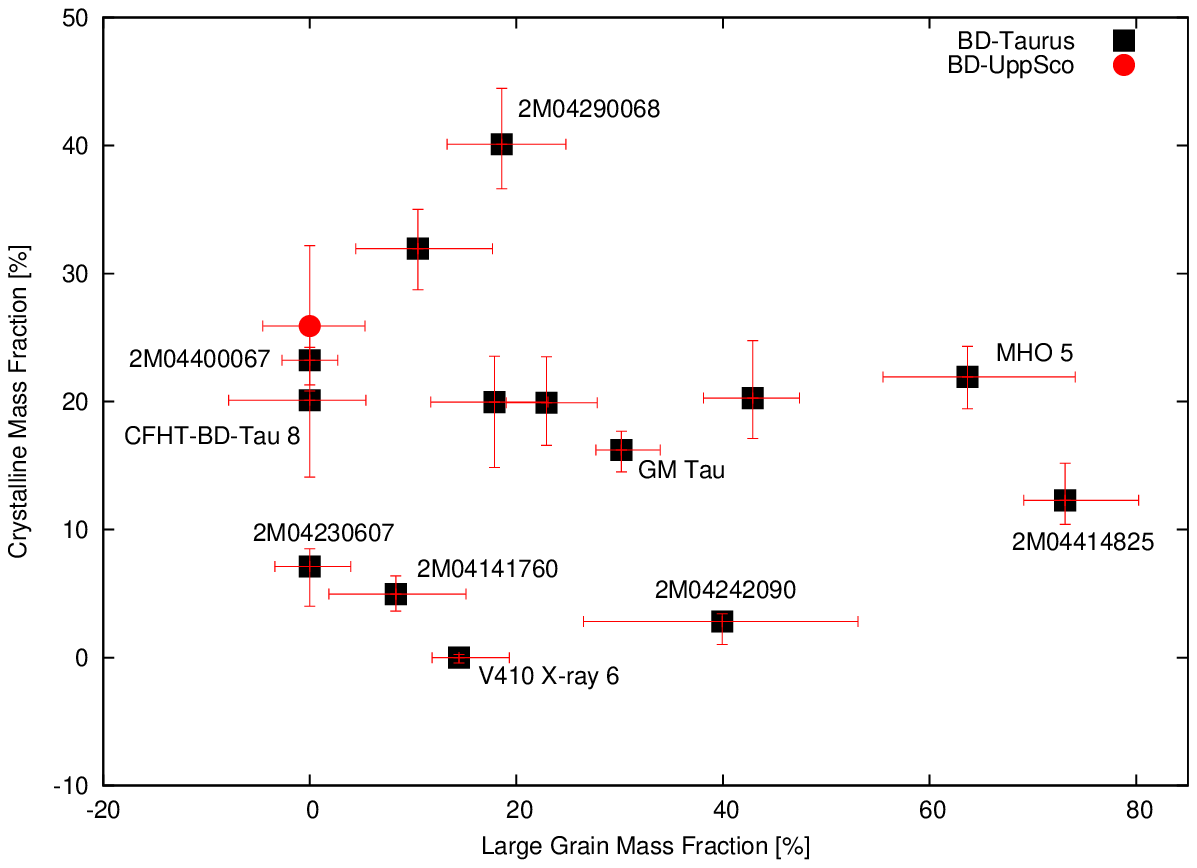}}   \\    
      \resizebox{80mm}{!}{\includegraphics[angle=0]{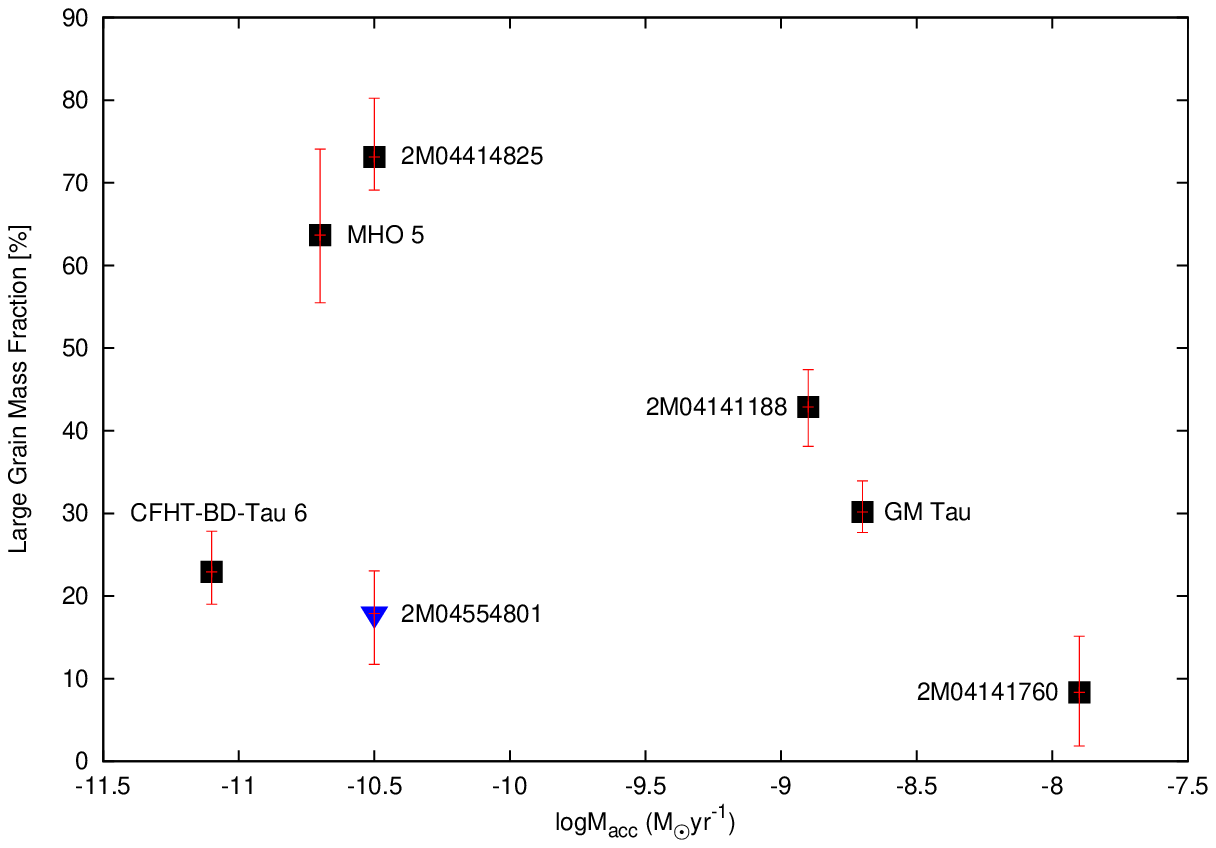}} &
     \resizebox{80mm}{!}{\includegraphics[angle=0]{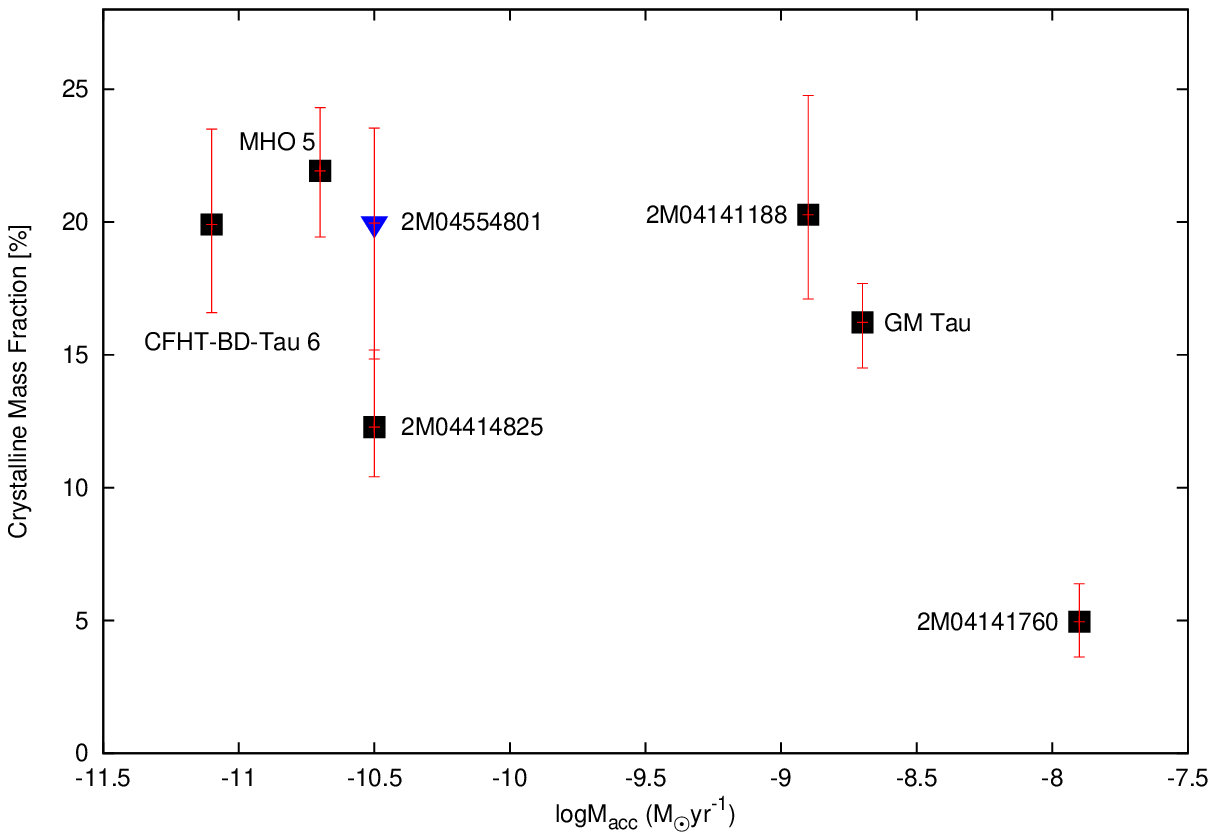}}   \\  
      \resizebox{80mm}{!}{\includegraphics[angle=0]{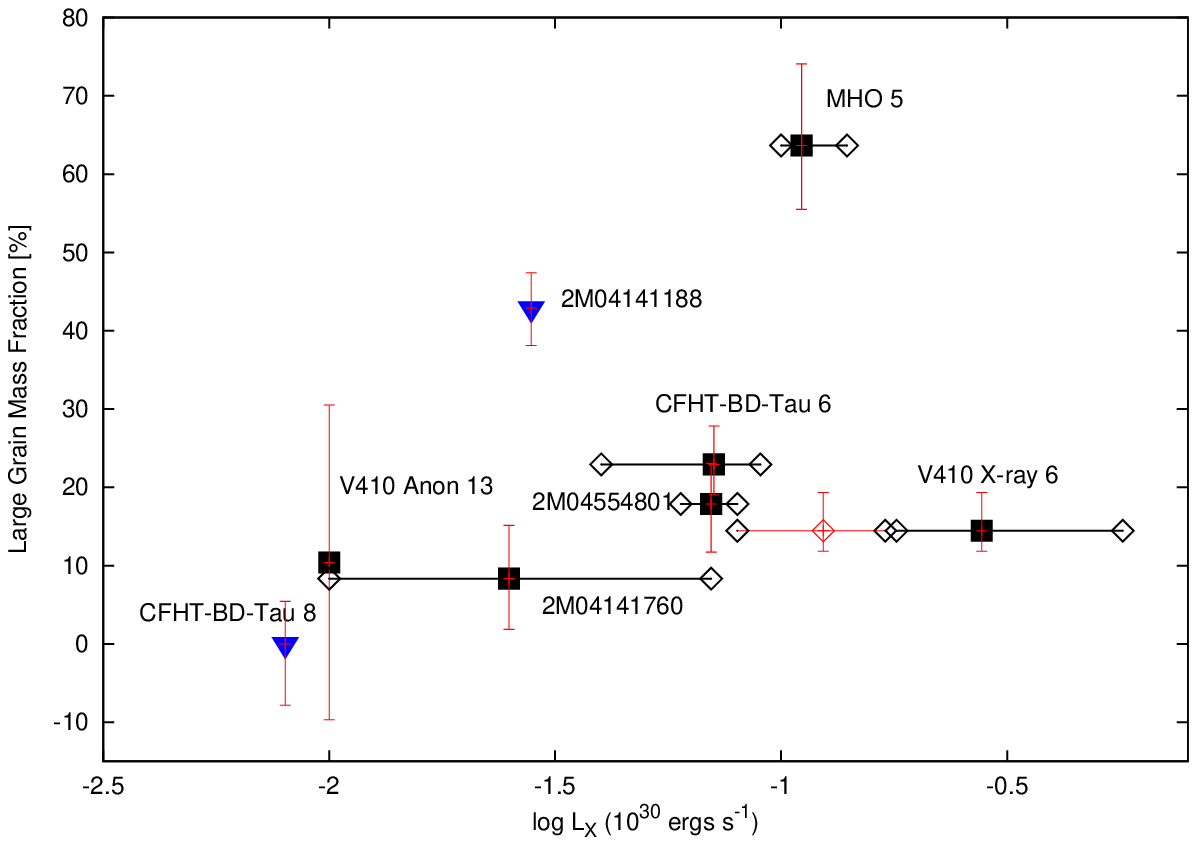}} &
     \resizebox{80mm}{!}{\includegraphics[angle=0]{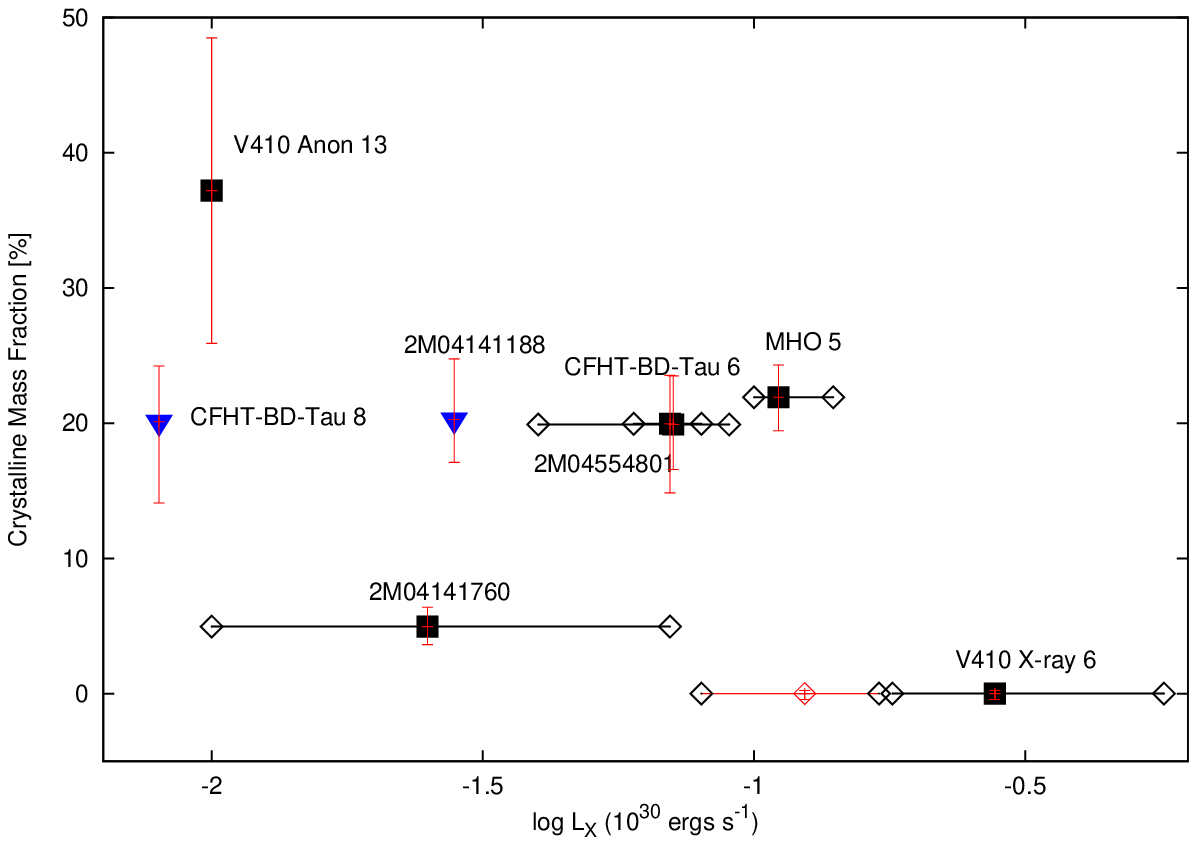}}   \\                       
    \end{tabular}
    \caption{ {\it Top, left}: (a) Disk geometry for the Taurus and UppSco brown dwarf disks. Also plotted are Cha I brown dwarfs (blue diamonds). The values next to the target names are the crystalline mass fractions. {\it Top, right}: (b) Large grain vs. the crystalline mass fractions. {\it Middle, left}: (c) Disk mass accretion rate vs. the large grain mass fraction. Upper limits are indicated by blue arrowheads. The accretion rates have been obtained from Muzerolle et al. (2003; 2005). {\it Middle, right}: (d) Disk mass accretion rate vs. the crystalline mass fraction.  {\it Bottom, left}: (e) The X-ray luminosity plotted against the large-grain mass fractions. Red open diamond denotes the quiescent state for V410 X-ray 6. Black open diamonds mark the range in X-ray emission. Upper limits are indicated by blue arrowheads. {\it Bottom, right}: (f) X-ray luminosity versus the crystalline mass fractions.     }
    \label{plots2}
  \end{center}
 \end{figure} 
 
  \begin{figure}
 \begin{center}
    \begin{tabular}{cc}                 
     \resizebox{80mm}{!}{\includegraphics[angle=0]{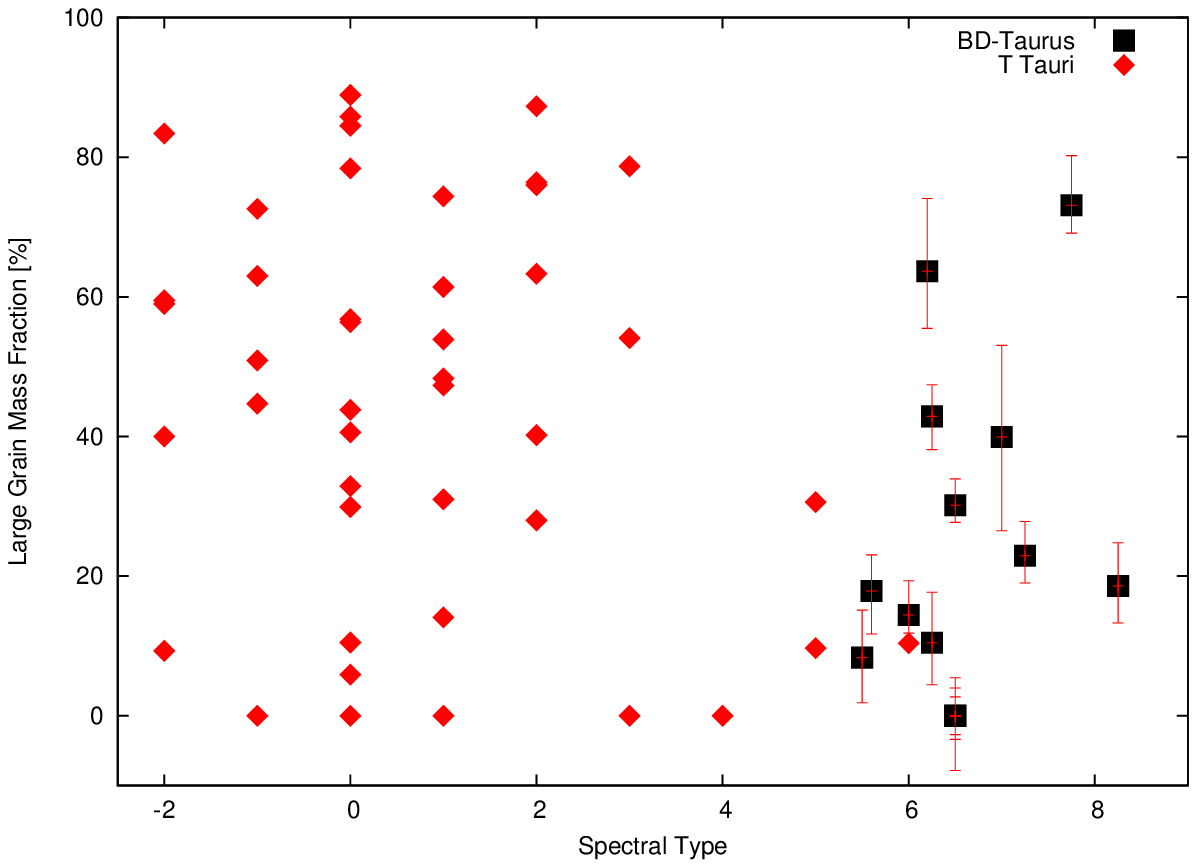}} &
     \resizebox{80mm}{!}{\includegraphics[angle=0]{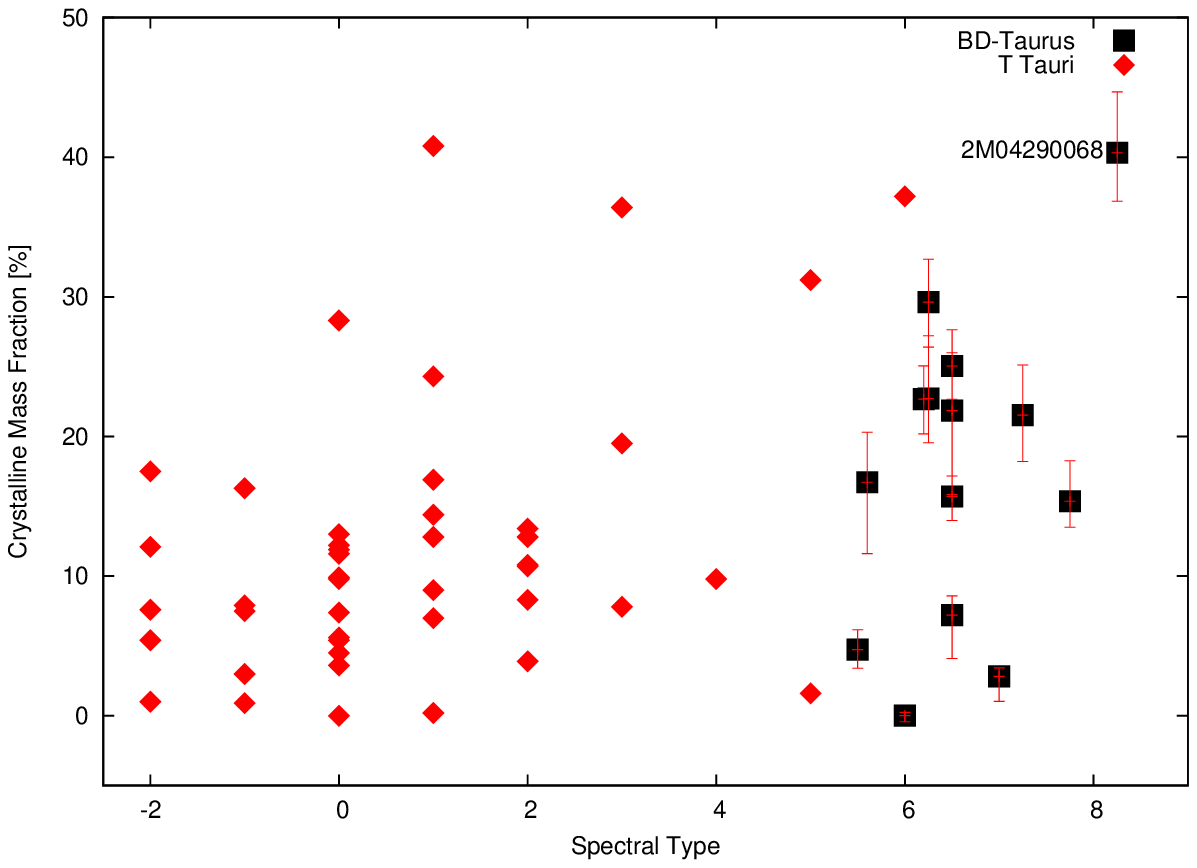}}   \\   
         \end{tabular}
    \caption{ {\it Left}: (a) SpT vs. the large grain mass fractions for T Tauri stars and brown dwarfs in Taurus. The value of -2 indicates a SpT of K7, -1 is K5, while 0-9 are M0-M9. {\it Right}: (b) SpT vs. the crystalline mass fractions. }
    \label{plots3}
  \end{center}
 \end{figure}

\end{document}